\documentclass[reprint,amsmath,amssymb,aps,twocolumn,prb]{revtex4-2}

\usepackage{graphicx}
\usepackage{babel}
\usepackage{color}
\usepackage{soul}

\begin{document}

\title{Creating a three dimensional intrinsic electric dipole on rotated CrI$_3$ bilayers}

\author{Shiva~P.~Poudel,$^1$ Juan~M.~Marmolejo-Tejada,$^2$ Joseph E. Roll,$^1$ Mart\'in A. Mosquera,$^2$ and Salvador Barraza-Lopez$^1$}
\affiliation{1. Department of Physics, University of Arkansas, Fayetteville, AR 72701, USA and MonArk NSF Quantum Foundry, University of Arkansas, Fayetteville, AR 72701, USA\\
2. Department of Chemistry and Biochemistry, Montana State University, Bozeman, MT 59717 USA and MonArk NSF Quantum Foundry, Montana State University, Bozeman, MT 59717 USA}

\date{\today}

\begin{abstract}
Two-dimensional (2D) materials are being explored as a novel multiferroic platform. One of the most studied magnetoelectric multiferroic 2D materials are antiferromagnetically-coupled (AFM) CrI$_3$ bilayers. Neglecting magnetism, those bilayers possess a crystalline point of inversion, which is only removed by the antiparallel spin configuration among its two constituent monolayers. The resultant intrinsic electric dipole on those bilayers has a magnitude no larger than 0.04 pC/m, it points out-of-plane, and it reverts direction when the--Ising-like--cromium spins are flipped (toward opposite layers {\em versus} away from opposite layers). The combined presence of antiferromagnetism and a weak intrinsic electric dipole makes this material a two-dimensional magnetoelectric multiferroic. Here, we remove the crystalline center of inversion of the bilayer by a relative $60^{\circ}$ rotation of its constituent monolayers. This process {\em enhances} the out-of-plane intrinsic electric dipole tenfold with respect to its magnitude in the non-rotated AFM bilayer and also creates an even stronger and switchable in-plane intrinsic electric dipole. The ability to create a three-dimensional electric dipole is important, because it enhances the magnetoelectric coupling on this experimentally accessible 2D material, {which is explicitly calculated here as well.}
\end{abstract}

\maketitle

\section{\label{sec:level1}Introduction}
{\em Ferroicity} is a symmetry-breaking quantum phenomenon \cite{Curie} that gives rise to domain physics. Magnetoelectric multiferroics \cite{Smolenskii_1982,fiebig2013magnetoelectric} lack spatial and time inversion symmetries \cite{Landau,Astrov,Kimura2003,Kimura2005,rev1,rev2,NicolaFewMultiferroic,NicolaMultiferroic2,Ramesh2011}, and the experimental exfoliation of an atomically-thin material (graphene) \cite{Graphene1,Graphene2,Graphene3} is now evolving into a field aiming to deliver advanced functional multiferroic materials at the atomic-thickness limit.

Effects from coupled magnetic and electric fields have been studied for over a century: R\"ontgen recorded magnetic fields when dielectric objects move within a constant electric field \cite{Rontgen}, Pierre Curie concluded that magnetoelectric effects are due to broken symmetries \cite{Curie}, and Debye coined the term ``magnetoelectric directional effect'' \cite{Debye}. The first multiferroic material was grown in 1959 \cite{Smolenskii}, and the demonstration of intrinsic magnetoelectric coupling in antiferromagnetically-coupled Cr$_2$O$_3$ was rationalized by Dzyaloshinskii \cite{Dzyaloshinskii}, and experimentally demonstrated by Astrov \cite{Astrov}. The main crystallographic families for bulk multiferroics \cite{Kimura2003,BFO_Neaton,rev2,NicolaFewMultiferroic,Rabe_1,Kimura2005,
SpaldinFiebigScience2005,NicolaMultiferroic2,Tsymbal2006,Fennie,Dagotto2006,Picozzi2007,
DagottoBook,PhysRevLett.107.067601,Bibes2011,TsymbalJunctions2012,Bellaiche2014,Spaldin2019,Ramesh2020,TopoPeroviskite2022,2022paper,Xiaofan} are perovskites \cite{Schmid_1973,Smolenskii_1982}, compounds with hexagonal structure \cite{Sugie}, boracites \cite{SCHMID1965327}, and BaMF$_4$ compounds \cite{Scott1980,PhysRevB.2.754}.

Initial forays into multiferroics within the 2D materials community are studies of quantum paraelastic/ferromagnetic SnO monolayers \cite{seixas,Tyler,Alejandro}, and of ferroelastic/ferroelectric (experimentally available) group-IV monochalcogenide monolayers \cite{Mehboudi,Qian,Kai2016,Barraza2018,RMP2021}. Magnetoelectric couplings induced by the proximal placement of ferroelectric and ferromagnetic 2D materials have been discussed, too \cite{EMproximal1,EMproximal2,EMproximal3,HeterostructuresToo}, along with theoretical and experimental reports on the change of magnetic ordering by charge doping \cite{MagneticOrderByDoping,EdaTransfer}, and proposals for atomistic defects as sources of multiferroic behavior \cite{2DMultiferroicsReview}. Analytical models exploring magnetoelectric couplings in 2D materials exist as well \cite{review2Dmultiferroics,2DanalyticalModel_2}.  Notably, layered multiferroic NiI$_2$ was experimentally demonstrated recently \cite{NiI2,Picozzi}.

{A recent and comprehensive study of ferroelectricity on bilayers based on an analysis of {\em non-magnetic groups} appeared recently \cite{New}.} Nevertheless, being one of the most studied magnetoelectric bilayer systems, the electric control of magnetization on AFM-coupled CrI$_3$ bilayers \cite{CrI3_bilayer1,CrI3_bilayer2,CrI3_bilayer3,MakRalphReview_2019,SuarezMorell_2019} {(this is, their {\em magnetoelectric coupling}) was experimentally demonstrated back in 2019 through second harmonic generation \cite{Sun2019}.} The coupling relies on the broken inversion symmetry induced by the alternating ({\em Ising}-like) out-of-plane magnetization at opposite monolayers \cite{sivadasKerr2016,Sun2019}{, which is furnished when considering the full magnetic group}. The magnetic ordering of monolayers \cite{YiAnYan} and bilayers \cite{Leon_2020} can also be tuned by uniaxial strain.

Here, we optimize the intrinsic electric dipole moment $\mathbf{P}$ of a CrI$_3$ bilayer by a relative $60^{\circ}$ rotation among layers (a concept recently applied to MnSe bilayers \cite{MnSe}). This procedure results on a ten-fold increase of the out of plane intrinsic electric dipole and it gives rise to an even larger in-plane intrinsic electric dipole {\cite{New}}. This effect is unlike what is reported on MnSe bilayers, which {\em only develop an out-of-plane intrinsic electric dipole} \cite{MnSe}. The three-dimensional intrinsic polarization {\em enhances} the magnetoelectric coupling of CrI$_3$ bilayers{, which is calculated here explicitly}.

The manuscript is organized as follows: The numerical methods are described in Sec.~\ref{sec:methods}. The results and discussion are provided in Sec.~\ref{sec:results}. Conclusions can be found in Sec.~\ref{sec:Conc}.

\section{Methods}\label{sec:methods}
This study is based on density functional theory with the PBE approximation for exchange-correlation \cite{pbe} as implemented in the VASP package \cite{vasp-1,vasp-2,vasp-3}. The implementation of spin-orbit-coupling is inconsistent with the use of {\em ab initio} van der Waals corrections on this code, and empirical van der Waals corrections due to Grimme \cite{df3-grimme} were utilized for that reason (see Fig.~S1 and Table SI in the Supplementary Material \cite{footnotes} for an extended discussion of structure {\em versus} exchange-correlation functionals). Structural optimizations were performed using a $15\times 15\times 1$ $k-$point mesh. We used an energy cutoff of 500 eV for the expansion of electronic states in terms of plane waves, an energy convergence criteria of $10^{-6}$ eV, and a force convergence limit of $10^{-4}$ eV/\AA{}. Structural optimizations were performed in configurations whereby the angle among in-plane lattice vectors $\mathbf{a}_1$ and $\mathbf{a}_2$ was kept fixed at 120$^{\circ}$. All calculations included dipole corrections, and the lattice constant parallel to the two-dimensional material's normal was set to $\mathbf{a}_3=50$ \AA{} to further reduce out-of-plane interactions among periodic copies (see Fig.~S2(a) in Ref.~\cite{footnotes} for an energy convergence test {\em versus} $a_3$). In order to treat the strong on-site Coulomb interaction of Cr \textit{d} orbitals, we used the DFT+U method introduced by Dudarev {\em et al.}~\cite{Dudarev}, with an effective value of $U$ equal to 3 eV ($J=0$) \cite{DFT-xc-analysis}. Structural optimizations were originally carried without spin-orbit coupling (SOC); SOC was turned on for structures at local minima, and to compute one-dimensional cuts of the energy landscape.

Having a well-converged electronic density is crucial to determine the intrinsic electric dipole moment $\mathbf{P}$, {which was calculated using the Berry phase theory of polarization} \cite{vanderbilt}. {$\mathbf{P}$} remained unchanged on a denser, $71\times 71\times 1$ $k-$point mesh. In addition, the convergence of $\mathbf{P}$ against the magnitude of $\mathbf{a}_3$ was guaranteed as well (see Fig.~S2(b) for a test of $P_z$ {\em versus} $a_3$ \cite{footnotes}).

Phonon dispersion calculations were performed with the PHONOPY tool \cite{phonopy} on a $3\times3\times1$ supercell. Atomic displacements were set at 0.005 $ \AA{}$. In those calculations, the $k-$point grid was set to $5\times5\times1$. Cutoff energy and energy convergence criteria remained at 500 eV and $10^{-4}$ eV/$ \AA{}$, respectively. The force constant (Hessian) matrix at the $\Gamma-$point was obtained as well; eigenvalues and eigenvectors of this Hessian were utilized to calculate the magnetoelectric coupling tensor $\alpha$ using the method developed by \'I\~niguez \cite{Jorge}.

\section{Results and discussion}\label{sec:results}
\subsection{AB and AB' CrI$_3$ bilayers}

\begin{figure*}[tb]
\begin{center}
\includegraphics[width=0.96\textwidth]{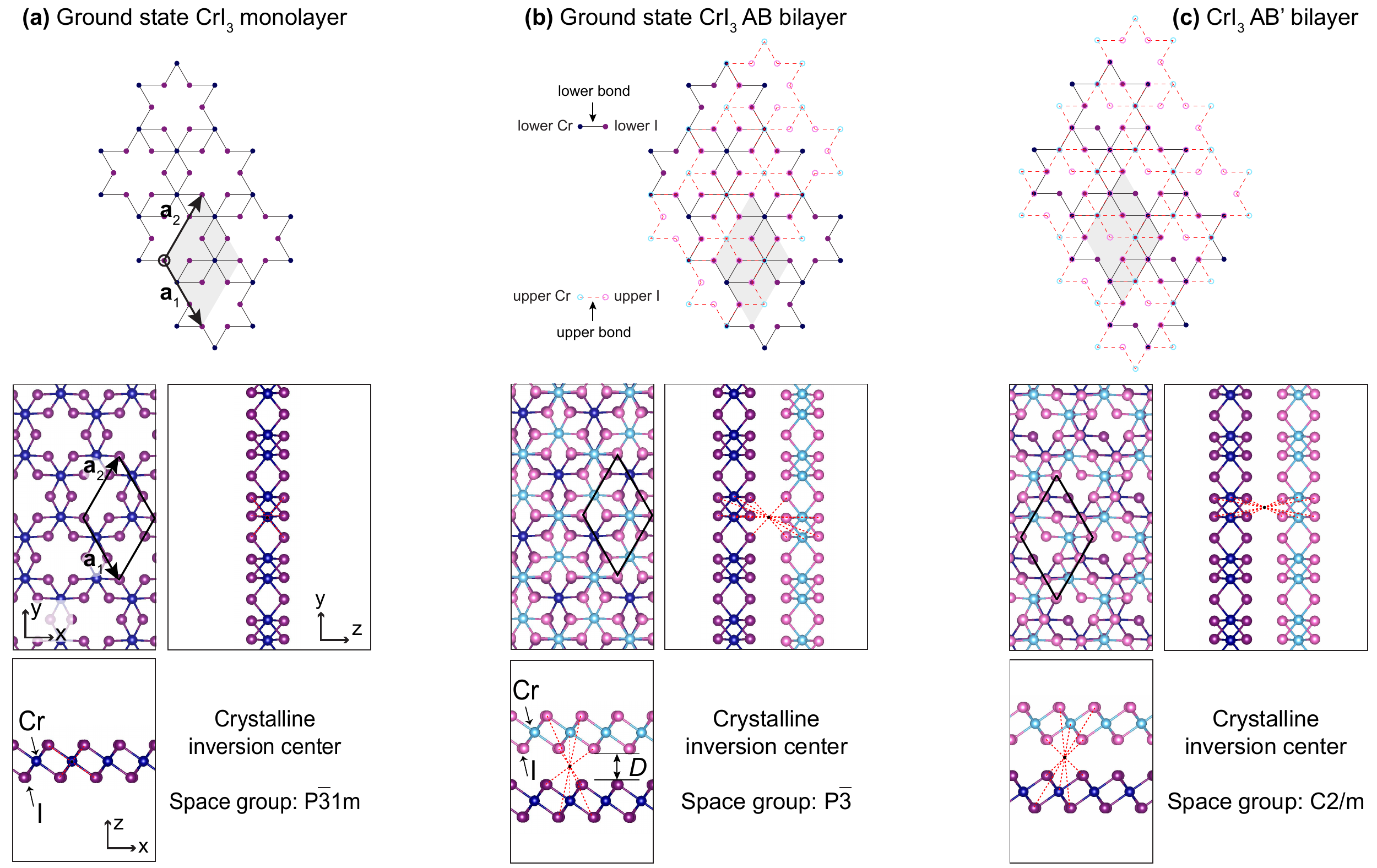}
\end{center}
\caption{(a) CrI$_3$ monolayer. (b) CrI$_3$ bilayer on its AB FM ($uuuu$) configuration prior to the relative rotation: FM coupling prevents the removal of a center of inversion and the development of a net intrinsic electric dipole; see Table \ref{table:t1}. (c) CrI$_3$ bilayer with AB'-stacking. The crystalline point of inversion--not considering the magnetic configuration--is shown.\label{fig:f1}}
\end{figure*}

The CrI$_3$ monolayers \cite{Huang2017,CrI3_bilayer1}  depicted on Fig.~\ref{fig:f1}(a) are one of many available two-dimensional ferromagnets and antiferromagnets \cite{other2Dmagnets_0,other2Dmagnets_1,other2Dmagnets_2,other2Dmagnets_2_2,other2Dmagnets_3,other2Dmagnets_4}. Crucial for an eventual creation of an in-plane electric dipole moment are the voids within star-of-David patterns on this structure. No such voids are present on hexagonal boron nitride nor transition metal dichalcogenide monolayers, for which no in-plane intrinsic electric dipoles occur on their (non-moir\'e, rotated by $60^{\circ}$) ferroelectric bilayer configurations \cite{hBNFerro0,hBNFerro1,hBNFerro2,hBNFerro2,TMDCFerro0,TMDCFerro1,FerroelectricWSe2,Juan}. This monolayer belongs to the crystalline space group $P\overline{3}1m$ (space group 162) \cite{sivadas}. (A listing of all its potential magnetic groups can be found in Table SII \cite{footnotes}.)

When analyzing CrI$_3$  bilayers, previous studies explore two stable structures (labeled AB and AB') that are formed by relative translations between layers \cite{CrI3_bilayer1,CrI3_bilayer2,CrI3_bilayer3,Leon_2020,sivadas,CrI3-DFT-stacking}:  Starting with an AA structure in which the two monolayers sit atop one another, translating the top layer by $2\mathbf{a}_1/3+\mathbf{a}_2/3$ yields the AB structure displayed in Fig.~\ref{fig:f1}(b). Not considering magnetism, it belongs to space group $P\overline{3}$ (space group 147). (Possible magnetic groups are listed in Table SII \cite{footnotes}.) If one translates the lower monolayer originally set on the AB conformation by $\mathbf{a}_1/3$, the resultant structure is AB' {[Fig.~\ref{fig:f1}(c)]}, which belongs to the space group $C2/m$ (12) \cite{sivadas} without the inclusion of magnetism (see Table SII for potential magnetic groups \cite{footnotes}).

The AB bilayer is observed in low-temperature samples ($T< 210$ K) \cite{R_CrI3_bilayer} and has the magnetic spins of Cr atoms ferromagnetically (FM) aligned in an Ising-like (out-of-plane) configuration. To set ideas, each monolayer has two Cr atoms on its unit cell (navy blue dots on Fig.~\ref{fig:f1}), and each Cr atom contributes a spin magnetic dipole moment $\mu$ of about 3 Bohr magnetons ($\mu_B=9.274\times 10^{-24}$ J/T) along the $z-$axis. Without loss of generality, we set each of the four Cr atoms on the unit cell (u.c.) to $\boldsymbol{\mu}=(0,0,+3\mu_B$), and label such FM setting $uuuu$ (for up-up-up-up), where the first two labels refer to the Cr atoms in the lower monolayer and the remaining two labels indicate the direction of spin on the two atoms at the upper monolayer. A displacement by either $\mathbf{a}_1/3$ or $\mathbf{a}_2/3$ takes the structure shown into the widely studied AB' configuration \cite{doi:10.1021/cm504242t,Leon_2020,sivadas,CrI3-DFT-stacking,DFT-xc-analysis,Sun2019,MakRalphReview_2019,Huang2017}, with a preferred AFM configuration which is explicitly labeled $uudd$ and in which Cr atoms at each monolayer have out-of-plane magnetic dipoles pointing toward the opposite monolayer ($d$ stands for spin ``down'' in this context). Without considering their magnetic moment, antiferromagnetically coupled (AB') CrI$_3$ bilayers stem from a bulk monoclinic phase and belong to space group $C2/m$ (space group 12). All possible magnetic groups with an AB' configuration are listed in Table SII \cite{footnotes}.

The lack of a center of inversion on AFM AB' bilayers was experimentally verified through second harmonic generation and it is driven by their magnetic configuration \cite{Sun2019,MakRalphReview_2019,Huang2017}. In other words, {its non-zero magnetoelectric coupling \cite{Sun2019}} originates from the removal of a center of inversion by an antiparallel {\em spin}. Theory indicates a FM coupling among CrI$_3$ monolayers in the AB (ground state) configuration \cite{Leon_2020,sivadas,CrI3-DFT-stacking,DFT-xc-analysis}, which is inconsistent with experiment, where an AFM coupling is observed. That theoretical result is independent of the exchange-correlation functional, the van der Waals correction scheme, and the Hubbard (U) term on the Cr atoms with and without spin-orbit coupling (SOC) \cite{DFT-xc-analysis}.  The ground state structure at bilayer thickness is still a matter of debate--it has been suggested that the AFM AB' structure is created at room temperature and ``trapped'' by the extra layers used to create experimental devices \cite{Leon_2020,sivadas,CrI3-DFT-stacking}. Nevertheless, switching among AFM and FM magnetic configurations can be experimentally performed at will \cite{CrI3_bilayer1,CrI3_bilayer2,CrI3_bilayer3}.

\subsection{Rotated CrI$_3$ bilayers ($s_1$, $s_{2,1}$, and $s_{2,2}$)}
While the electric control of magnetism in CrI$_3$ bilayers reported in the recent past \cite{CrI3_bilayer1,CrI3_bilayer2,CrI3_bilayer3} relies on an external electric field, one precept of this work is to engineer an intrinsic electric dipole by a relative rotation \cite{New} to enhance the {\em multiferroic} properties of those bilayers; such rotations are a new handle for magnetoelectric multiferroic behavior that is not available on non-layered multiferroics \cite{Kimura2003,BFO_Neaton,rev2,NicolaFewMultiferroic,Rabe_1,Kimura2005,
SpaldinFiebigScience2005,NicolaMultiferroic2,Tsymbal2006,Fennie,Dagotto2006,Picozzi2007,
DagottoBook,PhysRevLett.107.067601,Bibes2011,TsymbalJunctions2012,Bellaiche2014,Spaldin2019,
Ramesh2020,TopoPeroviskite2022,2022paper,Xiaofan,Schmid_1973,Smolenskii_1982,Sugie,SCHMID1965327,Scott1980,PhysRevB.2.754}.

Following a concept introduced by Wu and collaborators \cite{MoS2Ferro,WTe2ferro} (and experimentally verified for hexagonal boron nitride and transition metal dichalcogenide bilayers \cite{hBNFerro0,hBNFerro1,hBNFerro2,hBNFerro2,TMDCFerro0,TMDCFerro1,FerroelectricWSe2}), we rotated the upper monolayer on the AB structure by $60^{\circ}$. This removes the bilayer's crystalline center of inversion, and induces an intrinsic electric dipole.  The rotated bilayer structure--which we labeled $s_1$ (for {\em structure 1})--is displayed on Fig.~\ref{fig:f1}(c) and it belongs to the crystalline space group $Cm$ (space group 8) (see Table SII for its possible magnetic groups). The group $Cm$ does not forbid the creation of an intrinsic {\em in-plane} electric dipole that is parallel to the mirror plane depicted on Fig.~\ref{fig:f1}(c) {\cite{New}}.

\begin{figure*}[tb]
\begin{center}
\includegraphics[width=0.96\textwidth]{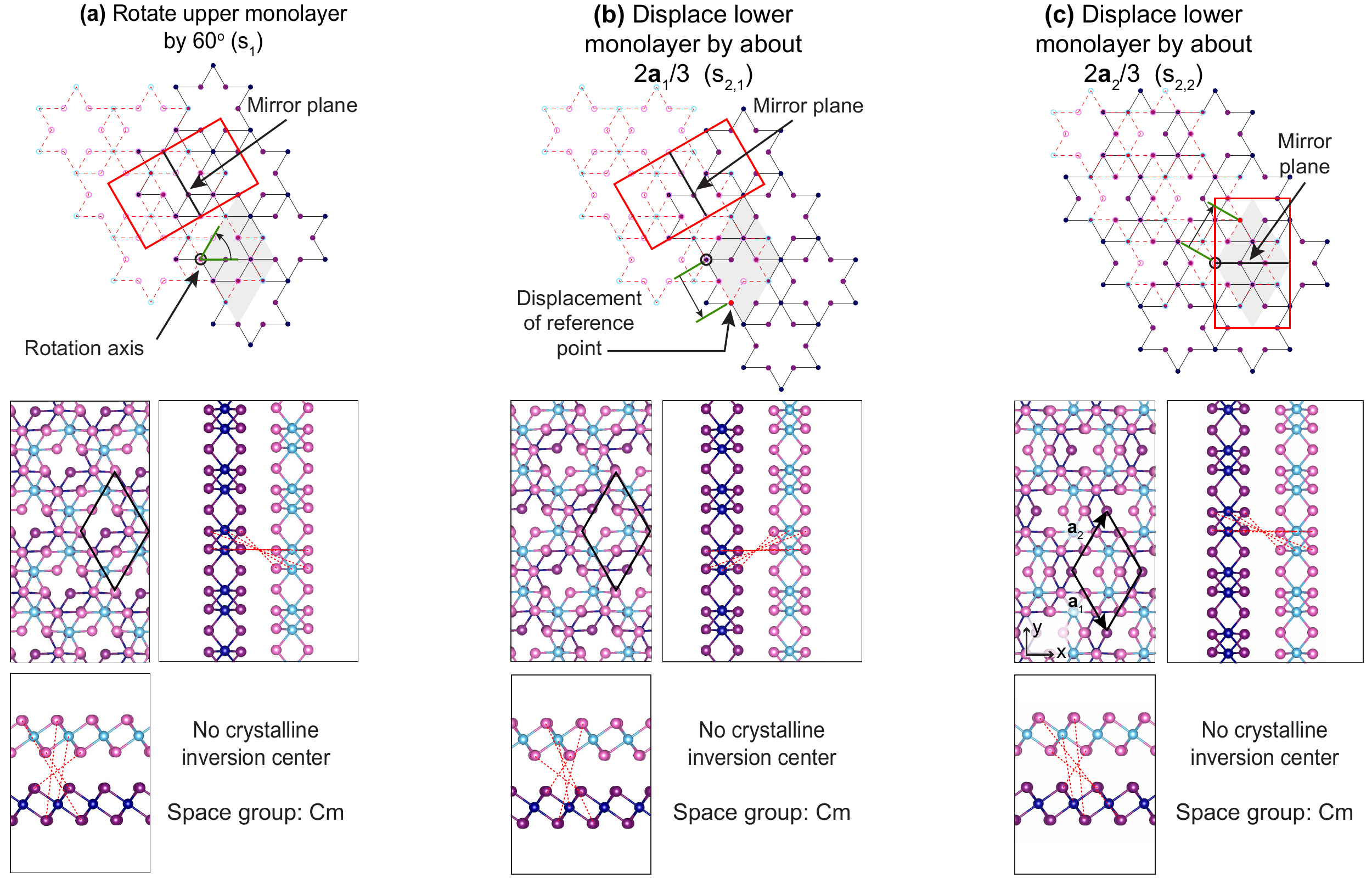}
\end{center}
\caption{(a) CrI$_3$ bilayer at a local minimum after a $60^{\circ}$ rotation with respect to the axis shown. This bilayer structure is dubbed $s_1$. The removal of a crystalline inversion center is made evident in the lowermost side views. $P_z$ can be reverted by sliding the lower monolayer by (c) about $2\mathbf{a}_1/3$ on a bilayer structure labelled $s_{2,1}$, or (d) by about $2\mathbf{a}_2/3$, on a structure labeled $s_{2,2}$. The different orientations of the mirror plane for structures $s_{2,1}$ and $s_{2,2}$ are highlighted.\label{fig:f1b}}
\end{figure*}

An additional displacement of the lower monolayer in bilayer $s_1$ [Fig.~\ref{fig:f1b}(a)] by about $2\mathbf{a}_1/3$ [as explicitly displayed on Fig.~\ref{fig:f1b}(b)] or by about $2\mathbf{a}_2/3$ {[Fig.~\ref{fig:f1b}(c)]} still results on a structure belonging to crystalline space group $Cm$ (8). We label $s_{2,1}$ the structure seen on Fig.~\ref{fig:f1b}(b)--in which the lower monolayer was originally displaced by $2\mathbf{a}_1/3$--to tell it apart from the bilayer $s_{2,2}$ shown in Fig.~\ref{fig:f1b}(c), in which the lower monolayer was initially displaced by about $2\mathbf{a}_2/3$ instead.

A comparison among Figs.~\ref{fig:f1b}(a) and \ref{fig:f1b}(b) shows that the mirror plane of bilayer $s_{2,1}$ remains parallel to that seen on bilayer $s_1$. Thus, besides the natural creation of an out-of-plane intrinsic electric polarization by the $60^{\circ}$ relative rotation, this mirror plane allows for a non-zero {\em in-plane} intrinsic electric dipole parallel to $\mathbf{a}_1$ on bilayer $s_{2,1}$ as well. Nevertheless, the mirror plane turns out to be parallel to the $x-$axis (this is, parallel to $\mathbf{a}_1+\mathbf{a}_2$) for the $s_{2,2}$ bilayer [Fig.~\ref{fig:f1b}(c)], allowing for a possible in-plane intrinsic electric dipole along the $x-$axis on the $s_{2,2}$ bilayer. This way, the rotated CrI$_3$ bilayers may develop an in-plane component of the intrinsic electric dipole that can be controlled by sliding. 
The creation of an {\em in-plane} intrinsic electric dipole through a relative $60^{\circ}$ rotation among monolayers lends additional handles for the engineering of magnetoelectric couplings by geometrical/mechanical means.

\begin{figure}[tb]
\begin{center}
\includegraphics[width=0.48\textwidth]{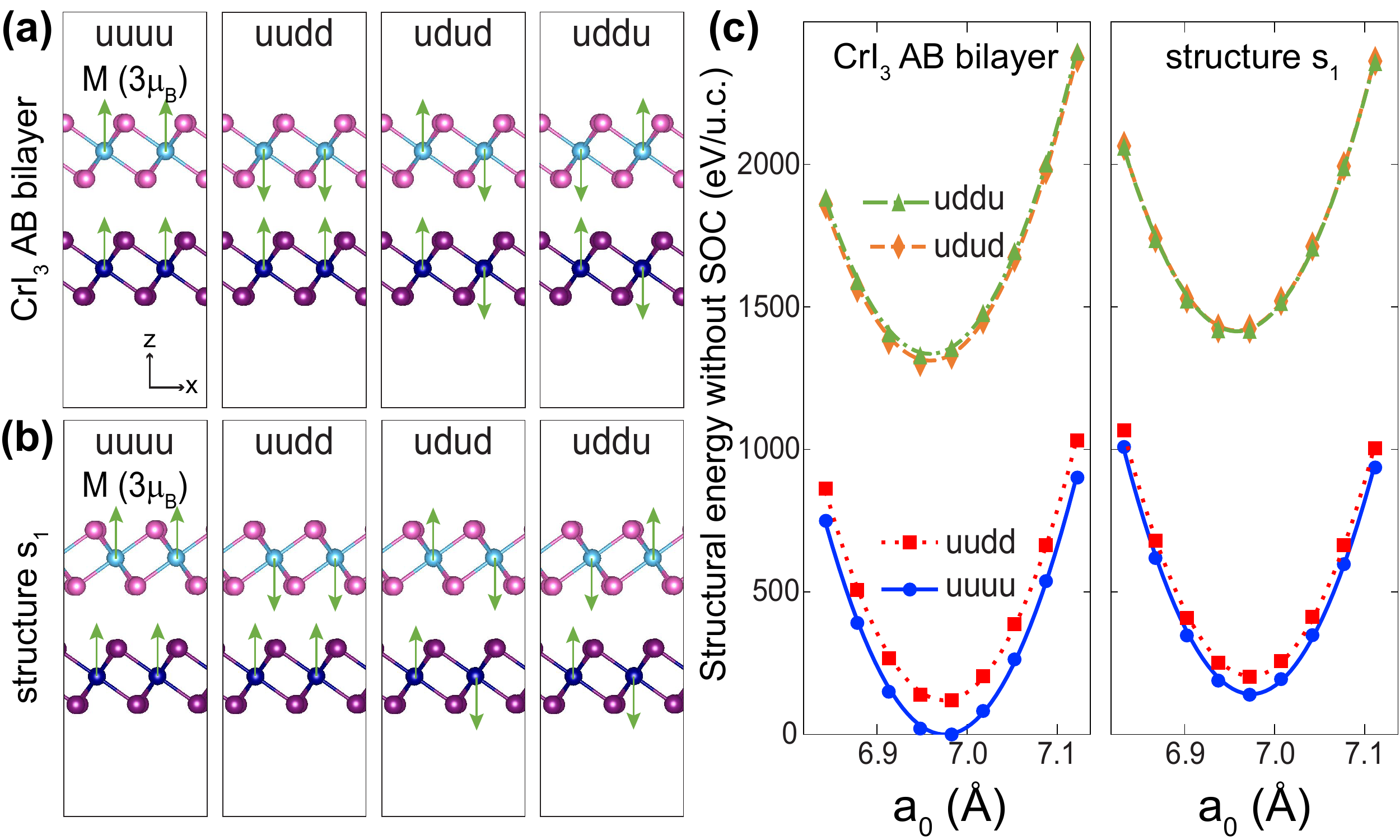}
\end{center}
\caption{Side views illustrating four possible (Ising-like) magnetic configurations for (a) the AB CrI$_3$ bilayer and for (b) the rotated bilayer on the $s_1$ configuration. The FM configuration is labeled $uuuu$, and three AFM configurations ($uudd$, $udud$, and $uddu$) were considered as well. (c) Total structural energy around global (AB) and local ($s_1$) minima for CrI$_3$ bilayers as a function of the lattice parameter $a_0$ and magnetic configuration, without SOC. The $s_1$ local minima on the FM configuration sits 139 K/u.c.~above the global minima, while the AFM $s_1$ structure with a $uudd$ magnetic configuration sits 202 K/u.c.~above that global minima (see Table SIII for numerical details \cite{footnotes}).\label{fig:f2}}
\end{figure}

Fig.~\ref{fig:f2} shows the total energy and the global minima for the AB and (rotated) $s_1$ FM bilayers, and for three additional AFM configurations ($uudd$, $udud$, and $uddu$) for both AB and $s_1$ structures. The energetics of the AB bilayer are consistent with previous results \cite{sivadas,Leon_2020}. The numerical results reported on Fig.~\ref{fig:f2} were obtained without SOC. (Electronic band structures for AB and $s_1$ bilayers show similar band gaps; those can be found in Fig.~S3 \cite{footnotes}.) Note that--though the $s_{1}$ structure sits higher in energy than the AB bilayer on Fig.~\ref{fig:f2}(c)--``falling back'' into the AB structure requires a macroscopic rotation of the entire upper layer in unison, an event that is statistically rare. In other words, structure $s_1$ sits at a stable local energy minimum.

\begin{figure*}[tb]
\begin{center}
\includegraphics[width=0.96\textwidth]{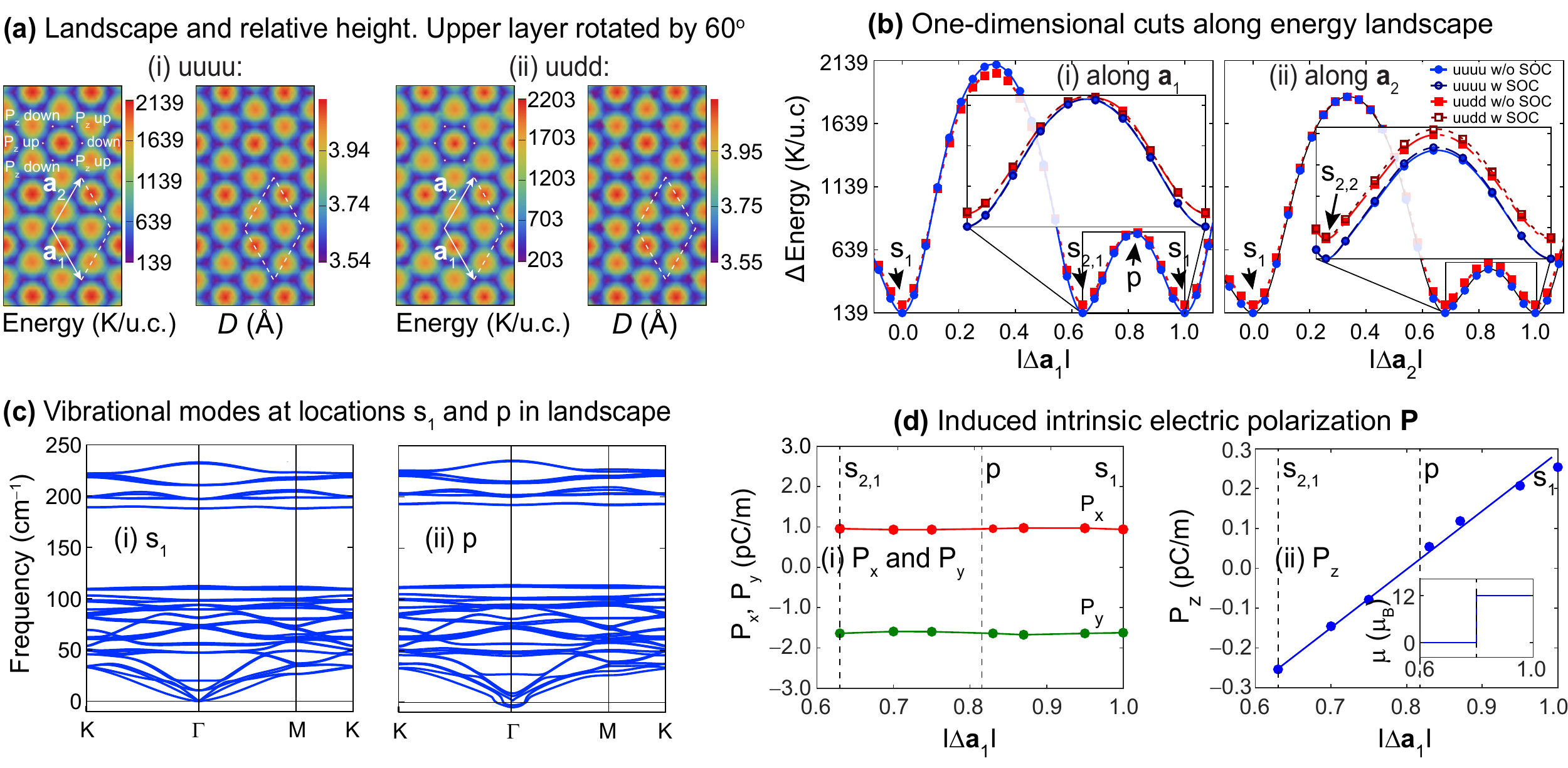}
\end{center}
\caption{(a) Energy landscape as a function of sliding among CrI$_3$ monolayers for FM ($uuuu$) and AFM ($uudd$) magnetic configurations (without SOC), and corresponding optimal vertical separation among monolayers $D$ (see Fig.~\ref{fig:f1}). Structure $s_1$ sits at the point from which lattice vectors $\mathbf{a}_1$ and $\mathbf{a}_2$ join, and it is periodic upon lattice translations. Structure $s_{2,1}$ ($s_{2,2}$) sits $0.6389\mathbf{a}_1$ ($0.6806\mathbf{a}_2$) away from $s_1$ locations. (b) One-dimensional cuts of the energy landscape along (i) $\mathbf{a}_1$ and (ii) $\mathbf{a}_2$ with and without SOC turned on. (c) Structural stability of $s_1$ bilayer, and vibrational spectra at saddle point $p$ displaying a few imaginary modes. (d) Induced intrinsic three-dimensional electric dipole along the straight path joining structures $s_1$ and $s_{2,1}$. The inset shows the change in magnetization at a nearly degenerate crossover among AFM and FM configurations at point $p$.\label{fig:f3}}
\end{figure*}

Fig.~\ref{fig:f3}(a) displays the energy landscape and the optimal relative height among monolayers upon relative sliding, on the bilayer configuration in which the upper monolayer was rotated by $60^{\circ}$. This was calculated for FM ($uuuu$) and AFM ($uudd$) magnetic configurations without SOC. The point at which lattice vectors $\mathbf{a}_1$ and $\mathbf{a}_2$ join in Fig.~\ref{fig:f3}(a) provides the energy and separation among monolayers for structure $s_1$ on Figs.~\ref{fig:f1}(c) and \ref{fig:f2}(c).

Working with a primitive u.c., the energy and structural sampling was carried out by displacing the upper layer by $n\mathbf{a}_1/12+m\mathbf{a}_2/12$--$n$ and $m$ ran from 0 to 11--with respect to structure $s_1$. So that bilayers do not fall back into local minima while computing the landscape, the atomic in-plane coordinates were kept fixed, and only the out-of-plane components were allowed to relax. The landscape and height profiles depicted upon multiple u.c.s on Fig.~\ref{fig:f3}(a) rely on periodicity, and the energies were expressed with respect to that obtained for the AB FM configuration (see Fig.~\ref{fig:f2} and Table SIII \cite{footnotes}).

A more detailed calculation was performed along one-dimensional cuts along $\mathbf{a}_1$ or $\mathbf{a}_2$ (Fig.~\ref{fig:f3}(b)). In those calculations, all atoms were allowed to move around the local minima, and SOC was turned on. Unlike the energy landscape of rotated hBN or transition metal dichalcogenide (TMDC) bilayers, which have identical maximum energies as the layers move either along the $\mathbf{a}_1$ or $\mathbf{a}_2$ axes (see, {\em e.g.}, Refs.~\cite{Juan} and \cite{landscapeGrapheneHBN}), the rotated CrI$_3$ bilayer displays barriers of dissimilar height, with the largest height assigned to a displacement along $\mathbf{a}_1$. {This anisotropy was not discussed in Ref.~\cite{New}.} The identical barriers observed on rotated hBN and TMDC bilayers \cite{Juan,landscapeGrapheneHBN} guarantees that the local minima be exactly at a distance $2\mathbf{a}_1/3$ or $2\mathbf{a}_2/3$. To the contrary, and as seen on Fig.~\ref{fig:f3}(b), the dissimilar heights of the barriers makes the $s_{2,1}$ structure to be located at 0.6389$\mathbf{a}_1$ instead, while the $s_{2,2}$ structure sits at 0.6806$\mathbf{a}_2$, and it justifies our choice of words (``about 2/3'') made earlier on.

Calculations in which SOC is turned on require copious additional computational resources, and those were only pursued for structures along the one-dimensional cuts of the landscape in Fig.~\ref{fig:f3}(b) for that reason. Surprisingly, SOC did not break the energy degeneracy of the minima $s_1$ and $s_{2,1}$ upon inversion of the magnetic moment significantly, as the total energy difference upon a magnetic swap from FM configurations $uuuu$ to $dddd$ was only 0.15 K/u.c. Calculations of magnetic anisotropy barriers can be found on Fig.~S4 \cite{footnotes}.

The structural stability of bilayer $s_1$ alluded to earlier is established with the vibrational spectra shown in Fig.~\ref{fig:f3}(c), which lacks imaginary modes. In addition, and in preparation for the calculation of the magnetoelectric coupling tensor, we also considered the bilayer structure situated at the (unstable) saddle point $p$ in the landscape (see imaginary modes), for which both FM and AFM magnetic configurations turn degenerate. Point $p$ is shown on Fig.~\ref{fig:f3}(b). Fig.~\ref{fig:f3}(d) shows $\mathbf{P}$ as the structure is turned from $s_{2,1}$ to $s_1$ along $\mathbf{a}_1$.  For comparison, the reported magnitudes of the out-of-plane intrinsic electric dipole are 1.0 pC/m for a MoS$_2$ bilayer \cite{MoS2Ferro}, 0.5 pC/m for a WSe$_2$ bilayer \cite{FerroelectricWSe2}, and 0.4 pC/m for a WTe$_2$ bilayer \cite{WTe2ferro,towards}.

\subsection{Three-dimensional intrinsic electric polarization of rotated CrI$_3$ bilayers}

While $P_z$ swaps sign as the bilayer evolves from structure $s_1$ into $s_{2,1}$ or $s_{2,2}$ as expected \cite{MoS2Ferro,WTe2ferro}, {\em there is an in-plane intrinsic electric dipole moment} in Fig.~\ref{fig:f3}(d). In the remainder of this work we will (i) justify this observation, (ii) argue that it can be switched, and (iii) calculate the magnetoelectric coupling of the rotated CrI$_3$ bilayer.

We start by reporting the magnitude of the intrinsic electric dipoles for the AB and AB' bilayers in Table \ref{table:t1}. For context, electric dipole moments as large as 5 pC/m have been recently reported experimentally~\cite{FerroelectricGrapheneBilayer2}.

\begin{table}[tb]
\begin{center}
\caption{Intrinsic electric dipole moments per u.c.~area for CrI$_3$ bilayers at AB and AB' configurations. FM configurations have all spins pointing up ($uuuu$) or down ($dddd$), while the studed AFM configurations were $uudd$ and $dduu$. Digits within parenthesis indicate the precision of calculations.} 
\label{table:t1}
\begin{tabular}{c|c}
\hline
\hline
Configuration     & $\mathbf{P}$ (pC/m) \\
\hline
AB FM ($uuuu$)      & $(-0.0(1),0.0(0),-0.0(1))$  \\
AB FM ($dddd$)      & $(-0.0(1),0.0(0),-0.0(1))$  \\
AB'$_1$ FM ($uuuu$) & $(-0.0(1),0.0(0),-0.0(1))$  \\ 
AB'$_1$ FM ($dddd$) & $(-0.0(1),0.0(0),-0.0(1))$  \\ 
AB'$_2$ FM ($uuuu$) & $(-0.0(1),0.0(0),-0.0(1))$  \\ 
AB'$_2$ FM ($dddd$) & $(-0.0(1),0.0(0),-0.0(1))$  \\ 
\hline
Configuration & $\mathbf{P}$ (pC/m) \\
\hline
AB AFM ($uudd$)    & $(-0.0(1),0.0(0),-0.0(3))$ \\
AB AFM ($dduu$)    & $(-0.0(1),0.0(0),+0.0(2))$ \\
AB'$_1$ AFM ($uudd$) & $(-0.0(1),0.0(0),-0.0(4))$ \\ 
AB'$_1$ AFM ($dduu$) & $(-0.0(1),0.0(0),+0.0(2))$ \\ 
AB'$_2$ AFM ($uudd$) & $(-0.0(1),0.0(0),-0.0(4))$ \\ 
AB'$_2$ AFM ($dduu$) & $(-0.0(1),0.0(1),+0.0(2))$ \\ 
\hline
\hline
\end{tabular}
\end{center}
\end{table}

Table \ref{table:t1} conveys the following information: (i) FM-coupled AB and AB' bilayers display negligible intrinsic electric dipole moments. Since there is a center of inversion in those structures, the reported $\mathbf{P}$ on for the FM-coupled structures conveys the numerical precision achieved here, which is of the order of $\pm 0.01$ pC/m.

On the other hand, the lack of a center of inversion facilitated by AFM magnetic configurations gives rise to an expected net $P_z$ \cite{Sun2019,MakRalphReview_2019,Huang2017} in Table \ref{table:t1}. In AFM structures, $P_z$ does not change sign in going from structure AB onto any of the two energy-degenerate structures AB', which are arrived at by a displacement of the upper monolayer by $\mathbf{a}_1/3$ (for a structure labelled AB'$_{1}$) or $\mathbf{a}_2/3$ (AB'$_2$). The two structures AB'$_1$ and AB'$_2$ were optimized independently, hence providing a cross-check of numerical results. In short, FM AB and AB' configurations have a center of inversion and do not develop an intrinsic electric dipole; AFM AB and AB' configurations, on the other hand, break inversion symmetry {due to their spin configuration} and give rise to a non-ionic out-of-plane net intrinsic electric dipole moment ($P_z$) that swaps sign upon spin reversal and regardless of the (AB or AB') atomistic configuration. {We are unaware of any previous computational study discussing the magnitude of the intrinsic electric dipole on experimentally available CrI$_3$ bilayer configurations as the one provided in Table \ref{table:t1}.}

The effectiveness of the relative $60^{\circ}$ rotation procedure in enhancing $P_z$ is demonstrated in Table \ref{table:t2}, where a ten-fold increase with respect to the experimentally available AB' AFM bilayer is registered. {Rounding up the reported values to 0.3 pm/C (a value close to the 0.2 pm/C reported in Ref.~\cite{CPL}), and dividing by $\simeq$12 \AA{}--which would correspond to the thickness of a u.c.~in a bulk 3D stack--one gets 0.025 $\mu$C/cm$^2$, which sits halfway among the values reported in Refs.~\cite{New} and \cite{PNAS21} and may be mostly due to the different exchange-correlation functionals employed in our works}. Furthermore, switching of $P_z$ is achieved by sliding from structure $s_1$ into either $s_{2,1}$ or $s_{2,2}$ \cite{MoS2Ferro,WTe2ferro}. A more remarkable result is the creation of {\em an even larger} in-plane intrinsic electric dipole like the one present on group-IV monochalcogenide monolayers \cite{Kai2016,doi:10.1021/acs.nanolett.0c02357,RMP2021}.

\begin{table}[tb]
\begin{center}
\caption{Intrinsic electric dipole moments per u.c.~area for CrI$_3$ bilayers at the $s_1$, $s_{2,1}$, and $s_{2,2}$ configurations. Digits within parenthesis indicate the precision of calculations. The removal of a center of inversion upon rotation makes $P_z$ to increase an order of magnitude with respect to its value on AFM AB and AB' configurations (Table \ref{table:t1}). In addition, $P_z$ does not change sign upon reversal of antiferromagnetic polarization ($uudd$ {\em versus} $dduu$), but only upon relative displacements among monolayers. A larger in-plane intrinsic electric dipole swapping direction at $s_{2,1}$ and $s_{2,2}$ is now created.}
\label{table:t2}
\begin{tabular}{c|c}
\hline
\hline
Configuration   & $\mathbf{P}$ (pC/m)  \\
\hline
$s_1$ FM ($uuuu$)    & $(+0.9(3),-1.6(0),+0.2(5))$ \\
$s_1$ FM ($dddd$)    & $(+0.9(3),-1.6(0),+0.3(0))$ \\
$s_{2,1}$ FM ($uuuu$) & $(+0.9(4),-1.6(0),-0.2(5))$ \\ 
$s_{2,1}$ FM ($dddd$) & $(+0.9(4),-1.6(1),-0.3(1))$ \\ 
$s_{2,2}$ FM ($uuuu$) & $(-1.8(4),+0.0(1),-0.2(5))$ \\ 
$s_{2,2}$ FM ($dddd$) & $(-1.8(7),+0.0(1),-0.3(0))$ \\ 
\hline
Configuration   & $\mathbf{P}$ (pC/m)  \\
\hline
$s_1$ AFM ($uudd$)     & $(+0.8(0),-1.3(7),+0.2(5))$ \\
$s_1$ AFM ($dduu$)     & $(+0.8(0),-1.3(7),+0.2(5))$ \\
$s_{2,1}$ AFM ($uudd$) & $(+0.8(0),-1.3(7),-0.2(5))$ \\ 
$s_{2,1}$ AFM ($dduu$) & $(+0.8(0),-1.3(7),-0.2(5))$ \\ 
$s_{2,2}$ AFM ($uudd$) & $(-1.6(1),+0.0(1),-0.2(5))$ \\ 
$s_{2,2}$ AFM ($dduu$) & $(-1.6(1),+0.0(1),-0.2(5))$ \\ 
\hline
\hline
\end{tabular}
\end{center}
\end{table}

We explain the existence of an in-plane intrinsic electric dipole (which provides new vistas for magnetoelectric couplings on this novel multiferroic) by recourse to symmetry: $s_1$, $s_{2,1}$, and $s_{2,2}$ CrI$_3$ bilayers all belong to crystalline space group $Cm$. Nevertheless, while the $s_1$ and $s_{2,1}$ bilayers have a mirror plane parallel to $\mathbf{a}_1=a_0\left(1/2,-\sqrt{3}/2,0\right)$ (Figures \ref{fig:f1}(c) and \ref{fig:f1}(d)), bilayer $s_{2,2}$ has a mirror plane parallel to $\mathbf{a}_1+\mathbf{a}_2$ ($x-$axis on Fig.~\ref{fig:f1b}). Any in-plane dipole for structures $s_1$ through $s_{2,1}$, if it is to exist, must be parallel to $\mathbf{a}_{1}$ (and this is the case for $P_x$ and $P_y$ on Fig.~\ref{fig:f3}(d)). In other words, its magnitude must satisfy the ratio $-\sqrt{3}\simeq -1.73$. This is precisely documented in Table \ref{ta:t3}. As for structure $s_{2,2}$, the in-plane electric dipole can only be parallel to the $x-$axis, which is what one sees in Table \ref{table:t2}. Furthermore, the magnitude of the in-plane dipole is nearly identical at $s_{1}$, $s_{2,1}$, and $s_{2,2}$, and it must switch in going from structure $s_{2,1}$ to structure $s_{2,2}$ through $s_1$.

\begin{figure}[tb]
\begin{center}
\includegraphics[width=0.48\textwidth]{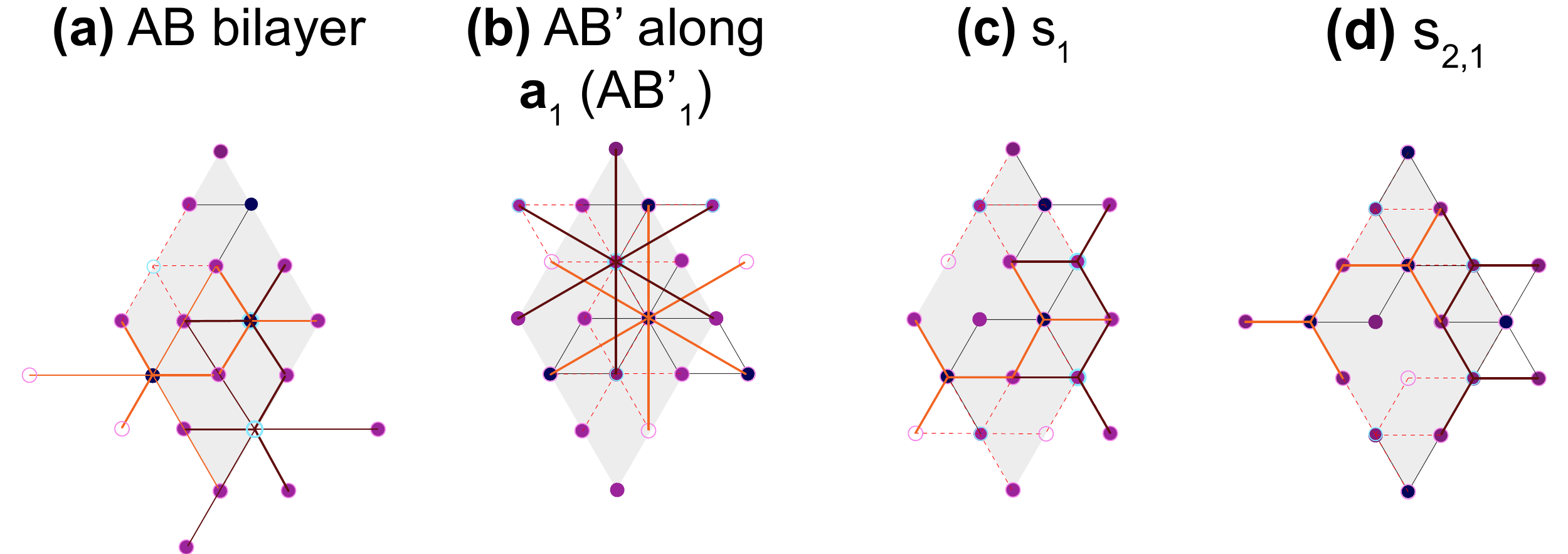}
\end{center}
\caption{Schematic top views of Cr atoms and nearest I atoms on the opposite monolayer; vectors joining those atoms are utilized to determine $(\Delta_x,\Delta_y,\Delta_z)$ in Table \ref{ta:t3}.\label{fig:f4}}
\end{figure}

The second check to justify the creation of in-plane components of intrinsic electric polarization is numerical: adding up all vectors that can be built from all four Cr atoms to the nearest I atoms on the opposite monolayer within the u.c.~($\Delta_x,\Delta_y,\Delta_z$), one gets the results listed in Table \ref{ta:t3}. A top view of the atoms involved is displayed in Fig.~\ref{fig:f4} for bilayers AB, AB'$_1$, $s_1$, and $s_{2,1}$. The observed relative displacements indeed respect the crystalline symmetry of group $Cm$: they are parallel to the mirror planes.

\begin{table}[tb]
\caption{\label{ta:t3}Sum of vectors pointing from Cr atoms to the first and second nearest I neighbors located in the opposite monolayer. This sum is carried out over all Cr atoms in the u.c.  $P_y/P_x$ as reported in Table \ref{table:t2} correlates with $\Delta_y/\Delta_x$, and hence justifies the existence of a non-zero in-plane intrinsic dipole moments on rotated CrI$_3$ bilayers. (${a_{1,y}}/{a_{1,x}}=-\sqrt{3}\simeq -1.73$.)}
\centering
\begin{tabular}{c|c|cc}\hline\hline
Configuration & ($\Delta_x,\Delta_y,\Delta_z$) ($\AA{}$/u.c.) & $P_y/P_x$ & $\Delta_y/\Delta_x$ \\
\hline
AB FM ($uuuu$)       &  $(0.000,0.000,0.000)$     & -       & -       \\
AB AFM ($uudd$)      &  $(0.000,0.000,0.000)$     & -       & -       \\
$s_1$ FM ($uuuu$)    &  $(+0.216,-0.388,+10.372)$ & $-1.79$ & $-1.72$ \\
$s_1$ AFM ($uudd$)   &  $(+0.216,-0.382,+10.374)$ & $-1.77$ & $-1.72$ \\
$s_{21}$ FM ($uuuu$) &  $(+0.221,-0.410,-10.369)$ & $-1.70$ & $-1.85$ \\
$s_{21}$ AFM ($uudd$)&  $(+0.247,-0.423,-10.306)$ & $-1.71$ & $-1.71$ \\
$s_{22}$ FM ($uuuu$) &  $(-0.453,-0.021,-10.373)$ & $ 0.00$ & $+0.05$ \\
$s_{22}$ AFM ($uudd$)&  $(-0.453,-0.021,-10.373)$ & $ 0.00$ & $+0.05$ \\
\hline\hline
\end{tabular}
\end{table}

\subsection{Calculation of the magnetoelectric coupling at point $p$ within the energy landscape}

We end this work with a calculation of the magnetoelectric tensor for the rotated CrI$_3$ bilayer. Following \'I\~niguez \cite{Jorge}, we seek a point in the landscape for which both intrinsic electric and magnetic dipoles are zero. As seen on Fig.~\ref{fig:f3}, the in-plane components of the intrinsic polarization remain constant, and so we used the {\em change} of $\mathbf{P}$ ($\Delta\mathbf{P}$) away from the polarization at point $p$ here. The eigenvalues $C_n$ of the bilayer's Hessian at $p$ {listed in Table \ref{ta:t4}} range in between $-$0.011 and 13.556 eV/\AA{}, which compare well with Ref.~\cite{Jorge}. We calculated the intrinsic electric dipole for eigenvalues greater than zero (IR modes get revealed by giving non-zero net electric polarizations), and set all associated eigenvectors to be $u=0.1$ \AA{} in magnitude.

\begin{table}[tb]
{
\caption{\label{ta:t4} Hessian eigenvalues for rotated CrI$_3$ bilayer at point $p$ (eV/\AA).}
\centering
\begin{tabular}{cccccc}
\hline\hline
$-$0.011 & 0.000 & 0.000 & 0.000 & 0.007 & 0.263\\
0.917 & 1.022 & 1.088 & 1.110 & 1.456 & 1.482\\
2.413 & 2.457 & 2.457 & 2.483 & 2.497 & 2.524\\
2.714 & 2.731 & 3.552 & 3.614 & 3.629 & 3.672\\
4.002 & 4.089 & 4.138 & 4.183 & 4.831 & 4.900\\
4.935 & 4.963 & 6.137 & 6.177 & 6.203 & 6.267\\
9.319 & 9.356 & 9.830 & 9.917 & 9.935 &10.047\\
10.836&10.944 &10.971 &11.098 &13.556 &13.556\\
\hline\hline
\end{tabular}
}
\end{table}

Since $\boldsymbol{\mu}$ changes stepwise from (0,0,0) to (0,0,12$\mu_B$) away from $p$, we used a value for the magnetic polarity equal to $\mathbf{p}^m=(0,0,\frac{12\mu_B\mu_0}{u})$, where  $\mu_0=1.2566\times 10^{-6}$ 	N/A$^2$ is the permeability of vacuum. This way, our expression for the magnetoelectric tensor takes the form (SI units):
\begin{eqnarray}
\alpha_{3,j}=\alpha_{j,3}=\frac{12 \mu_B\mu_0}{0.01 \text{\AA}^2}\sum_{n=5}^{48}\frac{\Delta P_{n,j}}{C_n}\\
=(-31.7,6.7,-4.3) \times 10^{-20}\text{ s},\nonumber
\end{eqnarray}
where the lack of 1/m reflects the fact that $\mathbf{P}$ is reported in units of charge per length here. Dividing by the bilayer thickness ($\simeq 10$ \AA) and multiplying by the speed of light \cite{Rivera}, one gets
\begin{equation}
\alpha=
\begin{pmatrix}
0.00 & 0.00 & -9.30\\
0.00 & 0.00 &  2.01\\
-9.30& 2.01 &-1.29
\end{pmatrix}\times 10^{-2}
\end{equation}
in gaussian units. This way, $\alpha$ here is one order of magnitude smaller than the values reported for Cr$_2$O$_3$ by \'I\~niguez.

{
CrI$_3$ bilayers are experimentally known to become paramagnetic at a few Kelvin. $AB'$ bilayers will lose their ferroelectric properties at that point, while rotated $s_{1}$, $s_{2,1}$ or $s_{2,2}$ bilayers may still maintain their ferroelectric properties at temperatures well above those for which their intrinsic magnetism is lost.}

{The magnetoelectric tensor thus found implies that the out-of-plane magnetic moment can be most easily switched by applying an electric field along the $x-$direction, or that the in-plane $x-$component of the intrinsic electric dipole will switch direction most easily as the Ising (out-of-plane) spin swaps sign. The explicit calculation of $\alpha$ brings the engineering of two-dimensional magnetoelectric multiferroics on rotated magnetic bilayers onto a more quantitative footing.}

\section{\label{sec:Conc}Conclusion}
Seeking to enhance the out-of-plane intrinsic electric dipole of CrI$_3$ bilayers by a relative rotation, we discovered that those bilayers develop an unexpected and sizeable in-plane intrinsic electric dipole moment as well. This is due to the unique atomistic structure of those magnetic materials that does not forbid the creation of in-plane intrinsic electric dipoles. The magnetoelectric coupling tensor $\alpha$ was estimated explicitly as well. Those results enhance our understanding, and provide new vistas into magnetoelectric multiferroics on two-dimensional material platforms.

\begin{acknowledgments}
Calculations from the Arkansas team were performed at Cori at NERSC, a DOE facility funded under Contract DE-AC02-05CH11231, and at the University of Arkansas’ Pinnacle supercomputer, funded by the U.S. National Science Foundation, the Arkansas Economic Development Commission, and the Office of the Vice Provost for Research and Innovation. J.M.M.T.~and M.A.M.~thank Montana State University, Bozeman, for startup support and computational resources within the Tempest Research Cluster. All authors acknowledge financial support from the MonArk NSF Quantum Foundry, supported by the National Science Foundation Q-AMASE-i program under NSF Award DMR-1906383.
\end{acknowledgments}


\begin{thebibliography}{109}%
\makeatletter
\providecommand \@ifxundefined [1]{%
 \@ifx{#1\undefined}
}%
\providecommand \@ifnum [1]{%
 \ifnum #1\expandafter \@firstoftwo
 \else \expandafter \@secondoftwo
 \fi
}%
\providecommand \@ifx [1]{%
 \ifx #1\expandafter \@firstoftwo
 \else \expandafter \@secondoftwo
 \fi
}%
\providecommand \natexlab [1]{#1}%
\providecommand \enquote  [1]{``#1''}%
\providecommand \bibnamefont  [1]{#1}%
\providecommand \bibfnamefont [1]{#1}%
\providecommand \citenamefont [1]{#1}%
\providecommand \href@noop [0]{\@secondoftwo}%
\providecommand \href [0]{\begingroup \@sanitize@url \@href}%
\providecommand \@href[1]{\@@startlink{#1}\@@href}%
\providecommand \@@href[1]{\endgroup#1\@@endlink}%
\providecommand \@sanitize@url [0]{\catcode `\\12\catcode `\$12\catcode
  `\&12\catcode `\#12\catcode `\^12\catcode `\_12\catcode `\%12\relax}%
\providecommand \@@startlink[1]{}%
\providecommand \@@endlink[0]{}%
\providecommand \url  [0]{\begingroup\@sanitize@url \@url }%
\providecommand \@url [1]{\endgroup\@href {#1}{\urlprefix }}%
\providecommand \urlprefix  [0]{URL }%
\providecommand \Eprint [0]{\href }%
\providecommand \doibase [0]{https://doi.org/}%
\providecommand \selectlanguage [0]{\@gobble}%
\providecommand \bibinfo  [0]{\@secondoftwo}%
\providecommand \bibfield  [0]{\@secondoftwo}%
\providecommand \translation [1]{[#1]}%
\providecommand \BibitemOpen [0]{}%
\providecommand \bibitemStop [0]{}%
\providecommand \bibitemNoStop [0]{.\EOS\space}%
\providecommand \EOS [0]{\spacefactor3000\relax}%
\providecommand \BibitemShut  [1]{\csname bibitem#1\endcsname}%
\let\auto@bib@innerbib\@empty
\bibitem [{\citenamefont {Curie}(1894)}]{Curie}%
  \BibitemOpen
  \bibfield  {author} {\bibinfo {author} {\bibfnamefont {M.~P.}\ \bibnamefont
  {Curie}},\ }\bibfield  {title} {\bibinfo {title} {Sur la sym{\'e}trie dans
  les ph{\'e}nom{\`e}nes physiques, sym{\'e}trie d'un champ {\'e}lectrique et
  d'un champ magn{\'e}tique},\ }\href
  {https://doi.org/10.1051/jphystap:018940030039300} {\bibfield  {journal}
  {\bibinfo  {journal} {J. Phys. Theor. Appl.}\ }\textbf {\bibinfo {volume}
  {3}},\ \bibinfo {pages} {393} (\bibinfo {year} {1894})}\BibitemShut {NoStop}%
\bibitem [{\citenamefont {Smolenskii}\ and\ \citenamefont
  {Chupis}(1982)}]{Smolenskii_1982}%
  \BibitemOpen
  \bibfield  {author} {\bibinfo {author} {\bibfnamefont {G.~A.}\ \bibnamefont
  {Smolenskii}}\ and\ \bibinfo {author} {\bibfnamefont {I.~E.}\ \bibnamefont
  {Chupis}},\ }\bibfield  {title} {\bibinfo {title} {Ferroelectromagnets},\
  }\href {https://doi.org/10.1070/PU1982v025n07ABEH004570} {\bibfield
  {journal} {\bibinfo  {journal} {Sov. phys., Usp.}\ }\textbf {\bibinfo
  {volume} {25}},\ \bibinfo {pages} {475} (\bibinfo {year} {1982})}\BibitemShut
  {NoStop}%
\bibitem [{\citenamefont {Fiebig}\ \emph {et~al.}(2013)\citenamefont {Fiebig},
  \citenamefont {Eremenko},\ and\ \citenamefont
  {Chupis}}]{fiebig2013magnetoelectric}%
  \BibitemOpen
  \bibfield  {author} {\bibinfo {author} {\bibfnamefont {M.}~\bibnamefont
  {Fiebig}}, \bibinfo {author} {\bibfnamefont {V.}~\bibnamefont {Eremenko}},\
  and\ \bibinfo {author} {\bibfnamefont {I.}~\bibnamefont {Chupis}},\
  }\href@noop {} {\emph {\bibinfo {title} {Magnetoelectric Interaction
  Phenomena in Crystals}}},\ NATO Science Series II: Mathematics, Physics and
  Chemistry\ (\bibinfo  {publisher} {Springer Netherlands},\ \bibinfo {year}
  {2013})\BibitemShut {NoStop}%
\bibitem [{\citenamefont {Landau}\ and\ \citenamefont
  {Lifshitz}(1984)}]{Landau}%
  \BibitemOpen
  \bibfield  {author} {\bibinfo {author} {\bibfnamefont {L.~D.}\ \bibnamefont
  {Landau}}\ and\ \bibinfo {author} {\bibfnamefont {E.~M.}\ \bibnamefont
  {Lifshitz}},\ }\href@noop {} {\emph {\bibinfo {title} {Electrodynamics of
  Continuous Media}}}\ (\bibinfo  {publisher} {Pergamon},\ \bibinfo {address}
  {New York},\ \bibinfo {year} {1984})\BibitemShut {NoStop}%
\bibitem [{\citenamefont {Astrov}(1960)}]{Astrov}%
  \BibitemOpen
  \bibfield  {author} {\bibinfo {author} {\bibfnamefont {D.~N.}\ \bibnamefont
  {Astrov}},\ }\bibfield  {title} {\bibinfo {title} {The magnetoelectric effect
  in antiferromagnets},\ }\href@noop {} {\bibfield  {journal} {\bibinfo
  {journal} {J. Exptl. Theoret. Phys. (U.S.S.R.)}\ }\textbf {\bibinfo {volume}
  {38}},\ \bibinfo {pages} {984} (\bibinfo {year} {1960})}\BibitemShut
  {NoStop}%
\bibitem [{\citenamefont {Kimura}\ \emph {et~al.}(2003)\citenamefont {Kimura},
  \citenamefont {Goto}, \citenamefont {Shintani}, \citenamefont {Ishizaka},
  \citenamefont {Arima},\ and\ \citenamefont {Tokura}}]{Kimura2003}%
  \BibitemOpen
  \bibfield  {author} {\bibinfo {author} {\bibfnamefont {T.}~\bibnamefont
  {Kimura}}, \bibinfo {author} {\bibfnamefont {T.}~\bibnamefont {Goto}},
  \bibinfo {author} {\bibfnamefont {H.}~\bibnamefont {Shintani}}, \bibinfo
  {author} {\bibfnamefont {K.}~\bibnamefont {Ishizaka}}, \bibinfo {author}
  {\bibfnamefont {T.}~\bibnamefont {Arima}},\ and\ \bibinfo {author}
  {\bibfnamefont {Y.}~\bibnamefont {Tokura}},\ }\bibfield  {title} {\bibinfo
  {title} {Magnetic control of ferroelectric polarization},\ }\href
  {https://doi.org/10.1038/nature02018} {\bibfield  {journal} {\bibinfo
  {journal} {Nature}\ }\textbf {\bibinfo {volume} {426}},\ \bibinfo {pages}
  {55} (\bibinfo {year} {2003})}\BibitemShut {NoStop}%
\bibitem [{\citenamefont {Kimura}\ \emph {et~al.}(2005)\citenamefont {Kimura},
  \citenamefont {Lawes}, \citenamefont {Goto}, \citenamefont {Tokura},\ and\
  \citenamefont {Ramirez}}]{Kimura2005}%
  \BibitemOpen
  \bibfield  {author} {\bibinfo {author} {\bibfnamefont {T.}~\bibnamefont
  {Kimura}}, \bibinfo {author} {\bibfnamefont {G.}~\bibnamefont {Lawes}},
  \bibinfo {author} {\bibfnamefont {T.}~\bibnamefont {Goto}}, \bibinfo {author}
  {\bibfnamefont {Y.}~\bibnamefont {Tokura}},\ and\ \bibinfo {author}
  {\bibfnamefont {A.~P.}\ \bibnamefont {Ramirez}},\ }\bibfield  {title}
  {\bibinfo {title} {Magnetoelectric phase diagrams of orthorhombic
  {RMnO}$_{3}$ ({R=Gd, Tb, and Dy})},\ }\href
  {https://doi.org/10.1103/PhysRevB.71.224425} {\bibfield  {journal} {\bibinfo
  {journal} {Phys. Rev. B}\ }\textbf {\bibinfo {volume} {71}},\ \bibinfo
  {pages} {224425} (\bibinfo {year} {2005})}\BibitemShut {NoStop}%
\bibitem [{\citenamefont {Eerenstein}\ \emph {et~al.}(2006)\citenamefont
  {Eerenstein}, \citenamefont {Mathur},\ and\ \citenamefont {Scott}}]{rev1}%
  \BibitemOpen
  \bibfield  {author} {\bibinfo {author} {\bibfnamefont {W.}~\bibnamefont
  {Eerenstein}}, \bibinfo {author} {\bibfnamefont {N.~D.}\ \bibnamefont
  {Mathur}},\ and\ \bibinfo {author} {\bibfnamefont {J.~F.}\ \bibnamefont
  {Scott}},\ }\bibfield  {title} {\bibinfo {title} {Multiferroic and
  magnetoelectric materials},\ }\href {https://doi.org/10.1038/nature05023}
  {\bibfield  {journal} {\bibinfo  {journal} {Nature}\ }\textbf {\bibinfo
  {volume} {442}},\ \bibinfo {pages} {759} (\bibinfo {year}
  {2006})}\BibitemShut {NoStop}%
\bibitem [{\citenamefont {Fiebig}(2005)}]{rev2}%
  \BibitemOpen
  \bibfield  {author} {\bibinfo {author} {\bibfnamefont {M.}~\bibnamefont
  {Fiebig}},\ }\bibfield  {title} {\bibinfo {title} {Revival of the
  magnetoelectric effect},\ }\href {https://doi.org/10.1088/0022-3727/38/8/R01}
  {\bibfield  {journal} {\bibinfo  {journal} {J. Phys. D: Appl. Phys.}\
  }\textbf {\bibinfo {volume} {38}},\ \bibinfo {pages} {R123} (\bibinfo {year}
  {2005})}\BibitemShut {NoStop}%
\bibitem [{\citenamefont {Hill}(2000)}]{NicolaFewMultiferroic}%
  \BibitemOpen
  \bibfield  {author} {\bibinfo {author} {\bibfnamefont {N.~A.}\ \bibnamefont
  {Hill}},\ }\bibfield  {title} {\bibinfo {title} {Why are there so few
  magnetic ferroelectrics?},\ }\href {https://doi.org/10.1021/jp000114x}
  {\bibfield  {journal} {\bibinfo  {journal} {J. Phys. Chem. B}\ }\textbf
  {\bibinfo {volume} {104}},\ \bibinfo {pages} {6694} (\bibinfo {year}
  {2000})}\BibitemShut {NoStop}%
\bibitem [{\citenamefont {Van~Aken}\ \emph {et~al.}(2004)\citenamefont
  {Van~Aken}, \citenamefont {Palstra}, \citenamefont {Filippetti},\ and\
  \citenamefont {Spaldin}}]{NicolaMultiferroic2}%
  \BibitemOpen
  \bibfield  {author} {\bibinfo {author} {\bibfnamefont {B.~B.}\ \bibnamefont
  {Van~Aken}}, \bibinfo {author} {\bibfnamefont {T.~T.}\ \bibnamefont
  {Palstra}}, \bibinfo {author} {\bibfnamefont {A.}~\bibnamefont
  {Filippetti}},\ and\ \bibinfo {author} {\bibfnamefont {N.~A.}\ \bibnamefont
  {Spaldin}},\ }\bibfield  {title} {\bibinfo {title} {The origin of
  ferroelectricity in magnetoelectric {YMnO$_3$}},\ }\href
  {https://doi.org/10.1038/nmat1080} {\bibfield  {journal} {\bibinfo  {journal}
  {Nat. Mater.}\ }\textbf {\bibinfo {volume} {3}},\ \bibinfo {pages} {164}
  (\bibinfo {year} {2004})}\BibitemShut {NoStop}%
\bibitem [{\citenamefont {Heron}\ \emph {et~al.}(2011)\citenamefont {Heron},
  \citenamefont {Trassin}, \citenamefont {Ashraf}, \citenamefont {Gajek},
  \citenamefont {He}, \citenamefont {Yang}, \citenamefont {Nikonov},
  \citenamefont {Chu}, \citenamefont {Salahuddin},\ and\ \citenamefont
  {Ramesh}}]{Ramesh2011}%
  \BibitemOpen
  \bibfield  {author} {\bibinfo {author} {\bibfnamefont {J.~T.}\ \bibnamefont
  {Heron}}, \bibinfo {author} {\bibfnamefont {M.}~\bibnamefont {Trassin}},
  \bibinfo {author} {\bibfnamefont {K.}~\bibnamefont {Ashraf}}, \bibinfo
  {author} {\bibfnamefont {M.}~\bibnamefont {Gajek}}, \bibinfo {author}
  {\bibfnamefont {Q.}~\bibnamefont {He}}, \bibinfo {author} {\bibfnamefont
  {S.~Y.}\ \bibnamefont {Yang}}, \bibinfo {author} {\bibfnamefont {D.~E.}\
  \bibnamefont {Nikonov}}, \bibinfo {author} {\bibfnamefont {Y.-H.}\
  \bibnamefont {Chu}}, \bibinfo {author} {\bibfnamefont {S.}~\bibnamefont
  {Salahuddin}},\ and\ \bibinfo {author} {\bibfnamefont {R.}~\bibnamefont
  {Ramesh}},\ }\bibfield  {title} {\bibinfo {title} {Electric-field-induced
  magnetization reversal in a ferromagnet-multiferroic heterostructure},\
  }\href {https://doi.org/10.1103/PhysRevLett.107.217202} {\bibfield  {journal}
  {\bibinfo  {journal} {Phys. Rev. Lett.}\ }\textbf {\bibinfo {volume} {107}},\
  \bibinfo {pages} {217202} (\bibinfo {year} {2011})}\BibitemShut {NoStop}%
\bibitem [{\citenamefont {Novoselov}\ \emph {et~al.}(2005)\citenamefont
  {Novoselov}, \citenamefont {Geim}, \citenamefont {Morozov}, \citenamefont
  {Jiang}, \citenamefont {Katsnelson}, \citenamefont {Grigorieva},
  \citenamefont {Dubonos},\ and\ \citenamefont {Firsov}}]{Graphene1}%
  \BibitemOpen
  \bibfield  {author} {\bibinfo {author} {\bibfnamefont {K.~S.}\ \bibnamefont
  {Novoselov}}, \bibinfo {author} {\bibfnamefont {A.~K.}\ \bibnamefont {Geim}},
  \bibinfo {author} {\bibfnamefont {S.~V.}\ \bibnamefont {Morozov}}, \bibinfo
  {author} {\bibfnamefont {D.}~\bibnamefont {Jiang}}, \bibinfo {author}
  {\bibfnamefont {M.~I.}\ \bibnamefont {Katsnelson}}, \bibinfo {author}
  {\bibfnamefont {I.~V.}\ \bibnamefont {Grigorieva}}, \bibinfo {author}
  {\bibfnamefont {S.~V.}\ \bibnamefont {Dubonos}},\ and\ \bibinfo {author}
  {\bibfnamefont {A.~A.}\ \bibnamefont {Firsov}},\ }\bibfield  {title}
  {\bibinfo {title} {Two-dimensional gas of massless {Dirac} fermions in
  graphene},\ }\href {https://doi.org/10.1038/nature04233} {\bibfield
  {journal} {\bibinfo  {journal} {Nature}\ }\textbf {\bibinfo {volume} {438}},\
  \bibinfo {pages} {197} (\bibinfo {year} {2005})}\BibitemShut {NoStop}%
\bibitem [{\citenamefont {Zhang}\ \emph {et~al.}(2005)\citenamefont {Zhang},
  \citenamefont {Tan}, \citenamefont {Stormer},\ and\ \citenamefont
  {Kim}}]{Graphene2}%
  \BibitemOpen
  \bibfield  {author} {\bibinfo {author} {\bibfnamefont {Y.}~\bibnamefont
  {Zhang}}, \bibinfo {author} {\bibfnamefont {Y.-W.}\ \bibnamefont {Tan}},
  \bibinfo {author} {\bibfnamefont {H.~L.}\ \bibnamefont {Stormer}},\ and\
  \bibinfo {author} {\bibfnamefont {P.}~\bibnamefont {Kim}},\ }\bibfield
  {title} {\bibinfo {title} {Experimental observation of the quantum {Hall
  effect and Berry's} phase in graphene},\ }\href
  {https://doi.org/10.1038/nature04235} {\bibfield  {journal} {\bibinfo
  {journal} {Nature}\ }\textbf {\bibinfo {volume} {438}},\ \bibinfo {pages}
  {201} (\bibinfo {year} {2005})}\BibitemShut {NoStop}%
\bibitem [{\citenamefont {Katsnelson}(2012)}]{Graphene3}%
  \BibitemOpen
  \bibfield  {author} {\bibinfo {author} {\bibfnamefont {M.~I.}\ \bibnamefont
  {Katsnelson}},\ }\href {https://doi.org/10.1017/CBO9781139031080} {\emph
  {\bibinfo {title} {Graphene: Carbon in Two Dimensions}}}\ (\bibinfo
  {publisher} {Cambridge University Press},\ \bibinfo {year}
  {2012})\BibitemShut {NoStop}%
\bibitem [{\citenamefont {R{\"o}ntgen}(1888)}]{Rontgen}%
  \BibitemOpen
  \bibfield  {author} {\bibinfo {author} {\bibfnamefont {W.~C.}\ \bibnamefont
  {R{\"o}ntgen}},\ }\bibfield  {title} {\bibinfo {title} {Ueber die durch
  bewegung eines im homogenen electrischen felde befindlichen dielectricums
  hervorgerufene electrodynamische kraft},\ }\href
  {https://doi.org/10.1002/andp.18882711003} {\bibfield  {journal} {\bibinfo
  {journal} {Ann. Phys.}\ }\textbf {\bibinfo {volume} {271}},\ \bibinfo {pages}
  {264} (\bibinfo {year} {1888})}\BibitemShut {NoStop}%
\bibitem [{\citenamefont {Debye}(1926)}]{Debye}%
  \BibitemOpen
  \bibfield  {author} {\bibinfo {author} {\bibfnamefont {P.}~\bibnamefont
  {Debye}},\ }\bibfield  {title} {\bibinfo {title} {Bemerkung zu einigen neuen
  versuchen {\"u}ber einen magneto-elektrischen richteffekt},\ }\href
  {https://doi.org/10.1007/BF01557844} {\bibfield  {journal} {\bibinfo
  {journal} {Z. Physik}\ }\textbf {\bibinfo {volume} {36}},\ \bibinfo {pages}
  {300} (\bibinfo {year} {1926})}\BibitemShut {NoStop}%
\bibitem [{\citenamefont {Smolenskii}\ and\ \citenamefont
  {Agranovskaya}(1960)}]{Smolenskii}%
  \BibitemOpen
  \bibfield  {author} {\bibinfo {author} {\bibfnamefont {G.~A.}\ \bibnamefont
  {Smolenskii}}\ and\ \bibinfo {author} {\bibfnamefont {A.~I.}\ \bibnamefont
  {Agranovskaya}},\ }\bibfield  {title} {\bibinfo {title} {Dielectric
  polarization of a number of complex compounds},\ }\href@noop {} {\bibfield
  {journal} {\bibinfo  {journal} {Sov. phys., Solid state}\ }\textbf {\bibinfo
  {volume} {1}},\ \bibinfo {pages} {1429} (\bibinfo {year} {1960})}\BibitemShut
  {NoStop}%
\bibitem [{\citenamefont {Dzyaloshinskii}(1959)}]{Dzyaloshinskii}%
  \BibitemOpen
  \bibfield  {author} {\bibinfo {author} {\bibfnamefont {I.~E.}\ \bibnamefont
  {Dzyaloshinskii}},\ }\bibfield  {title} {\bibinfo {title} {On the
  magneto-electrical effects in antiferromagnets},\ }\href@noop {} {\bibfield
  {journal} {\bibinfo  {journal} {J. Exptl. Theoret. Phys. (U.S.S.R.)}\
  }\textbf {\bibinfo {volume} {10}},\ \bibinfo {pages} {628} (\bibinfo {year}
  {1959})}\BibitemShut {NoStop}%
\bibitem [{\citenamefont {Wang}\ \emph {et~al.}(2003)\citenamefont {Wang},
  \citenamefont {Neaton}, \citenamefont {Zheng}, \citenamefont {Nagarajan},
  \citenamefont {Ogale}, \citenamefont {Liu}, \citenamefont {Viehland},
  \citenamefont {Vaithyanathan}, \citenamefont {Schlom}, \citenamefont
  {Waghmare}, \citenamefont {Spaldin}, \citenamefont {Rabe}, \citenamefont
  {Wuttig},\ and\ \citenamefont {Ramesh}}]{BFO_Neaton}%
  \BibitemOpen
  \bibfield  {author} {\bibinfo {author} {\bibfnamefont {J.}~\bibnamefont
  {Wang}}, \bibinfo {author} {\bibfnamefont {J.~B.}\ \bibnamefont {Neaton}},
  \bibinfo {author} {\bibfnamefont {H.}~\bibnamefont {Zheng}}, \bibinfo
  {author} {\bibfnamefont {V.}~\bibnamefont {Nagarajan}}, \bibinfo {author}
  {\bibfnamefont {S.~B.}\ \bibnamefont {Ogale}}, \bibinfo {author}
  {\bibfnamefont {B.}~\bibnamefont {Liu}}, \bibinfo {author} {\bibfnamefont
  {D.}~\bibnamefont {Viehland}}, \bibinfo {author} {\bibfnamefont
  {V.}~\bibnamefont {Vaithyanathan}}, \bibinfo {author} {\bibfnamefont {D.~G.}\
  \bibnamefont {Schlom}}, \bibinfo {author} {\bibfnamefont {U.~V.}\
  \bibnamefont {Waghmare}}, \bibinfo {author} {\bibfnamefont {N.~A.}\
  \bibnamefont {Spaldin}}, \bibinfo {author} {\bibfnamefont {K.~M.}\
  \bibnamefont {Rabe}}, \bibinfo {author} {\bibfnamefont {M.}~\bibnamefont
  {Wuttig}},\ and\ \bibinfo {author} {\bibfnamefont {R.}~\bibnamefont
  {Ramesh}},\ }\bibfield  {title} {\bibinfo {title} {Epitaxial {BiFeO$_3$}
  multiferroic thin film heterostructures},\ }\href@noop {} {\bibfield
  {journal} {\bibinfo  {journal} {Science}\ }\textbf {\bibinfo {volume}
  {299}},\ \bibinfo {pages} {1719} (\bibinfo {year} {2003})}\BibitemShut
  {NoStop}%
\bibitem [{\citenamefont {Neaton}\ \emph {et~al.}(2005)\citenamefont {Neaton},
  \citenamefont {Ederer}, \citenamefont {Waghmare}, \citenamefont {Spaldin},\
  and\ \citenamefont {Rabe}}]{Rabe_1}%
  \BibitemOpen
  \bibfield  {author} {\bibinfo {author} {\bibfnamefont {J.~B.}\ \bibnamefont
  {Neaton}}, \bibinfo {author} {\bibfnamefont {C.}~\bibnamefont {Ederer}},
  \bibinfo {author} {\bibfnamefont {U.~V.}\ \bibnamefont {Waghmare}}, \bibinfo
  {author} {\bibfnamefont {N.~A.}\ \bibnamefont {Spaldin}},\ and\ \bibinfo
  {author} {\bibfnamefont {K.~M.}\ \bibnamefont {Rabe}},\ }\bibfield  {title}
  {\bibinfo {title} {First-principles study of spontaneous polarization in
  multiferroic {$\mathrm{Bi}\mathrm{Fe}{\mathrm{O}}_{3}$}},\ }\href
  {https://doi.org/10.1103/PhysRevB.71.014113} {\bibfield  {journal} {\bibinfo
  {journal} {Phys. Rev. B}\ }\textbf {\bibinfo {volume} {71}},\ \bibinfo
  {pages} {014113} (\bibinfo {year} {2005})}\BibitemShut {NoStop}%
\bibitem [{\citenamefont {Spaldin}\ and\ \citenamefont
  {Fiebig}(2005)}]{SpaldinFiebigScience2005}%
  \BibitemOpen
  \bibfield  {author} {\bibinfo {author} {\bibfnamefont {N.~A.}\ \bibnamefont
  {Spaldin}}\ and\ \bibinfo {author} {\bibfnamefont {M.}~\bibnamefont
  {Fiebig}},\ }\bibfield  {title} {\bibinfo {title} {The renaissance of
  magnetoelectric multiferroics},\ }\href
  {https://doi.org/10.1126/science.1113357} {\bibfield  {journal} {\bibinfo
  {journal} {Science}\ }\textbf {\bibinfo {volume} {309}},\ \bibinfo {pages}
  {391} (\bibinfo {year} {2005})}\BibitemShut {NoStop}%
\bibitem [{\citenamefont {Duan}\ \emph {et~al.}(2006)\citenamefont {Duan},
  \citenamefont {Jaswal},\ and\ \citenamefont {Tsymbal}}]{Tsymbal2006}%
  \BibitemOpen
  \bibfield  {author} {\bibinfo {author} {\bibfnamefont {C.-G.}\ \bibnamefont
  {Duan}}, \bibinfo {author} {\bibfnamefont {S.~S.}\ \bibnamefont {Jaswal}},\
  and\ \bibinfo {author} {\bibfnamefont {E.~Y.}\ \bibnamefont {Tsymbal}},\
  }\bibfield  {title} {\bibinfo {title} {Predicted magnetoelectric effect in
  {Fe/BaTiO}$_{3}$ multilayers: Ferroelectric control of magnetism},\ }\href
  {https://doi.org/10.1103/PhysRevLett.97.047201} {\bibfield  {journal}
  {\bibinfo  {journal} {Phys. Rev. Lett.}\ }\textbf {\bibinfo {volume} {97}},\
  \bibinfo {pages} {047201} (\bibinfo {year} {2006})}\BibitemShut {NoStop}%
\bibitem [{\citenamefont {Fennie}\ and\ \citenamefont {Rabe}(2006)}]{Fennie}%
  \BibitemOpen
  \bibfield  {author} {\bibinfo {author} {\bibfnamefont {C.~J.}\ \bibnamefont
  {Fennie}}\ and\ \bibinfo {author} {\bibfnamefont {K.~M.}\ \bibnamefont
  {Rabe}},\ }\bibfield  {title} {\bibinfo {title} {Magnetic and electric phase
  control in epitaxial {${\mathrm{EuTiO}}_{3}$} from first principles},\ }\href
  {https://doi.org/10.1103/PhysRevLett.97.267602} {\bibfield  {journal}
  {\bibinfo  {journal} {Phys. Rev. Lett.}\ }\textbf {\bibinfo {volume} {97}},\
  \bibinfo {pages} {267602} (\bibinfo {year} {2006})}\BibitemShut {NoStop}%
\bibitem [{\citenamefont {Sergienko}\ and\ \citenamefont
  {Dagotto}(2006)}]{Dagotto2006}%
  \BibitemOpen
  \bibfield  {author} {\bibinfo {author} {\bibfnamefont {I.~A.}\ \bibnamefont
  {Sergienko}}\ and\ \bibinfo {author} {\bibfnamefont {E.}~\bibnamefont
  {Dagotto}},\ }\bibfield  {title} {\bibinfo {title} {Role of the
  {Dzyaloshinskii-Moriya} interaction in multiferroic perovskites},\ }\href
  {https://doi.org/10.1103/PhysRevB.73.094434} {\bibfield  {journal} {\bibinfo
  {journal} {Phys. Rev. B}\ }\textbf {\bibinfo {volume} {73}},\ \bibinfo
  {pages} {094434} (\bibinfo {year} {2006})}\BibitemShut {NoStop}%
\bibitem [{\citenamefont {Picozzi}\ \emph {et~al.}(2007)\citenamefont
  {Picozzi}, \citenamefont {Yamauchi}, \citenamefont {Sanyal}, \citenamefont
  {Sergienko},\ and\ \citenamefont {Dagotto}}]{Picozzi2007}%
  \BibitemOpen
  \bibfield  {author} {\bibinfo {author} {\bibfnamefont {S.}~\bibnamefont
  {Picozzi}}, \bibinfo {author} {\bibfnamefont {K.}~\bibnamefont {Yamauchi}},
  \bibinfo {author} {\bibfnamefont {B.}~\bibnamefont {Sanyal}}, \bibinfo
  {author} {\bibfnamefont {I.~A.}\ \bibnamefont {Sergienko}},\ and\ \bibinfo
  {author} {\bibfnamefont {E.}~\bibnamefont {Dagotto}},\ }\bibfield  {title}
  {\bibinfo {title} {Dual nature of improper ferroelectricity in a
  magnetoelectric multiferroic},\ }\href
  {https://doi.org/10.1103/PhysRevLett.99.227201} {\bibfield  {journal}
  {\bibinfo  {journal} {Phys. Rev. Lett.}\ }\textbf {\bibinfo {volume} {99}},\
  \bibinfo {pages} {227201} (\bibinfo {year} {2007})}\BibitemShut {NoStop}%
\bibitem [{\citenamefont {Tsymbal}\ \emph
  {et~al.}(2012{\natexlab{a}})\citenamefont {Tsymbal}, \citenamefont {Dagotto},
  \citenamefont {Eom},\ and\ \citenamefont {Ramesh}}]{DagottoBook}%
  \BibitemOpen
  \bibinfo {editor} {\bibfnamefont {E.~Y.}\ \bibnamefont {Tsymbal}}, \bibinfo
  {editor} {\bibfnamefont {E.~R.~A.}\ \bibnamefont {Dagotto}}, \bibinfo
  {editor} {\bibfnamefont {C.-B.}\ \bibnamefont {Eom}},\ and\ \bibinfo {editor}
  {\bibfnamefont {R.}~\bibnamefont {Ramesh}},\ eds.,\ \href@noop {} {\emph
  {\bibinfo {title} {Multifunctional Oxide Heterostructures}}}\ (\bibinfo
  {publisher} {Oxford University Press},\ \bibinfo {year} {2012})\BibitemShut
  {NoStop}%
\bibitem [{\citenamefont {Lee}\ and\ \citenamefont
  {Rabe}(2011)}]{PhysRevLett.107.067601}%
  \BibitemOpen
  \bibfield  {author} {\bibinfo {author} {\bibfnamefont {J.~H.}\ \bibnamefont
  {Lee}}\ and\ \bibinfo {author} {\bibfnamefont {K.~M.}\ \bibnamefont {Rabe}},\
  }\bibfield  {title} {\bibinfo {title} {Coupled magnetic-ferroelectric
  metal-insulator transition in epitaxially strained ${\mathrm{srcoo}}_{3}$
  from first principles},\ }\href
  {https://doi.org/10.1103/PhysRevLett.107.067601} {\bibfield  {journal}
  {\bibinfo  {journal} {Phys. Rev. Lett.}\ }\textbf {\bibinfo {volume} {107}},\
  \bibinfo {pages} {067601} (\bibinfo {year} {2011})}\BibitemShut {NoStop}%
\bibitem [{\citenamefont {Valencia}\ \emph {et~al.}(2011)\citenamefont
  {Valencia}, \citenamefont {Crassous}, \citenamefont {Bocher}, \citenamefont
  {Garcia}, \citenamefont {Moya}, \citenamefont {Cherifi}, \citenamefont
  {Deranlot}, \citenamefont {Bouzehouane}, \citenamefont {Fusil}, \citenamefont
  {Zobelli}, \citenamefont {Gloter}, \citenamefont {Mathur}, \citenamefont
  {Gaupp}, \citenamefont {Abrudan}, \citenamefont {Radu}, \citenamefont
  {Barth{\'e}l{\'e}my},\ and\ \citenamefont {Bibes}}]{Bibes2011}%
  \BibitemOpen
  \bibfield  {author} {\bibinfo {author} {\bibfnamefont {S.}~\bibnamefont
  {Valencia}}, \bibinfo {author} {\bibfnamefont {A.}~\bibnamefont {Crassous}},
  \bibinfo {author} {\bibfnamefont {L.}~\bibnamefont {Bocher}}, \bibinfo
  {author} {\bibfnamefont {V.}~\bibnamefont {Garcia}}, \bibinfo {author}
  {\bibfnamefont {X.}~\bibnamefont {Moya}}, \bibinfo {author} {\bibfnamefont
  {R.~O.}\ \bibnamefont {Cherifi}}, \bibinfo {author} {\bibfnamefont
  {C.}~\bibnamefont {Deranlot}}, \bibinfo {author} {\bibfnamefont
  {K.}~\bibnamefont {Bouzehouane}}, \bibinfo {author} {\bibfnamefont
  {S.}~\bibnamefont {Fusil}}, \bibinfo {author} {\bibfnamefont
  {A.}~\bibnamefont {Zobelli}}, \bibinfo {author} {\bibfnamefont
  {A.}~\bibnamefont {Gloter}}, \bibinfo {author} {\bibfnamefont {N.~D.}\
  \bibnamefont {Mathur}}, \bibinfo {author} {\bibfnamefont {A.}~\bibnamefont
  {Gaupp}}, \bibinfo {author} {\bibfnamefont {R.}~\bibnamefont {Abrudan}},
  \bibinfo {author} {\bibfnamefont {F.}~\bibnamefont {Radu}}, \bibinfo {author}
  {\bibfnamefont {A.}~\bibnamefont {Barth{\'e}l{\'e}my}},\ and\ \bibinfo
  {author} {\bibfnamefont {M.}~\bibnamefont {Bibes}},\ }\bibfield  {title}
  {\bibinfo {title} {Interface-induced room-temperature multiferroicity in
  {BaTiO$_3$}},\ }\href {https://doi.org/10.1038/nmat3098} {\bibfield
  {journal} {\bibinfo  {journal} {Nat. Mater.}\ }\textbf {\bibinfo {volume}
  {10}},\ \bibinfo {pages} {753} (\bibinfo {year} {2011})}\BibitemShut
  {NoStop}%
\bibitem [{\citenamefont {Tsymbal}\ \emph
  {et~al.}(2012{\natexlab{b}})\citenamefont {Tsymbal}, \citenamefont
  {Gruverman}, \citenamefont {Garcia}, \citenamefont {Bibes}, ,\ and\
  \citenamefont {Barth\'el\'emy}}]{TsymbalJunctions2012}%
  \BibitemOpen
  \bibfield  {author} {\bibinfo {author} {\bibfnamefont {E.~Y.}\ \bibnamefont
  {Tsymbal}}, \bibinfo {author} {\bibfnamefont {A.}~\bibnamefont {Gruverman}},
  \bibinfo {author} {\bibfnamefont {V.}~\bibnamefont {Garcia}}, \bibinfo
  {author} {\bibfnamefont {M.}~\bibnamefont {Bibes}}, ,\ and\ \bibinfo {author}
  {\bibfnamefont {A.}~\bibnamefont {Barth\'el\'emy}},\ }\bibfield  {title}
  {\bibinfo {title} {Ferroelectric and multiferroic tunnel junctions},\ }\href
  {https://doi.org/10.1557/mrs.2011.358} {\bibfield  {journal} {\bibinfo
  {journal} {MRS Bulletin}\ }\textbf {\bibinfo {volume} {37}},\ \bibinfo
  {pages} {138} (\bibinfo {year} {2012}{\natexlab{b}})}\BibitemShut {NoStop}%
\bibitem [{\citenamefont {Zhao}\ \emph {et~al.}(2014)\citenamefont {Zhao},
  \citenamefont {Ren}, \citenamefont {Yang}, \citenamefont {I{\~n}iguez},
  \citenamefont {Chen},\ and\ \citenamefont {Bellaiche}}]{Bellaiche2014}%
  \BibitemOpen
  \bibfield  {author} {\bibinfo {author} {\bibfnamefont {H.~J.}\ \bibnamefont
  {Zhao}}, \bibinfo {author} {\bibfnamefont {W.}~\bibnamefont {Ren}}, \bibinfo
  {author} {\bibfnamefont {Y.}~\bibnamefont {Yang}}, \bibinfo {author}
  {\bibfnamefont {J.}~\bibnamefont {I{\~n}iguez}}, \bibinfo {author}
  {\bibfnamefont {X.~M.}\ \bibnamefont {Chen}},\ and\ \bibinfo {author}
  {\bibfnamefont {L.}~\bibnamefont {Bellaiche}},\ }\bibfield  {title} {\bibinfo
  {title} {Near room-temperature multiferroic materials with tunable
  ferromagnetic and electrical properties},\ }\href
  {https://doi.org/10.1038/ncomms5021} {\bibfield  {journal} {\bibinfo
  {journal} {Nat. Commun.}\ }\textbf {\bibinfo {volume} {5}},\ \bibinfo {pages}
  {4021} (\bibinfo {year} {2014})}\BibitemShut {NoStop}%
\bibitem [{\citenamefont {Spaldin}\ and\ \citenamefont
  {Ramesh}(2019)}]{Spaldin2019}%
  \BibitemOpen
  \bibfield  {author} {\bibinfo {author} {\bibfnamefont {N.}~\bibnamefont
  {Spaldin}}\ and\ \bibinfo {author} {\bibfnamefont {R.}~\bibnamefont
  {Ramesh}},\ }\bibfield  {title} {\bibinfo {title} {Advances in
  magnetoelectric multiferroics},\ }\href
  {https://doi.org/10.1038/s41563-018-0275-2} {\bibfield  {journal} {\bibinfo
  {journal} {Nat. Mater.}\ }\textbf {\bibinfo {volume} {18}},\ \bibinfo {pages}
  {203} (\bibinfo {year} {2019})}\BibitemShut {NoStop}%
\bibitem [{\citenamefont {Huang}\ \emph {et~al.}(2020)\citenamefont {Huang},
  \citenamefont {Nikonov}, \citenamefont {Addiego}, \citenamefont {Chopdekar},
  \citenamefont {Prasad}, \citenamefont {Zhang}, \citenamefont {Chatterjee},
  \citenamefont {Liu}, \citenamefont {Farhan}, \citenamefont {Chu},
  \citenamefont {Yang}, \citenamefont {Ramesh}, \citenamefont {Qiu},
  \citenamefont {Huey}, \citenamefont {Lin}, \citenamefont {Gosavi},
  \citenamefont {I{\~n}iguez}, \citenamefont {Bokor}, \citenamefont {Pan},
  \citenamefont {Young}, \citenamefont {Martin},\ and\ \citenamefont
  {Ramesh}}]{Ramesh2020}%
  \BibitemOpen
  \bibfield  {author} {\bibinfo {author} {\bibfnamefont {Y.-L.}\ \bibnamefont
  {Huang}}, \bibinfo {author} {\bibfnamefont {D.}~\bibnamefont {Nikonov}},
  \bibinfo {author} {\bibfnamefont {C.}~\bibnamefont {Addiego}}, \bibinfo
  {author} {\bibfnamefont {R.~V.}\ \bibnamefont {Chopdekar}}, \bibinfo {author}
  {\bibfnamefont {B.}~\bibnamefont {Prasad}}, \bibinfo {author} {\bibfnamefont
  {L.}~\bibnamefont {Zhang}}, \bibinfo {author} {\bibfnamefont
  {J.}~\bibnamefont {Chatterjee}}, \bibinfo {author} {\bibfnamefont {H.-J.}\
  \bibnamefont {Liu}}, \bibinfo {author} {\bibfnamefont {A.}~\bibnamefont
  {Farhan}}, \bibinfo {author} {\bibfnamefont {Y.-H.}\ \bibnamefont {Chu}},
  \bibinfo {author} {\bibfnamefont {M.}~\bibnamefont {Yang}}, \bibinfo {author}
  {\bibfnamefont {M.}~\bibnamefont {Ramesh}}, \bibinfo {author} {\bibfnamefont
  {Z.~Q.}\ \bibnamefont {Qiu}}, \bibinfo {author} {\bibfnamefont {B.~D.}\
  \bibnamefont {Huey}}, \bibinfo {author} {\bibfnamefont {C.-C.}\ \bibnamefont
  {Lin}}, \bibinfo {author} {\bibfnamefont {T.}~\bibnamefont {Gosavi}},
  \bibinfo {author} {\bibfnamefont {J.}~\bibnamefont {I{\~n}iguez}}, \bibinfo
  {author} {\bibfnamefont {J.}~\bibnamefont {Bokor}}, \bibinfo {author}
  {\bibfnamefont {X.}~\bibnamefont {Pan}}, \bibinfo {author} {\bibfnamefont
  {I.}~\bibnamefont {Young}}, \bibinfo {author} {\bibfnamefont {L.~W.}\
  \bibnamefont {Martin}},\ and\ \bibinfo {author} {\bibfnamefont
  {R.}~\bibnamefont {Ramesh}},\ }\bibfield  {title} {\bibinfo {title}
  {Manipulating magnetoelectric energy landscape in multiferroics},\ }\href
  {https://doi.org/10.1038/s41467-020-16727-2} {\bibfield  {journal} {\bibinfo
  {journal} {Nat. Commun.}\ }\textbf {\bibinfo {volume} {11}},\ \bibinfo
  {pages} {2836} (\bibinfo {year} {2020})}\BibitemShut {NoStop}%
\bibitem [{\citenamefont {Ponet}\ \emph {et~al.}(2022)\citenamefont {Ponet},
  \citenamefont {Artyukhin}, \citenamefont {Kain}, \citenamefont {Wettstein},
  \citenamefont {Pimenov}, \citenamefont {Shuvaev}, \citenamefont {Wang},
  \citenamefont {Cheong}, \citenamefont {Mostovoy},\ and\ \citenamefont
  {Pimenov}}]{TopoPeroviskite2022}%
  \BibitemOpen
  \bibfield  {author} {\bibinfo {author} {\bibfnamefont {L.}~\bibnamefont
  {Ponet}}, \bibinfo {author} {\bibfnamefont {S.}~\bibnamefont {Artyukhin}},
  \bibinfo {author} {\bibfnamefont {T.}~\bibnamefont {Kain}}, \bibinfo {author}
  {\bibfnamefont {J.}~\bibnamefont {Wettstein}}, \bibinfo {author}
  {\bibfnamefont {A.}~\bibnamefont {Pimenov}}, \bibinfo {author} {\bibfnamefont
  {A.}~\bibnamefont {Shuvaev}}, \bibinfo {author} {\bibfnamefont
  {X.}~\bibnamefont {Wang}}, \bibinfo {author} {\bibfnamefont {S.-W.}\
  \bibnamefont {Cheong}}, \bibinfo {author} {\bibfnamefont {M.}~\bibnamefont
  {Mostovoy}},\ and\ \bibinfo {author} {\bibfnamefont {A.}~\bibnamefont
  {Pimenov}},\ }\bibfield  {title} {\bibinfo {title} {Topologically protected
  magnetoelectric switching in a multiferroic},\ }\href
  {https://doi.org/10.1038/s41586-022-04851-6} {\bibfield  {journal} {\bibinfo
  {journal} {Nature}\ }\textbf {\bibinfo {volume} {607}},\ \bibinfo {pages}
  {81} (\bibinfo {year} {2022})}\BibitemShut {NoStop}%
\bibitem [{\citenamefont {Li}\ \emph {et~al.}(2022)\citenamefont {Li},
  \citenamefont {Zhou},\ and\ \citenamefont {Guo}}]{2022paper}%
  \BibitemOpen
  \bibfield  {author} {\bibinfo {author} {\bibfnamefont {P.}~\bibnamefont
  {Li}}, \bibinfo {author} {\bibfnamefont {X.}~\bibnamefont {Zhou}},\ and\
  \bibinfo {author} {\bibfnamefont {Z.}~\bibnamefont {Guo}},\ }\bibfield
  {title} {\bibinfo {title} {Intriguing magnetoelectric effect in
  two-dimensional ferromagnetic/perovskite oxide ferroelectric
  heterostructure},\ }\href {https://doi.org/10.1038/s41524-022-00706-w}
  {\bibfield  {journal} {\bibinfo  {journal} {npj Comput. Mater.}\ }\textbf
  {\bibinfo {volume} {8}},\ \bibinfo {pages} {20} (\bibinfo {year}
  {2022})}\BibitemShut {NoStop}%
\bibitem [{\citenamefont {Shen}\ \emph {et~al.}(2023)\citenamefont {Shen},
  \citenamefont {Wang}, \citenamefont {Lu},\ and\ \citenamefont
  {Zhang}}]{Xiaofan}%
  \BibitemOpen
  \bibfield  {author} {\bibinfo {author} {\bibfnamefont {X.}~\bibnamefont
  {Shen}}, \bibinfo {author} {\bibfnamefont {F.}~\bibnamefont {Wang}}, \bibinfo
  {author} {\bibfnamefont {X.}~\bibnamefont {Lu}},\ and\ \bibinfo {author}
  {\bibfnamefont {J.}~\bibnamefont {Zhang}},\ }\bibfield  {title} {\bibinfo
  {title} {Two-dimensional multiferroics with intrinsic magnetoelectric
  coupling in {A-Site} ordered perovskite monolayers},\ }\href
  {https://doi.org/10.1021/acs.nanolett.2c03457} {\bibfield  {journal}
  {\bibinfo  {journal} {Nano Lett.}\ }\textbf {\bibinfo {volume} {23}},\
  \bibinfo {pages} {735} (\bibinfo {year} {2023})}\BibitemShut {NoStop}%
\bibitem [{\citenamefont {Schmid}(1973)}]{Schmid_1973}%
  \BibitemOpen
  \bibfield  {author} {\bibinfo {author} {\bibfnamefont {H.}~\bibnamefont
  {Schmid}},\ }\bibfield  {title} {\bibinfo {title} {On a magnetoelectric
  classification of materials},\ }\href@noop {} {\bibfield  {journal} {\bibinfo
   {journal} {Int. J. Magnetism}\ }\textbf {\bibinfo {volume} {4}},\ \bibinfo
  {pages} {337} (\bibinfo {year} {1973})}\BibitemShut {NoStop}%
\bibitem [{\citenamefont {Sugie}\ \emph {et~al.}(2002)\citenamefont {Sugie},
  \citenamefont {Iwata},\ and\ \citenamefont {Kohn}}]{Sugie}%
  \BibitemOpen
  \bibfield  {author} {\bibinfo {author} {\bibfnamefont {H.}~\bibnamefont
  {Sugie}}, \bibinfo {author} {\bibfnamefont {N.}~\bibnamefont {Iwata}},\ and\
  \bibinfo {author} {\bibfnamefont {K.}~\bibnamefont {Kohn}},\ }\bibfield
  {title} {\bibinfo {title} {Magnetic ordering of rare earth ions and
  magnetic-electric interaction of hexagonal {RMnO$_3$ (R=Ho, Er, Yb or Lu)}},\
  }\href {https://doi.org/10.1143/JPSJ.71.1558} {\bibfield  {journal} {\bibinfo
   {journal} {J. Phys. Soc. Japan}\ }\textbf {\bibinfo {volume} {71}},\
  \bibinfo {pages} {1558} (\bibinfo {year} {2002})}\BibitemShut {NoStop}%
\bibitem [{\citenamefont {Schmid}\ \emph {et~al.}(1965)\citenamefont {Schmid},
  \citenamefont {Rieder},\ and\ \citenamefont {Ascher}}]{SCHMID1965327}%
  \BibitemOpen
  \bibfield  {author} {\bibinfo {author} {\bibfnamefont {H.}~\bibnamefont
  {Schmid}}, \bibinfo {author} {\bibfnamefont {H.}~\bibnamefont {Rieder}},\
  and\ \bibinfo {author} {\bibfnamefont {E.}~\bibnamefont {Ascher}},\
  }\bibfield  {title} {\bibinfo {title} {Magnetic susceptibilities of some 3d
  transition metal boracites},\ }\href
  {https://doi.org/10.1016/0038-1098(65)90088-8} {\bibfield  {journal}
  {\bibinfo  {journal} {Solid State Commun.}\ }\textbf {\bibinfo {volume}
  {3}},\ \bibinfo {pages} {327} (\bibinfo {year} {1965})}\BibitemShut {NoStop}%
\bibitem [{\citenamefont {Scott}(1980)}]{Scott1980}%
  \BibitemOpen
  \bibfield  {author} {\bibinfo {author} {\bibfnamefont {J.~F.}\ \bibnamefont
  {Scott}},\ }\bibfield  {title} {\bibinfo {title} {Spectroscopy of
  incommensurate ferroelectrics},\ }\href
  {https://doi.org/10.1080/00150198008238631} {\bibfield  {journal} {\bibinfo
  {journal} {Ferroelectrics}\ }\textbf {\bibinfo {volume} {24}},\ \bibinfo
  {pages} {127} (\bibinfo {year} {1980})}\BibitemShut {NoStop}%
\bibitem [{\citenamefont {Aizu}(1970)}]{PhysRevB.2.754}%
  \BibitemOpen
  \bibfield  {author} {\bibinfo {author} {\bibfnamefont {K.}~\bibnamefont
  {Aizu}},\ }\bibfield  {title} {\bibinfo {title} {Possible species of
  ferromagnetic, ferroelectric, and ferroelastic crystals},\ }\href
  {https://doi.org/10.1103/PhysRevB.2.754} {\bibfield  {journal} {\bibinfo
  {journal} {Phys. Rev. B}\ }\textbf {\bibinfo {volume} {2}},\ \bibinfo {pages}
  {754} (\bibinfo {year} {1970})}\BibitemShut {NoStop}%
\bibitem [{\citenamefont {Seixas}\ \emph {et~al.}(2016)\citenamefont {Seixas},
  \citenamefont {Rodin}, \citenamefont {Carvalho},\ and\ \citenamefont
  {Castro~Neto}}]{seixas}%
  \BibitemOpen
  \bibfield  {author} {\bibinfo {author} {\bibfnamefont {L.}~\bibnamefont
  {Seixas}}, \bibinfo {author} {\bibfnamefont {A.~S.}\ \bibnamefont {Rodin}},
  \bibinfo {author} {\bibfnamefont {A.}~\bibnamefont {Carvalho}},\ and\
  \bibinfo {author} {\bibfnamefont {A.~H.}\ \bibnamefont {Castro~Neto}},\
  }\bibfield  {title} {\bibinfo {title} {Multiferroic two-dimensional
  materials},\ }\href {https://doi.org/10.1103/PhysRevLett.116.206803}
  {\bibfield  {journal} {\bibinfo  {journal} {Phys. Rev. Lett.}\ }\textbf
  {\bibinfo {volume} {116}},\ \bibinfo {pages} {206803} (\bibinfo {year}
  {2016})}\BibitemShut {NoStop}%
\bibitem [{\citenamefont {Bishop}\ \emph {et~al.}(2019)\citenamefont {Bishop},
  \citenamefont {Farmer}, \citenamefont {Sharmin}, \citenamefont
  {Pacheco-Sanjuan}, \citenamefont {Darancet},\ and\ \citenamefont
  {Barraza-Lopez}}]{Tyler}%
  \BibitemOpen
  \bibfield  {author} {\bibinfo {author} {\bibfnamefont {T.~B.}\ \bibnamefont
  {Bishop}}, \bibinfo {author} {\bibfnamefont {E.~E.}\ \bibnamefont {Farmer}},
  \bibinfo {author} {\bibfnamefont {A.}~\bibnamefont {Sharmin}}, \bibinfo
  {author} {\bibfnamefont {A.}~\bibnamefont {Pacheco-Sanjuan}}, \bibinfo
  {author} {\bibfnamefont {P.}~\bibnamefont {Darancet}},\ and\ \bibinfo
  {author} {\bibfnamefont {S.}~\bibnamefont {Barraza-Lopez}},\ }\bibfield
  {title} {\bibinfo {title} {Quantum paraelastic two-dimensional materials},\
  }\href {https://doi.org/10.1103/PhysRevLett.122.015703} {\bibfield  {journal}
  {\bibinfo  {journal} {Phys. Rev. Lett.}\ }\textbf {\bibinfo {volume} {122}},\
  \bibinfo {pages} {015703} (\bibinfo {year} {2019})}\BibitemShut {NoStop}%
\bibitem [{\citenamefont {Pacheco-Sanjuan}\ \emph {et~al.}(2019)\citenamefont
  {Pacheco-Sanjuan}, \citenamefont {Bishop}, \citenamefont {Farmer},
  \citenamefont {Kumar},\ and\ \citenamefont {Barraza-Lopez}}]{Alejandro}%
  \BibitemOpen
  \bibfield  {author} {\bibinfo {author} {\bibfnamefont {A.}~\bibnamefont
  {Pacheco-Sanjuan}}, \bibinfo {author} {\bibfnamefont {T.~B.}\ \bibnamefont
  {Bishop}}, \bibinfo {author} {\bibfnamefont {E.~E.}\ \bibnamefont {Farmer}},
  \bibinfo {author} {\bibfnamefont {P.}~\bibnamefont {Kumar}},\ and\ \bibinfo
  {author} {\bibfnamefont {S.}~\bibnamefont {Barraza-Lopez}},\ }\bibfield
  {title} {\bibinfo {title} {Evolution of elastic moduli through a
  two-dimensional structural transformation},\ }\href
  {https://doi.org/10.1103/PhysRevB.99.104108} {\bibfield  {journal} {\bibinfo
  {journal} {Phys. Rev. B}\ }\textbf {\bibinfo {volume} {99}},\ \bibinfo
  {pages} {104108} (\bibinfo {year} {2019})}\BibitemShut {NoStop}%
\bibitem [{\citenamefont {Mehboudi}\ \emph {et~al.}(2016)\citenamefont
  {Mehboudi}, \citenamefont {Dorio}, \citenamefont {Zhu}, \citenamefont
  {van~der Zande}, \citenamefont {Churchill}, \citenamefont {Pacheco-Sanjuan},
  \citenamefont {Harriss}, \citenamefont {Kumar},\ and\ \citenamefont
  {Barraza-Lopez}}]{Mehboudi}%
  \BibitemOpen
  \bibfield  {author} {\bibinfo {author} {\bibfnamefont {M.}~\bibnamefont
  {Mehboudi}}, \bibinfo {author} {\bibfnamefont {A.~M.}\ \bibnamefont {Dorio}},
  \bibinfo {author} {\bibfnamefont {W.}~\bibnamefont {Zhu}}, \bibinfo {author}
  {\bibfnamefont {A.}~\bibnamefont {van~der Zande}}, \bibinfo {author}
  {\bibfnamefont {H.~O.~H.}\ \bibnamefont {Churchill}}, \bibinfo {author}
  {\bibfnamefont {A.~A.}\ \bibnamefont {Pacheco-Sanjuan}}, \bibinfo {author}
  {\bibfnamefont {E.~O.}\ \bibnamefont {Harriss}}, \bibinfo {author}
  {\bibfnamefont {P.}~\bibnamefont {Kumar}},\ and\ \bibinfo {author}
  {\bibfnamefont {S.}~\bibnamefont {Barraza-Lopez}},\ }\bibfield  {title}
  {\bibinfo {title} {Two-dimensional disorder in black phosphorus and
  monochalcogenide monolayers},\ }\href
  {https://doi.org/10.1021/acs.nanolett.5b04613} {\bibfield  {journal}
  {\bibinfo  {journal} {Nano Lett.}\ }\textbf {\bibinfo {volume} {16}},\
  \bibinfo {pages} {1704} (\bibinfo {year} {2016})}\BibitemShut {NoStop}%
\bibitem [{\citenamefont {Wang}\ and\ \citenamefont {Qian}(2017)}]{Qian}%
  \BibitemOpen
  \bibfield  {author} {\bibinfo {author} {\bibfnamefont {H.}~\bibnamefont
  {Wang}}\ and\ \bibinfo {author} {\bibfnamefont {X.}~\bibnamefont {Qian}},\
  }\bibfield  {title} {\bibinfo {title} {Two-dimensional multiferroics in
  monolayer group {IV} monochalcogenides},\ }\href
  {https://doi.org/10.1088/2053-1583/4/1/015042} {\bibfield  {journal}
  {\bibinfo  {journal} {2D Mater.}\ }\textbf {\bibinfo {volume} {4}},\ \bibinfo
  {pages} {015042} (\bibinfo {year} {2017})}\BibitemShut {NoStop}%
\bibitem [{\citenamefont {Chang}\ \emph {et~al.}(2016)\citenamefont {Chang},
  \citenamefont {Liu}, \citenamefont {Lin}, \citenamefont {Wang}, \citenamefont
  {Zhao}, \citenamefont {Zhang}, \citenamefont {Jin}, \citenamefont {Zhong},
  \citenamefont {Hu}, \citenamefont {Duan}, \citenamefont {Zhang},
  \citenamefont {Fu}, \citenamefont {Xue}, \citenamefont {Chen},\ and\
  \citenamefont {Ji}}]{Kai2016}%
  \BibitemOpen
  \bibfield  {author} {\bibinfo {author} {\bibfnamefont {K.}~\bibnamefont
  {Chang}}, \bibinfo {author} {\bibfnamefont {J.}~\bibnamefont {Liu}}, \bibinfo
  {author} {\bibfnamefont {H.}~\bibnamefont {Lin}}, \bibinfo {author}
  {\bibfnamefont {N.}~\bibnamefont {Wang}}, \bibinfo {author} {\bibfnamefont
  {K.}~\bibnamefont {Zhao}}, \bibinfo {author} {\bibfnamefont {A.}~\bibnamefont
  {Zhang}}, \bibinfo {author} {\bibfnamefont {F.}~\bibnamefont {Jin}}, \bibinfo
  {author} {\bibfnamefont {Y.}~\bibnamefont {Zhong}}, \bibinfo {author}
  {\bibfnamefont {X.}~\bibnamefont {Hu}}, \bibinfo {author} {\bibfnamefont
  {W.}~\bibnamefont {Duan}}, \bibinfo {author} {\bibfnamefont {Q.}~\bibnamefont
  {Zhang}}, \bibinfo {author} {\bibfnamefont {L.}~\bibnamefont {Fu}}, \bibinfo
  {author} {\bibfnamefont {Q.-K.}\ \bibnamefont {Xue}}, \bibinfo {author}
  {\bibfnamefont {X.}~\bibnamefont {Chen}},\ and\ \bibinfo {author}
  {\bibfnamefont {S.-H.}\ \bibnamefont {Ji}},\ }\bibfield  {title} {\bibinfo
  {title} {Discovery of robust in-plane ferroelectricity in atomic-thick
  {SnTe}},\ }\href {https://doi.org/10.1126/science.aad8609} {\bibfield
  {journal} {\bibinfo  {journal} {Science}\ }\textbf {\bibinfo {volume}
  {353}},\ \bibinfo {pages} {274} (\bibinfo {year} {2016})}\BibitemShut
  {NoStop}%
\bibitem [{\citenamefont {Barraza-Lopez}\ \emph {et~al.}(2018)\citenamefont
  {Barraza-Lopez}, \citenamefont {Kaloni}, \citenamefont {Poudel},\ and\
  \citenamefont {Kumar}}]{Barraza2018}%
  \BibitemOpen
  \bibfield  {author} {\bibinfo {author} {\bibfnamefont {S.}~\bibnamefont
  {Barraza-Lopez}}, \bibinfo {author} {\bibfnamefont {T.~P.}\ \bibnamefont
  {Kaloni}}, \bibinfo {author} {\bibfnamefont {S.~P.}\ \bibnamefont {Poudel}},\
  and\ \bibinfo {author} {\bibfnamefont {P.}~\bibnamefont {Kumar}},\ }\bibfield
   {title} {\bibinfo {title} {Tuning the ferroelectric-to-paraelectric
  transition temperature and dipole orientation of {group-IV} monochalcogenide
  monolayers},\ }\href {https://doi.org/10.1103/PhysRevB.97.024110} {\bibfield
  {journal} {\bibinfo  {journal} {Phys. Rev. B}\ }\textbf {\bibinfo {volume}
  {97}},\ \bibinfo {pages} {024110} (\bibinfo {year} {2018})}\BibitemShut
  {NoStop}%
\bibitem [{\citenamefont {Barraza-Lopez}\ \emph {et~al.}(2021)\citenamefont
  {Barraza-Lopez}, \citenamefont {Fregoso}, \citenamefont {Villanova},
  \citenamefont {Parkin},\ and\ \citenamefont {Chang}}]{RMP2021}%
  \BibitemOpen
  \bibfield  {author} {\bibinfo {author} {\bibfnamefont {S.}~\bibnamefont
  {Barraza-Lopez}}, \bibinfo {author} {\bibfnamefont {B.~M.}\ \bibnamefont
  {Fregoso}}, \bibinfo {author} {\bibfnamefont {J.~W.}\ \bibnamefont
  {Villanova}}, \bibinfo {author} {\bibfnamefont {S.~S.~P.}\ \bibnamefont
  {Parkin}},\ and\ \bibinfo {author} {\bibfnamefont {K.}~\bibnamefont
  {Chang}},\ }\bibfield  {title} {\bibinfo {title} {Colloquium: Physical
  properties of {group-IV} monochalcogenide monolayers},\ }\href
  {https://doi.org/10.1103/RevModPhys.93.011001} {\bibfield  {journal}
  {\bibinfo  {journal} {Rev. Mod. Phys.}\ }\textbf {\bibinfo {volume} {93}},\
  \bibinfo {pages} {011001} (\bibinfo {year} {2021})}\BibitemShut {NoStop}%
\bibitem [{\citenamefont {Cheng}\ \emph {et~al.}(2021)\citenamefont {Cheng},
  \citenamefont {Zhou}, \citenamefont {Wang}, \citenamefont {Ji},\ and\
  \citenamefont {Zhang}}]{EMproximal1}%
  \BibitemOpen
  \bibfield  {author} {\bibinfo {author} {\bibfnamefont {H.-X.}\ \bibnamefont
  {Cheng}}, \bibinfo {author} {\bibfnamefont {J.}~\bibnamefont {Zhou}},
  \bibinfo {author} {\bibfnamefont {C.}~\bibnamefont {Wang}}, \bibinfo {author}
  {\bibfnamefont {W.}~\bibnamefont {Ji}},\ and\ \bibinfo {author}
  {\bibfnamefont {Y.-N.}\ \bibnamefont {Zhang}},\ }\bibfield  {title} {\bibinfo
  {title} {Nonvolatile electric field control of magnetism in bilayer
  {CrI$_{3}$ on monolayer In$_{2}$Se$_{3}$}},\ }\href
  {https://doi.org/10.1103/PhysRevB.104.064443} {\bibfield  {journal} {\bibinfo
   {journal} {Phys. Rev. B}\ }\textbf {\bibinfo {volume} {104}},\ \bibinfo
  {pages} {064443} (\bibinfo {year} {2021})}\BibitemShut {NoStop}%
\bibitem [{\citenamefont {Hu}\ \emph {et~al.}(2022)\citenamefont {Hu},
  \citenamefont {Chen}, \citenamefont {Du}, \citenamefont {Ju}, \citenamefont
  {An},\ and\ \citenamefont {Gong}}]{EMproximal2}%
  \BibitemOpen
  \bibfield  {author} {\bibinfo {author} {\bibfnamefont {C.}~\bibnamefont
  {Hu}}, \bibinfo {author} {\bibfnamefont {J.}~\bibnamefont {Chen}}, \bibinfo
  {author} {\bibfnamefont {E.}~\bibnamefont {Du}}, \bibinfo {author}
  {\bibfnamefont {W.}~\bibnamefont {Ju}}, \bibinfo {author} {\bibfnamefont
  {Y.}~\bibnamefont {An}},\ and\ \bibinfo {author} {\bibfnamefont {S.-J.}\
  \bibnamefont {Gong}},\ }\bibfield  {title} {\bibinfo {title} {Ferroelectric
  control of band alignments and magnetic properties in the two-dimensional
  multiferroic {VSe$_2$/In$_2$Se$_3$}},\ }\href
  {https://doi.org/10.1088/1361-648X/ac8406} {\bibfield  {journal} {\bibinfo
  {journal} {J. Phys.: Condens. Matter}\ }\textbf {\bibinfo {volume} {34}},\
  \bibinfo {pages} {425801} (\bibinfo {year} {2022})}\BibitemShut {NoStop}%
\bibitem [{\citenamefont {Zhu}\ \emph {et~al.}(2022)\citenamefont {Zhu},
  \citenamefont {Gao}, \citenamefont {Hou}, \citenamefont {Gui},\ and\
  \citenamefont {Huang}}]{EMproximal3}%
  \BibitemOpen
  \bibfield  {author} {\bibinfo {author} {\bibfnamefont {H.}~\bibnamefont
  {Zhu}}, \bibinfo {author} {\bibfnamefont {Y.}~\bibnamefont {Gao}}, \bibinfo
  {author} {\bibfnamefont {Y.}~\bibnamefont {Hou}}, \bibinfo {author}
  {\bibfnamefont {Z.}~\bibnamefont {Gui}},\ and\ \bibinfo {author}
  {\bibfnamefont {L.}~\bibnamefont {Huang}},\ }\bibfield  {title} {\bibinfo
  {title} {Tunable magnetic anisotropy in two-dimensional {CrX$_3$/AlN (X=I,
  Br, Cl)} heterostructures},\ }\href
  {https://doi.org/10.1103/PhysRevB.106.134412} {\bibfield  {journal} {\bibinfo
   {journal} {Phys. Rev. B}\ }\textbf {\bibinfo {volume} {106}},\ \bibinfo
  {pages} {134412} (\bibinfo {year} {2022})}\BibitemShut {NoStop}%
\bibitem [{\citenamefont {Gong}\ \emph {et~al.}(2019)\citenamefont {Gong},
  \citenamefont {Kim}, \citenamefont {Wang}, \citenamefont {Lee},\ and\
  \citenamefont {Zhang}}]{HeterostructuresToo}%
  \BibitemOpen
  \bibfield  {author} {\bibinfo {author} {\bibfnamefont {C.}~\bibnamefont
  {Gong}}, \bibinfo {author} {\bibfnamefont {E.~M.}\ \bibnamefont {Kim}},
  \bibinfo {author} {\bibfnamefont {Y.}~\bibnamefont {Wang}}, \bibinfo {author}
  {\bibfnamefont {G.}~\bibnamefont {Lee}},\ and\ \bibinfo {author}
  {\bibfnamefont {X.}~\bibnamefont {Zhang}},\ }\bibfield  {title} {\bibinfo
  {title} {Multiferroicity in atomic van der {Waals} heterostructures},\ }\href
  {https://doi.org/10.1038/s41467-019-10693-0} {\bibfield  {journal} {\bibinfo
  {journal} {Nat. Commun.}\ }\textbf {\bibinfo {volume} {10}},\ \bibinfo
  {pages} {2657} (\bibinfo {year} {2019})}\BibitemShut {NoStop}%
\bibitem [{\citenamefont {Lu}\ \emph {et~al.}(2022)\citenamefont {Lu},
  \citenamefont {Wang}, \citenamefont {Wang},\ and\ \citenamefont
  {Yang}}]{MagneticOrderByDoping}%
  \BibitemOpen
  \bibfield  {author} {\bibinfo {author} {\bibfnamefont {Y.}~\bibnamefont
  {Lu}}, \bibinfo {author} {\bibfnamefont {H.}~\bibnamefont {Wang}}, \bibinfo
  {author} {\bibfnamefont {L.}~\bibnamefont {Wang}},\ and\ \bibinfo {author}
  {\bibfnamefont {L.}~\bibnamefont {Yang}},\ }\bibfield  {title} {\bibinfo
  {title} {Mechanism of carrier doping induced magnetic phase transitions in
  two-dimensional materials},\ }\href
  {https://doi.org/10.1103/PhysRevB.106.205403} {\bibfield  {journal} {\bibinfo
   {journal} {Phys. Rev. B}\ }\textbf {\bibinfo {volume} {106}},\ \bibinfo
  {pages} {205403} (\bibinfo {year} {2022})}\BibitemShut {NoStop}%
\bibitem [{\citenamefont {Verzhbitskiy}\ \emph {et~al.}(2020)\citenamefont
  {Verzhbitskiy}, \citenamefont {Kurebayashi}, \citenamefont {Cheng},
  \citenamefont {Zhou}, \citenamefont {Khan}, \citenamefont {Feng},\ and\
  \citenamefont {Eda}}]{EdaTransfer}%
  \BibitemOpen
  \bibfield  {author} {\bibinfo {author} {\bibfnamefont {I.~A.}\ \bibnamefont
  {Verzhbitskiy}}, \bibinfo {author} {\bibfnamefont {H.}~\bibnamefont
  {Kurebayashi}}, \bibinfo {author} {\bibfnamefont {H.}~\bibnamefont {Cheng}},
  \bibinfo {author} {\bibfnamefont {J.}~\bibnamefont {Zhou}}, \bibinfo {author}
  {\bibfnamefont {S.}~\bibnamefont {Khan}}, \bibinfo {author} {\bibfnamefont
  {Y.~P.}\ \bibnamefont {Feng}},\ and\ \bibinfo {author} {\bibfnamefont
  {G.}~\bibnamefont {Eda}},\ }\bibfield  {title} {\bibinfo {title}
  {{Controlling the magnetic anisotropy in Cr$_2$Ge$_2$Te$_6$ by electrostatic
  gating}},\ }\href {https://doi.org/10.1038/s41928-020-0427-7} {\bibfield
  {journal} {\bibinfo  {journal} {Nat. Electron.}\ }\textbf {\bibinfo {volume}
  {3}},\ \bibinfo {pages} {460} (\bibinfo {year} {2020})}\BibitemShut {NoStop}%
\bibitem [{\citenamefont {Gao}\ \emph {et~al.}(2021)\citenamefont {Gao},
  \citenamefont {Gao},\ and\ \citenamefont {Lu}}]{2DMultiferroicsReview}%
  \BibitemOpen
  \bibfield  {author} {\bibinfo {author} {\bibfnamefont {Y.}~\bibnamefont
  {Gao}}, \bibinfo {author} {\bibfnamefont {M.}~\bibnamefont {Gao}},\ and\
  \bibinfo {author} {\bibfnamefont {Y.}~\bibnamefont {Lu}},\ }\bibfield
  {title} {\bibinfo {title} {Two-dimensional multiferroics},\ }\href
  {https://doi.org/10.1039/D1NR06598J} {\bibfield  {journal} {\bibinfo
  {journal} {Nanoscale}\ }\textbf {\bibinfo {volume} {13}},\ \bibinfo {pages}
  {19324} (\bibinfo {year} {2021})}\BibitemShut {NoStop}%
\bibitem [{\citenamefont {Yīng}\ and\ \citenamefont
  {Z\"{u}licke}(2022)}]{review2Dmultiferroics}%
  \BibitemOpen
  \bibfield  {author} {\bibinfo {author} {\bibfnamefont {Y.}~\bibnamefont
  {Yīng}}\ and\ \bibinfo {author} {\bibfnamefont {U.}~\bibnamefont
  {Z\"{u}licke}},\ }\bibfield  {title} {\bibinfo {title} {Magnetoelectricity in
  two-dimensional materials},\ }\href
  {https://doi.org/10.1080/23746149.2022.2032343} {\bibfield  {journal}
  {\bibinfo  {journal} {Adv. Phys.: X}\ }\textbf {\bibinfo {volume} {7}},\
  \bibinfo {pages} {2032343} (\bibinfo {year} {2022})}\BibitemShut {NoStop}%
\bibitem [{\citenamefont {Fumega}\ and\ \citenamefont
  {Lado}(2022)}]{2DanalyticalModel_2}%
  \BibitemOpen
  \bibfield  {author} {\bibinfo {author} {\bibfnamefont {A.~O.}\ \bibnamefont
  {Fumega}}\ and\ \bibinfo {author} {\bibfnamefont {J.~L.}\ \bibnamefont
  {Lado}},\ }\bibfield  {title} {\bibinfo {title} {{Moir{\'e}-driven
  multiferroic order in twisted CrCl$_3$, CrBr$_3$ and CrI$_3$ bilayers}}}
  (\bibinfo {year} {2022}),\ \bibinfo {note} {arXiv:2207.01416}\BibitemShut
  {NoStop}%
\bibitem [{\citenamefont {Ju}\ \emph {et~al.}(2021)\citenamefont {Ju},
  \citenamefont {Lee}, \citenamefont {Kim}, \citenamefont {Choi}, \citenamefont
  {Roh}, \citenamefont {Son}, \citenamefont {Park}, \citenamefont {Kim},
  \citenamefont {Jung}, \citenamefont {Kim}, \citenamefont {Kim}, \citenamefont
  {Park},\ and\ \citenamefont {Lee}}]{NiI2}%
  \BibitemOpen
  \bibfield  {author} {\bibinfo {author} {\bibfnamefont {H.}~\bibnamefont
  {Ju}}, \bibinfo {author} {\bibfnamefont {Y.}~\bibnamefont {Lee}}, \bibinfo
  {author} {\bibfnamefont {K.-T.}\ \bibnamefont {Kim}}, \bibinfo {author}
  {\bibfnamefont {I.~H.}\ \bibnamefont {Choi}}, \bibinfo {author}
  {\bibfnamefont {C.~J.}\ \bibnamefont {Roh}}, \bibinfo {author} {\bibfnamefont
  {S.}~\bibnamefont {Son}}, \bibinfo {author} {\bibfnamefont {P.}~\bibnamefont
  {Park}}, \bibinfo {author} {\bibfnamefont {J.~H.}\ \bibnamefont {Kim}},
  \bibinfo {author} {\bibfnamefont {T.~S.}\ \bibnamefont {Jung}}, \bibinfo
  {author} {\bibfnamefont {J.~H.}\ \bibnamefont {Kim}}, \bibinfo {author}
  {\bibfnamefont {K.~H.}\ \bibnamefont {Kim}}, \bibinfo {author} {\bibfnamefont
  {J.-G.}\ \bibnamefont {Park}},\ and\ \bibinfo {author} {\bibfnamefont
  {J.~S.}\ \bibnamefont {Lee}},\ }\bibfield  {title} {\bibinfo {title}
  {Possible persistence of multiferroic order down to bilayer limit of van der
  waals material {NiI$_2$}},\ }\href
  {https://doi.org/10.1021/acs.nanolett.1c01095} {\bibfield  {journal}
  {\bibinfo  {journal} {Nano Lett.}\ }\textbf {\bibinfo {volume} {21}},\
  \bibinfo {pages} {5126} (\bibinfo {year} {2021})}\BibitemShut {NoStop}%
\bibitem [{\citenamefont {Song}\ \emph {et~al.}(2022)\citenamefont {Song},
  \citenamefont {Occhialini}, \citenamefont {Erge\c{c}en}, \citenamefont
  {Ilyas}, \citenamefont {Amoroso}, \citenamefont {Barone}, \citenamefont
  {Kapeghian}, \citenamefont {Watanabe}, \citenamefont {Taniguchi},
  \citenamefont {Botana}, \citenamefont {Picozzi}, \citenamefont {Gedik},\ and\
  \citenamefont {Comin}}]{Picozzi}%
  \BibitemOpen
  \bibfield  {author} {\bibinfo {author} {\bibfnamefont {Q.}~\bibnamefont
  {Song}}, \bibinfo {author} {\bibfnamefont {C.~A.}\ \bibnamefont
  {Occhialini}}, \bibinfo {author} {\bibfnamefont {E.}~\bibnamefont
  {Erge\c{c}en}}, \bibinfo {author} {\bibfnamefont {B.}~\bibnamefont {Ilyas}},
  \bibinfo {author} {\bibfnamefont {D.}~\bibnamefont {Amoroso}}, \bibinfo
  {author} {\bibfnamefont {P.}~\bibnamefont {Barone}}, \bibinfo {author}
  {\bibfnamefont {J.}~\bibnamefont {Kapeghian}}, \bibinfo {author}
  {\bibfnamefont {K.}~\bibnamefont {Watanabe}}, \bibinfo {author}
  {\bibfnamefont {T.}~\bibnamefont {Taniguchi}}, \bibinfo {author}
  {\bibfnamefont {A.~S.}\ \bibnamefont {Botana}}, \bibinfo {author}
  {\bibfnamefont {S.}~\bibnamefont {Picozzi}}, \bibinfo {author} {\bibfnamefont
  {N.}~\bibnamefont {Gedik}},\ and\ \bibinfo {author} {\bibfnamefont
  {R.}~\bibnamefont {Comin}},\ }\bibfield  {title} {\bibinfo {title} {Evidence
  for a single-layer van der {Waals} multiferroic},\ }\href
  {https://doi.org/10.1038/s41586-021-04337-x} {\bibfield  {journal} {\bibinfo
  {journal} {Nature}\ }\textbf {\bibinfo {volume} {602}},\ \bibinfo {pages}
  {601} (\bibinfo {year} {2022})}\BibitemShut {NoStop}%
\bibitem [{\citenamefont {Ji}\ \emph {et~al.}(2023)\citenamefont {Ji},
  \citenamefont {Yu}, \citenamefont {Xu},\ and\ \citenamefont {Xiang}}]{New}%
  \BibitemOpen
  \bibfield  {author} {\bibinfo {author} {\bibfnamefont {J.}~\bibnamefont
  {Ji}}, \bibinfo {author} {\bibfnamefont {G.}~\bibnamefont {Yu}}, \bibinfo
  {author} {\bibfnamefont {C.}~\bibnamefont {Xu}},\ and\ \bibinfo {author}
  {\bibfnamefont {H.~J.}\ \bibnamefont {Xiang}},\ }\bibfield  {title} {\bibinfo
  {title} {General theory for bilayer stacking ferroelectricity},\ }\href
  {https://doi.org/10.1103/PhysRevLett.130.146801} {\bibfield  {journal}
  {\bibinfo  {journal} {Phys. Rev. Lett.}\ }\textbf {\bibinfo {volume} {130}},\
  \bibinfo {pages} {146801} (\bibinfo {year} {2023})}\BibitemShut {NoStop}%
\bibitem [{\citenamefont {Huang}\ \emph {et~al.}(2018)\citenamefont {Huang},
  \citenamefont {Clark}, \citenamefont {Klein}, \citenamefont {MacNeill},
  \citenamefont {Navarro-Moratalla}, \citenamefont {Seyler}, \citenamefont
  {Wilson}, \citenamefont {McGuire}, \citenamefont {Cobden}, \citenamefont
  {Xiao}, \citenamefont {Yao}, \citenamefont {Jarillo-Herrero},\ and\
  \citenamefont {Xu}}]{CrI3_bilayer1}%
  \BibitemOpen
  \bibfield  {author} {\bibinfo {author} {\bibfnamefont {B.}~\bibnamefont
  {Huang}}, \bibinfo {author} {\bibfnamefont {G.}~\bibnamefont {Clark}},
  \bibinfo {author} {\bibfnamefont {D.~R.}\ \bibnamefont {Klein}}, \bibinfo
  {author} {\bibfnamefont {D.}~\bibnamefont {MacNeill}}, \bibinfo {author}
  {\bibfnamefont {E.}~\bibnamefont {Navarro-Moratalla}}, \bibinfo {author}
  {\bibfnamefont {K.~L.}\ \bibnamefont {Seyler}}, \bibinfo {author}
  {\bibfnamefont {N.}~\bibnamefont {Wilson}}, \bibinfo {author} {\bibfnamefont
  {M.~A.}\ \bibnamefont {McGuire}}, \bibinfo {author} {\bibfnamefont {D.~H.}\
  \bibnamefont {Cobden}}, \bibinfo {author} {\bibfnamefont {D.}~\bibnamefont
  {Xiao}}, \bibinfo {author} {\bibfnamefont {W.}~\bibnamefont {Yao}}, \bibinfo
  {author} {\bibfnamefont {P.}~\bibnamefont {Jarillo-Herrero}},\ and\ \bibinfo
  {author} {\bibfnamefont {X.}~\bibnamefont {Xu}},\ }\bibfield  {title}
  {\bibinfo {title} {Electrical control of {2D} magnetism in bilayer
  {CrI$_3$}},\ }\href {https://doi.org/10.1038/s41565-018-0135-x} {\bibfield
  {journal} {\bibinfo  {journal} {Nat. Nanotech.}\ }\textbf {\bibinfo {volume}
  {13}},\ \bibinfo {pages} {544} (\bibinfo {year} {2018})}\BibitemShut
  {NoStop}%
\bibitem [{\citenamefont {Jiang}\ \emph
  {et~al.}(2018{\natexlab{a}})\citenamefont {Jiang}, \citenamefont {Li},
  \citenamefont {Wang}, \citenamefont {Mak},\ and\ \citenamefont
  {Shan}}]{CrI3_bilayer2}%
  \BibitemOpen
  \bibfield  {author} {\bibinfo {author} {\bibfnamefont {S.}~\bibnamefont
  {Jiang}}, \bibinfo {author} {\bibfnamefont {L.}~\bibnamefont {Li}}, \bibinfo
  {author} {\bibfnamefont {Z.}~\bibnamefont {Wang}}, \bibinfo {author}
  {\bibfnamefont {K.~F.}\ \bibnamefont {Mak}},\ and\ \bibinfo {author}
  {\bibfnamefont {J.}~\bibnamefont {Shan}},\ }\bibfield  {title} {\bibinfo
  {title} {Controlling magnetism in {2D CrI$_3$} by electrostatic doping},\
  }\href {https://doi.org/10.1038/s41565-018-0135-x} {\bibfield  {journal}
  {\bibinfo  {journal} {Nat. Nanotech.}\ }\textbf {\bibinfo {volume} {13}},\
  \bibinfo {pages} {549} (\bibinfo {year} {2018}{\natexlab{a}})}\BibitemShut
  {NoStop}%
\bibitem [{\citenamefont {Jiang}\ \emph
  {et~al.}(2018{\natexlab{b}})\citenamefont {Jiang}, \citenamefont {Shan},\
  and\ \citenamefont {Mak}}]{CrI3_bilayer3}%
  \BibitemOpen
  \bibfield  {author} {\bibinfo {author} {\bibfnamefont {S.}~\bibnamefont
  {Jiang}}, \bibinfo {author} {\bibfnamefont {J.}~\bibnamefont {Shan}},\ and\
  \bibinfo {author} {\bibfnamefont {K.~F.}\ \bibnamefont {Mak}},\ }\bibfield
  {title} {\bibinfo {title} {Electric-field switching of two-dimensional van
  der {Waals} magnets},\ }\href {https://doi.org/10.1038/s41563-018-0040-6}
  {\bibfield  {journal} {\bibinfo  {journal} {Nat. Mater.}\ }\textbf {\bibinfo
  {volume} {17}},\ \bibinfo {pages} {406} (\bibinfo {year}
  {2018}{\natexlab{b}})}\BibitemShut {NoStop}%
\bibitem [{\citenamefont {Mak}\ \emph {et~al.}(2019)\citenamefont {Mak},
  \citenamefont {Shan},\ and\ \citenamefont {Ralph}}]{MakRalphReview_2019}%
  \BibitemOpen
  \bibfield  {author} {\bibinfo {author} {\bibfnamefont {K.~F.}\ \bibnamefont
  {Mak}}, \bibinfo {author} {\bibfnamefont {J.}~\bibnamefont {Shan}},\ and\
  \bibinfo {author} {\bibfnamefont {D.}~\bibnamefont {Ralph}},\ }\bibfield
  {title} {\bibinfo {title} {Probing and controlling magnetic states in {2D}
  layered magnetic materials},\ }\href
  {https://doi.org/10.1038/s42254-019-0110-y} {\bibfield  {journal} {\bibinfo
  {journal} {Nat. Rev. Phys.}\ }\textbf {\bibinfo {volume} {1}},\ \bibinfo
  {pages} {646} (\bibinfo {year} {2019})}\BibitemShut {NoStop}%
\bibitem [{\citenamefont {Su{\'a}rez~Morell}\ \emph {et~al.}(2019)\citenamefont
  {Su{\'a}rez~Morell}, \citenamefont {Le{\'o}n}, \citenamefont {Hiroki~Miwa},\
  and\ \citenamefont {Vargas}}]{SuarezMorell_2019}%
  \BibitemOpen
  \bibfield  {author} {\bibinfo {author} {\bibfnamefont {E.}~\bibnamefont
  {Su{\'a}rez~Morell}}, \bibinfo {author} {\bibfnamefont {A.}~\bibnamefont
  {Le{\'o}n}}, \bibinfo {author} {\bibfnamefont {R.}~\bibnamefont
  {Hiroki~Miwa}},\ and\ \bibinfo {author} {\bibfnamefont {P.}~\bibnamefont
  {Vargas}},\ }\bibfield  {title} {\bibinfo {title} {{Control of magnetism in
  bilayer {CrI$_3$} by an external electric field}},\ }\href
  {https://doi.org/10.1088/2053-1583/ab04fb} {\bibfield  {journal} {\bibinfo
  {journal} {2D Mater.}\ }\textbf {\bibinfo {volume} {6}},\ \bibinfo {pages}
  {025020} (\bibinfo {year} {2019})}\BibitemShut {NoStop}%
\bibitem [{\citenamefont {Sun}\ \emph {et~al.}(2019)\citenamefont {Sun},
  \citenamefont {Yi}, \citenamefont {Song}, \citenamefont {Clark},
  \citenamefont {Huang}, \citenamefont {Shan}, \citenamefont {Wu},
  \citenamefont {Huang}, \citenamefont {Gao}, \citenamefont {Chen},
  \citenamefont {McGuire}, \citenamefont {Cao}, \citenamefont {Xiao},
  \citenamefont {Liu}, \citenamefont {Yao}, \citenamefont {Xu},\ and\
  \citenamefont {Wu}}]{Sun2019}%
  \BibitemOpen
  \bibfield  {author} {\bibinfo {author} {\bibfnamefont {Z.}~\bibnamefont
  {Sun}}, \bibinfo {author} {\bibfnamefont {Y.}~\bibnamefont {Yi}}, \bibinfo
  {author} {\bibfnamefont {T.}~\bibnamefont {Song}}, \bibinfo {author}
  {\bibfnamefont {G.}~\bibnamefont {Clark}}, \bibinfo {author} {\bibfnamefont
  {B.}~\bibnamefont {Huang}}, \bibinfo {author} {\bibfnamefont
  {Y.}~\bibnamefont {Shan}}, \bibinfo {author} {\bibfnamefont {S.}~\bibnamefont
  {Wu}}, \bibinfo {author} {\bibfnamefont {D.}~\bibnamefont {Huang}}, \bibinfo
  {author} {\bibfnamefont {C.}~\bibnamefont {Gao}}, \bibinfo {author}
  {\bibfnamefont {Z.}~\bibnamefont {Chen}}, \bibinfo {author} {\bibfnamefont
  {M.}~\bibnamefont {McGuire}}, \bibinfo {author} {\bibfnamefont
  {T.}~\bibnamefont {Cao}}, \bibinfo {author} {\bibfnamefont {D.}~\bibnamefont
  {Xiao}}, \bibinfo {author} {\bibfnamefont {W.-T.}\ \bibnamefont {Liu}},
  \bibinfo {author} {\bibfnamefont {W.}~\bibnamefont {Yao}}, \bibinfo {author}
  {\bibfnamefont {X.}~\bibnamefont {Xu}},\ and\ \bibinfo {author}
  {\bibfnamefont {S.}~\bibnamefont {Wu}},\ }\bibfield  {title} {\bibinfo
  {title} {Giant nonreciprocal second-harmonic generation from
  antiferromagnetic bilayer {CrI$_3$}},\ }\href
  {https://doi.org/10.1038/s41586-019-1445-3} {\bibfield  {journal} {\bibinfo
  {journal} {Nature}\ }\textbf {\bibinfo {volume} {572}},\ \bibinfo {pages}
  {497} (\bibinfo {year} {2019})}\BibitemShut {NoStop}%
\bibitem [{\citenamefont {Sivadas}\ \emph {et~al.}(2016)\citenamefont
  {Sivadas}, \citenamefont {Okamoto},\ and\ \citenamefont
  {Xiao}}]{sivadasKerr2016}%
  \BibitemOpen
  \bibfield  {author} {\bibinfo {author} {\bibfnamefont {N.}~\bibnamefont
  {Sivadas}}, \bibinfo {author} {\bibfnamefont {S.}~\bibnamefont {Okamoto}},\
  and\ \bibinfo {author} {\bibfnamefont {D.}~\bibnamefont {Xiao}},\ }\bibfield
  {title} {\bibinfo {title} {Gate-controllable magneto-optic kerr effect in
  layered collinear antiferromagnets},\ }\href
  {https://doi.org/10.1103/PhysRevLett.117.267203} {\bibfield  {journal}
  {\bibinfo  {journal} {Phys. Rev. Lett.}\ }\textbf {\bibinfo {volume} {117}},\
  \bibinfo {pages} {267203} (\bibinfo {year} {2016})}\BibitemShut {NoStop}%
\bibitem [{\citenamefont {Webster}\ and\ \citenamefont {Yan}(2018)}]{YiAnYan}%
  \BibitemOpen
  \bibfield  {author} {\bibinfo {author} {\bibfnamefont {L.}~\bibnamefont
  {Webster}}\ and\ \bibinfo {author} {\bibfnamefont {J.-A.}\ \bibnamefont
  {Yan}},\ }\bibfield  {title} {\bibinfo {title} {Strain-tunable magnetic
  anisotropy in monolayer {CrCl$_3$, CrBr$_3$, and CrI$_3$}},\ }\href
  {https://doi.org/10.1103/PhysRevB.98.144411} {\bibfield  {journal} {\bibinfo
  {journal} {Phys. Rev. B}\ }\textbf {\bibinfo {volume} {98}},\ \bibinfo
  {pages} {144411} (\bibinfo {year} {2018})}\BibitemShut {NoStop}%
\bibitem [{\citenamefont {Le{\'o}n}\ \emph {et~al.}(2020)\citenamefont
  {Le{\'o}n}, \citenamefont {Gonz{\'a}lez}, \citenamefont
  {Mej{\'\i}a-L{\'o}pez}, \citenamefont {Crasto~de Lima},\ and\ \citenamefont
  {Su{\'a}rez~Morell}}]{Leon_2020}%
  \BibitemOpen
  \bibfield  {author} {\bibinfo {author} {\bibfnamefont {A.~M.}\ \bibnamefont
  {Le{\'o}n}}, \bibinfo {author} {\bibfnamefont {J.~W.}\ \bibnamefont
  {Gonz{\'a}lez}}, \bibinfo {author} {\bibfnamefont {J.}~\bibnamefont
  {Mej{\'\i}a-L{\'o}pez}}, \bibinfo {author} {\bibfnamefont {F.}~\bibnamefont
  {Crasto~de Lima}},\ and\ \bibinfo {author} {\bibfnamefont {E.}~\bibnamefont
  {Su{\'a}rez~Morell}},\ }\bibfield  {title} {\bibinfo {title} {Strain-induced
  phase transition in {CrI$_3$} bilayers},\ }\href
  {https://doi.org/10.1088/2053-1583/ab8268} {\bibfield  {journal} {\bibinfo
  {journal} {2D Mater.}\ }\textbf {\bibinfo {volume} {7}},\ \bibinfo {pages}
  {035008} (\bibinfo {year} {2020})}\BibitemShut {NoStop}%
\bibitem [{\citenamefont {Liu}\ \emph {et~al.}(2023)\citenamefont {Liu},
  \citenamefont {Ma}, \citenamefont {Xu}, \citenamefont {Li},\ and\
  \citenamefont {Zhao}}]{MnSe}%
  \BibitemOpen
  \bibfield  {author} {\bibinfo {author} {\bibfnamefont {K.}~\bibnamefont
  {Liu}}, \bibinfo {author} {\bibfnamefont {X.}~\bibnamefont {Ma}}, \bibinfo
  {author} {\bibfnamefont {S.}~\bibnamefont {Xu}}, \bibinfo {author}
  {\bibfnamefont {Y.}~\bibnamefont {Li}},\ and\ \bibinfo {author}
  {\bibfnamefont {M.}~\bibnamefont {Zhao}},\ }\bibfield  {title} {\bibinfo
  {title} {Tunable sliding ferroelectricity and magnetoelectric coupling in
  two-dimensional multiferroic {MnSe} materials},\ }\href
  {https://doi.org/10.1038/s41524-023-00972-2} {\bibfield  {journal} {\bibinfo
  {journal} {npj Comput. Mater.}\ }\textbf {\bibinfo {volume} {9}},\ \bibinfo
  {pages} {16} (\bibinfo {year} {2023})}\BibitemShut {NoStop}%
\bibitem [{\citenamefont {Perdew}\ \emph {et~al.}(1996)\citenamefont {Perdew},
  \citenamefont {Burke},\ and\ \citenamefont {Ernzerhof}}]{pbe}%
  \BibitemOpen
  \bibfield  {author} {\bibinfo {author} {\bibfnamefont {J.~P.}\ \bibnamefont
  {Perdew}}, \bibinfo {author} {\bibfnamefont {K.}~\bibnamefont {Burke}},\ and\
  \bibinfo {author} {\bibfnamefont {M.}~\bibnamefont {Ernzerhof}},\ }\bibfield
  {title} {\bibinfo {title} {Generalized gradient approximation made simple},\
  }\href {https://doi.org/10.1103/PhysRevLett.77.3865} {\bibfield  {journal}
  {\bibinfo  {journal} {Phys. Rev. Lett.}\ }\textbf {\bibinfo {volume} {77}},\
  \bibinfo {pages} {3865} (\bibinfo {year} {1996})}\BibitemShut {NoStop}%
\bibitem [{\citenamefont {Kresse}\ and\ \citenamefont {Hafner}(1993)}]{vasp-1}%
  \BibitemOpen
  \bibfield  {author} {\bibinfo {author} {\bibfnamefont {G.}~\bibnamefont
  {Kresse}}\ and\ \bibinfo {author} {\bibfnamefont {J.}~\bibnamefont
  {Hafner}},\ }\bibfield  {title} {\bibinfo {title} {{\em Ab initio} molecular
  dynamics for liquid metals},\ }\href
  {https://doi.org/10.1103/PhysRevB.47.558} {\bibfield  {journal} {\bibinfo
  {journal} {Phys. Rev. B}\ }\textbf {\bibinfo {volume} {47}},\ \bibinfo
  {pages} {558} (\bibinfo {year} {1993})}\BibitemShut {NoStop}%
\bibitem [{\citenamefont {Kresse}\ and\ \citenamefont {Hafner}(1994)}]{vasp-2}%
  \BibitemOpen
  \bibfield  {author} {\bibinfo {author} {\bibfnamefont {G.}~\bibnamefont
  {Kresse}}\ and\ \bibinfo {author} {\bibfnamefont {J.}~\bibnamefont
  {Hafner}},\ }\bibfield  {title} {\bibinfo {title} {{\em Ab initio}
  molecular-dynamics simulation of the liquid-metal--amorphous-semiconductor
  transition in germanium},\ }\href {https://doi.org/10.1103/PhysRevB.49.14251}
  {\bibfield  {journal} {\bibinfo  {journal} {Phys. Rev. B}\ }\textbf {\bibinfo
  {volume} {49}},\ \bibinfo {pages} {14251} (\bibinfo {year}
  {1994})}\BibitemShut {NoStop}%
\bibitem [{\citenamefont {Kresse}\ and\ \citenamefont
  {Furthm\"uller}(1996)}]{vasp-3}%
  \BibitemOpen
  \bibfield  {author} {\bibinfo {author} {\bibfnamefont {G.}~\bibnamefont
  {Kresse}}\ and\ \bibinfo {author} {\bibfnamefont {J.}~\bibnamefont
  {Furthm\"uller}},\ }\bibfield  {title} {\bibinfo {title} {Efficient iterative
  schemes for {\em ab initio} total-energy calculations using a plane-wave
  basis set},\ }\href {https://doi.org/10.1103/PhysRevB.54.11169} {\bibfield
  {journal} {\bibinfo  {journal} {Phys. Rev. B}\ }\textbf {\bibinfo {volume}
  {54}},\ \bibinfo {pages} {11169} (\bibinfo {year} {1996})}\BibitemShut
  {NoStop}%
\bibitem [{\citenamefont {Grimme}\ \emph {et~al.}(2010)\citenamefont {Grimme},
  \citenamefont {Antony}, \citenamefont {Ehrlich},\ and\ \citenamefont
  {Krieg}}]{df3-grimme}%
  \BibitemOpen
  \bibfield  {author} {\bibinfo {author} {\bibfnamefont {S.}~\bibnamefont
  {Grimme}}, \bibinfo {author} {\bibfnamefont {J.}~\bibnamefont {Antony}},
  \bibinfo {author} {\bibfnamefont {S.}~\bibnamefont {Ehrlich}},\ and\ \bibinfo
  {author} {\bibfnamefont {H.}~\bibnamefont {Krieg}},\ }\bibfield  {title}
  {\bibinfo {title} {A consistent and accurate {\em ab initio} parametrization
  of density functional dispersion correction {(DFT-D) for the 94 elements
  H-Pu}},\ }\href {https://doi.org/10.1063/1.3382344} {\bibfield  {journal}
  {\bibinfo  {journal} {J. Chem. Phys.}\ }\textbf {\bibinfo {volume} {132}},\
  \bibinfo {pages} {154104} (\bibinfo {year} {2010})}\BibitemShut {NoStop}%
\bibitem [{foo()}]{footnotes}%
  \BibitemOpen
  \bibinfo {note} {Supplementary materials contain: (i) an analysis of
  structural parameters against exchange correlation functionals and SOC, (ii)
  a table of possible magnetic groups, (iii) tests of energy and intrinsic
  electric dipole convergence against the vacuum in the unit cell and $k-$point
  meshes, (iv) values of lattice parameters and energies at global and local
  minima, (v) electronic band structures for AB and $s_1$ bilayers, and (vi) a
  calculation of magnetic anisotropy barriers.}\BibitemShut {Stop}%
\bibitem [{\citenamefont {Dudarev}\ \emph {et~al.}(1998)\citenamefont
  {Dudarev}, \citenamefont {Botton}, \citenamefont {Savrasov}, \citenamefont
  {Humphreys},\ and\ \citenamefont {Sutton}}]{Dudarev}%
  \BibitemOpen
  \bibfield  {author} {\bibinfo {author} {\bibfnamefont {S.~L.}\ \bibnamefont
  {Dudarev}}, \bibinfo {author} {\bibfnamefont {G.~A.}\ \bibnamefont {Botton}},
  \bibinfo {author} {\bibfnamefont {S.~Y.}\ \bibnamefont {Savrasov}}, \bibinfo
  {author} {\bibfnamefont {C.~J.}\ \bibnamefont {Humphreys}},\ and\ \bibinfo
  {author} {\bibfnamefont {A.~P.}\ \bibnamefont {Sutton}},\ }\bibfield  {title}
  {\bibinfo {title} {Electron-energy-loss spectra and the structural stability
  of nickel oxide: An {LSDA+U} study},\ }\href
  {https://doi.org/10.1103/PhysRevB.57.1505} {\bibfield  {journal} {\bibinfo
  {journal} {Phys. Rev. B}\ }\textbf {\bibinfo {volume} {57}},\ \bibinfo
  {pages} {1505} (\bibinfo {year} {1998})}\BibitemShut {NoStop}%
\bibitem [{\citenamefont {Jiang}\ \emph {et~al.}(2019)\citenamefont {Jiang},
  \citenamefont {Wang}, \citenamefont {Chen}, \citenamefont {Zhong},
  \citenamefont {Yuan}, \citenamefont {Lu},\ and\ \citenamefont
  {Ji}}]{DFT-xc-analysis}%
  \BibitemOpen
  \bibfield  {author} {\bibinfo {author} {\bibfnamefont {P.}~\bibnamefont
  {Jiang}}, \bibinfo {author} {\bibfnamefont {C.}~\bibnamefont {Wang}},
  \bibinfo {author} {\bibfnamefont {D.}~\bibnamefont {Chen}}, \bibinfo {author}
  {\bibfnamefont {Z.}~\bibnamefont {Zhong}}, \bibinfo {author} {\bibfnamefont
  {Z.}~\bibnamefont {Yuan}}, \bibinfo {author} {\bibfnamefont {Z.-Y.}\
  \bibnamefont {Lu}},\ and\ \bibinfo {author} {\bibfnamefont {W.}~\bibnamefont
  {Ji}},\ }\bibfield  {title} {\bibinfo {title} {Stacking tunable interlayer
  magnetism in bilayer {CrI$_3$}},\ }\href
  {https://doi.org/10.1103/PhysRevB.99.144401} {\bibfield  {journal} {\bibinfo
  {journal} {Phys. Rev. B}\ }\textbf {\bibinfo {volume} {99}},\ \bibinfo
  {pages} {144401} (\bibinfo {year} {2019})}\BibitemShut {NoStop}%
\bibitem [{\citenamefont {Vanderbilt}(2018)}]{vanderbilt}%
  \BibitemOpen
  \bibfield  {author} {\bibinfo {author} {\bibfnamefont {D.}~\bibnamefont
  {Vanderbilt}},\ }\href {https://doi.org/10.1017/9781316662205} {\emph
  {\bibinfo {title} {Berry Phases in Electronic Structure Theory: Electric
  Polarization, Orbital Magnetization and Topological Insulators}}}\ (\bibinfo
  {publisher} {Cambridge University Press},\ \bibinfo {year}
  {2018})\BibitemShut {NoStop}%
\bibitem [{\citenamefont {Togo}\ and\ \citenamefont {Tanaka}(2015)}]{phonopy}%
  \BibitemOpen
  \bibfield  {author} {\bibinfo {author} {\bibfnamefont {A.}~\bibnamefont
  {Togo}}\ and\ \bibinfo {author} {\bibfnamefont {I.}~\bibnamefont {Tanaka}},\
  }\bibfield  {title} {\bibinfo {title} {First principles phonon calculations
  in materials science},\ }\href
  {https://doi.org/10.1016/j.scriptamat.2015.07.021} {\bibfield  {journal}
  {\bibinfo  {journal} {Scr. Mater.}\ }\textbf {\bibinfo {volume} {108}},\
  \bibinfo {pages} {1} (\bibinfo {year} {2015})}\BibitemShut {NoStop}%
\bibitem [{\citenamefont {\'I\~niguez}(2008)}]{Jorge}%
  \BibitemOpen
  \bibfield  {author} {\bibinfo {author} {\bibfnamefont {J.}~\bibnamefont
  {\'I\~niguez}},\ }\bibfield  {title} {\bibinfo {title} {First-principles
  approach to lattice-mediated magnetoelectric effects},\ }\href
  {https://doi.org/10.1103/PhysRevLett.101.117201} {\bibfield  {journal}
  {\bibinfo  {journal} {Phys. Rev. Lett.}\ }\textbf {\bibinfo {volume} {101}},\
  \bibinfo {pages} {117201} (\bibinfo {year} {2008})}\BibitemShut {NoStop}%
\bibitem [{\citenamefont {Huang}\ \emph {et~al.}(2017)\citenamefont {Huang},
  \citenamefont {Clark}, \citenamefont {Navarro-Moratalla}, \citenamefont
  {Klein}, \citenamefont {Cheng}, \citenamefont {Seyler}, \citenamefont
  {Zhong}, \citenamefont {Schmidgall}, \citenamefont {McGuire}, \citenamefont
  {Cobden}, \citenamefont {Yao}, \citenamefont {Xiao}, \citenamefont
  {Jarillo-Herrero},\ and\ \citenamefont {Xu}}]{Huang2017}%
  \BibitemOpen
  \bibfield  {author} {\bibinfo {author} {\bibfnamefont {B.}~\bibnamefont
  {Huang}}, \bibinfo {author} {\bibfnamefont {G.}~\bibnamefont {Clark}},
  \bibinfo {author} {\bibfnamefont {E.}~\bibnamefont {Navarro-Moratalla}},
  \bibinfo {author} {\bibfnamefont {D.~R.}\ \bibnamefont {Klein}}, \bibinfo
  {author} {\bibfnamefont {R.}~\bibnamefont {Cheng}}, \bibinfo {author}
  {\bibfnamefont {K.~L.}\ \bibnamefont {Seyler}}, \bibinfo {author}
  {\bibfnamefont {D.}~\bibnamefont {Zhong}}, \bibinfo {author} {\bibfnamefont
  {E.}~\bibnamefont {Schmidgall}}, \bibinfo {author} {\bibfnamefont {M.~A.}\
  \bibnamefont {McGuire}}, \bibinfo {author} {\bibfnamefont {D.~H.}\
  \bibnamefont {Cobden}}, \bibinfo {author} {\bibfnamefont {W.}~\bibnamefont
  {Yao}}, \bibinfo {author} {\bibfnamefont {D.}~\bibnamefont {Xiao}}, \bibinfo
  {author} {\bibfnamefont {P.}~\bibnamefont {Jarillo-Herrero}},\ and\ \bibinfo
  {author} {\bibfnamefont {X.}~\bibnamefont {Xu}},\ }\bibfield  {title}
  {\bibinfo {title} {Layer-dependent ferromagnetism in a van der {Waals}
  crystal down to the monolayer limit},\ }\href
  {https://doi.org/10.1038/nature22391} {\bibfield  {journal} {\bibinfo
  {journal} {Nature}\ }\textbf {\bibinfo {volume} {546}},\ \bibinfo {pages}
  {270} (\bibinfo {year} {2017})}\BibitemShut {NoStop}%
\bibitem [{\citenamefont {Gong}\ \emph {et~al.}(2017)\citenamefont {Gong},
  \citenamefont {Li}, \citenamefont {Li}, \citenamefont {Ji}, \citenamefont
  {Stern}, \citenamefont {Xia}, \citenamefont {Cao}, \citenamefont {Bao},
  \citenamefont {Wang}, \citenamefont {Wang}, \citenamefont {Qiu},
  \citenamefont {Cava}, \citenamefont {Louie}, \citenamefont {Xia},\ and\
  \citenamefont {Zhang}}]{other2Dmagnets_0}%
  \BibitemOpen
  \bibfield  {author} {\bibinfo {author} {\bibfnamefont {C.}~\bibnamefont
  {Gong}}, \bibinfo {author} {\bibfnamefont {L.}~\bibnamefont {Li}}, \bibinfo
  {author} {\bibfnamefont {Z.}~\bibnamefont {Li}}, \bibinfo {author}
  {\bibfnamefont {H.}~\bibnamefont {Ji}}, \bibinfo {author} {\bibfnamefont
  {A.}~\bibnamefont {Stern}}, \bibinfo {author} {\bibfnamefont
  {Y.}~\bibnamefont {Xia}}, \bibinfo {author} {\bibfnamefont {T.}~\bibnamefont
  {Cao}}, \bibinfo {author} {\bibfnamefont {W.}~\bibnamefont {Bao}}, \bibinfo
  {author} {\bibfnamefont {C.}~\bibnamefont {Wang}}, \bibinfo {author}
  {\bibfnamefont {Y.}~\bibnamefont {Wang}}, \bibinfo {author} {\bibfnamefont
  {Z.~Q.}\ \bibnamefont {Qiu}}, \bibinfo {author} {\bibfnamefont {R.~J.}\
  \bibnamefont {Cava}}, \bibinfo {author} {\bibfnamefont {S.~G.}\ \bibnamefont
  {Louie}}, \bibinfo {author} {\bibfnamefont {J.}~\bibnamefont {Xia}},\ and\
  \bibinfo {author} {\bibfnamefont {X.}~\bibnamefont {Zhang}},\ }\bibfield
  {title} {\bibinfo {title} {Discovery of intrinsic ferromagnetism in
  two-dimensional van der {Waals} crystals},\ }\href
  {https://doi.org/10.1038/nature22060} {\bibfield  {journal} {\bibinfo
  {journal} {Nature}\ }\textbf {\bibinfo {volume} {546}},\ \bibinfo {pages}
  {265} (\bibinfo {year} {2017})}\BibitemShut {NoStop}%
\bibitem [{\citenamefont {Kim}\ \emph {et~al.}(2019)\citenamefont {Kim},
  \citenamefont {Kumaravadivel}, \citenamefont {Birkbeck}, \citenamefont
  {Kuang}, \citenamefont {Xu}, \citenamefont {Hopkinson}, \citenamefont
  {Knolle}, \citenamefont {McClarty}, \citenamefont {Berdyugin}, \citenamefont
  {Ben~Shalom}, \citenamefont {Gorbachev}, \citenamefont {Haigh}, \citenamefont
  {Liu}, \citenamefont {Edgar}, \citenamefont {Novoselov}, \citenamefont
  {Grigorieva},\ and\ \citenamefont {Geim}}]{other2Dmagnets_1}%
  \BibitemOpen
  \bibfield  {author} {\bibinfo {author} {\bibfnamefont {M.}~\bibnamefont
  {Kim}}, \bibinfo {author} {\bibfnamefont {P.}~\bibnamefont {Kumaravadivel}},
  \bibinfo {author} {\bibfnamefont {J.}~\bibnamefont {Birkbeck}}, \bibinfo
  {author} {\bibfnamefont {W.}~\bibnamefont {Kuang}}, \bibinfo {author}
  {\bibfnamefont {S.~G.}\ \bibnamefont {Xu}}, \bibinfo {author} {\bibfnamefont
  {D.~G.}\ \bibnamefont {Hopkinson}}, \bibinfo {author} {\bibfnamefont
  {J.}~\bibnamefont {Knolle}}, \bibinfo {author} {\bibfnamefont {P.~A.}\
  \bibnamefont {McClarty}}, \bibinfo {author} {\bibfnamefont {A.~I.}\
  \bibnamefont {Berdyugin}}, \bibinfo {author} {\bibfnamefont {M.}~\bibnamefont
  {Ben~Shalom}}, \bibinfo {author} {\bibfnamefont {R.~V.}\ \bibnamefont
  {Gorbachev}}, \bibinfo {author} {\bibfnamefont {S.~J.}\ \bibnamefont
  {Haigh}}, \bibinfo {author} {\bibfnamefont {S.}~\bibnamefont {Liu}}, \bibinfo
  {author} {\bibfnamefont {J.~H.}\ \bibnamefont {Edgar}}, \bibinfo {author}
  {\bibfnamefont {K.~S.}\ \bibnamefont {Novoselov}}, \bibinfo {author}
  {\bibfnamefont {I.~V.}\ \bibnamefont {Grigorieva}},\ and\ \bibinfo {author}
  {\bibfnamefont {A.~K.}\ \bibnamefont {Geim}},\ }\bibfield  {title} {\bibinfo
  {title} {Micromagnetometry of two-dimensional ferromagnets},\ }\href
  {https://doi.org/10.1038/s41928-019-0302-6} {\bibfield  {journal} {\bibinfo
  {journal} {Nat. Electron.}\ }\textbf {\bibinfo {volume} {2}},\ \bibinfo
  {pages} {457} (\bibinfo {year} {2019})}\BibitemShut {NoStop}%
\bibitem [{\citenamefont {Cai}\ \emph {et~al.}(2019)\citenamefont {Cai},
  \citenamefont {Song}, \citenamefont {Wilson}, \citenamefont {Clark},
  \citenamefont {He}, \citenamefont {Zhang}, \citenamefont {Taniguchi},
  \citenamefont {Watanabe}, \citenamefont {Yao}, \citenamefont {Xiao},
  \citenamefont {McGuire}, \citenamefont {Cobden},\ and\ \citenamefont
  {Xu}}]{other2Dmagnets_2}%
  \BibitemOpen
  \bibfield  {author} {\bibinfo {author} {\bibfnamefont {X.}~\bibnamefont
  {Cai}}, \bibinfo {author} {\bibfnamefont {T.}~\bibnamefont {Song}}, \bibinfo
  {author} {\bibfnamefont {N.~P.}\ \bibnamefont {Wilson}}, \bibinfo {author}
  {\bibfnamefont {G.}~\bibnamefont {Clark}}, \bibinfo {author} {\bibfnamefont
  {M.}~\bibnamefont {He}}, \bibinfo {author} {\bibfnamefont {X.}~\bibnamefont
  {Zhang}}, \bibinfo {author} {\bibfnamefont {T.}~\bibnamefont {Taniguchi}},
  \bibinfo {author} {\bibfnamefont {K.}~\bibnamefont {Watanabe}}, \bibinfo
  {author} {\bibfnamefont {W.}~\bibnamefont {Yao}}, \bibinfo {author}
  {\bibfnamefont {D.}~\bibnamefont {Xiao}}, \bibinfo {author} {\bibfnamefont
  {M.~A.}\ \bibnamefont {McGuire}}, \bibinfo {author} {\bibfnamefont {D.~H.}\
  \bibnamefont {Cobden}},\ and\ \bibinfo {author} {\bibfnamefont
  {X.}~\bibnamefont {Xu}},\ }\bibfield  {title} {\bibinfo {title} {Atomically
  thin {CrCl$_3$}: An in-plane layered antiferromagnetic insulator},\ }\href
  {https://doi.org/10.1021/acs.nanolett.9b01317} {\bibfield  {journal}
  {\bibinfo  {journal} {Nano Lett.}\ }\textbf {\bibinfo {volume} {19}},\
  \bibinfo {pages} {3993} (\bibinfo {year} {2019})}\BibitemShut {NoStop}%
\bibitem [{\citenamefont {Klein}\ \emph {et~al.}(2019)\citenamefont {Klein},
  \citenamefont {MacNeill}, \citenamefont {Song}, \citenamefont {Larson},
  \citenamefont {Fang}, \citenamefont {Xu}, \citenamefont {Ribeiro},
  \citenamefont {Canfield}, \citenamefont {Kaxiras}, \citenamefont {Comin},\
  and\ \citenamefont {Jarillo-Herrero}}]{other2Dmagnets_2_2}%
  \BibitemOpen
  \bibfield  {author} {\bibinfo {author} {\bibfnamefont {D.~R.}\ \bibnamefont
  {Klein}}, \bibinfo {author} {\bibfnamefont {D.}~\bibnamefont {MacNeill}},
  \bibinfo {author} {\bibfnamefont {Q.}~\bibnamefont {Song}}, \bibinfo {author}
  {\bibfnamefont {D.~T.}\ \bibnamefont {Larson}}, \bibinfo {author}
  {\bibfnamefont {S.}~\bibnamefont {Fang}}, \bibinfo {author} {\bibfnamefont
  {M.}~\bibnamefont {Xu}}, \bibinfo {author} {\bibfnamefont {R.~A.}\
  \bibnamefont {Ribeiro}}, \bibinfo {author} {\bibfnamefont {P.~C.}\
  \bibnamefont {Canfield}}, \bibinfo {author} {\bibfnamefont {E.}~\bibnamefont
  {Kaxiras}}, \bibinfo {author} {\bibfnamefont {R.}~\bibnamefont {Comin}},\
  and\ \bibinfo {author} {\bibfnamefont {P.}~\bibnamefont {Jarillo-Herrero}},\
  }\bibfield  {title} {\bibinfo {title} {Enhancement of interlayer exchange in
  an ultrathin two-dimensional magnet},\ }\href
  {https://doi.org/10.1038/s41567-019-0651-0} {\bibfield  {journal} {\bibinfo
  {journal} {Nat. Phys.}\ }\textbf {\bibinfo {volume} {15}},\ \bibinfo {pages}
  {1255} (\bibinfo {year} {2019})}\BibitemShut {NoStop}%
\bibitem [{\citenamefont {Bedoya-Pinto}\ \emph {et~al.}(2021)\citenamefont
  {Bedoya-Pinto}, \citenamefont {Ji}, \citenamefont {Pandeya}, \citenamefont
  {Gargiani}, \citenamefont {Valvidares}, \citenamefont {Sessi}, \citenamefont
  {Taylor}, \citenamefont {Radu}, \citenamefont {Chang},\ and\ \citenamefont
  {Parkin}}]{other2Dmagnets_3}%
  \BibitemOpen
  \bibfield  {author} {\bibinfo {author} {\bibfnamefont {A.}~\bibnamefont
  {Bedoya-Pinto}}, \bibinfo {author} {\bibfnamefont {J.-R.}\ \bibnamefont
  {Ji}}, \bibinfo {author} {\bibfnamefont {A.~K.}\ \bibnamefont {Pandeya}},
  \bibinfo {author} {\bibfnamefont {P.}~\bibnamefont {Gargiani}}, \bibinfo
  {author} {\bibfnamefont {M.}~\bibnamefont {Valvidares}}, \bibinfo {author}
  {\bibfnamefont {P.}~\bibnamefont {Sessi}}, \bibinfo {author} {\bibfnamefont
  {J.~M.}\ \bibnamefont {Taylor}}, \bibinfo {author} {\bibfnamefont
  {F.}~\bibnamefont {Radu}}, \bibinfo {author} {\bibfnamefont {K.}~\bibnamefont
  {Chang}},\ and\ \bibinfo {author} {\bibfnamefont {S.~S.~P.}\ \bibnamefont
  {Parkin}},\ }\bibfield  {title} {\bibinfo {title} {Intrinsic {2D-XY}
  ferromagnetism in a van der {Waals} monolayer},\ }\href
  {https://doi.org/10.1126/science.abd5146} {\bibfield  {journal} {\bibinfo
  {journal} {Science}\ }\textbf {\bibinfo {volume} {374}},\ \bibinfo {pages}
  {616} (\bibinfo {year} {2021})}\BibitemShut {NoStop}%
\bibitem [{\citenamefont {Ghazaryan}\ \emph {et~al.}(2018)\citenamefont
  {Ghazaryan}, \citenamefont {Greenaway}, \citenamefont {Wang}, \citenamefont
  {Guarochico-Moreira}, \citenamefont {Vera-Marun}, \citenamefont {Yin},
  \citenamefont {Liao}, \citenamefont {Morozov}, \citenamefont {Kristanovski},
  \citenamefont {Lichtenstein}, \citenamefont {Katsnelson}, \citenamefont
  {Withers}, \citenamefont {Mishchenko}, \citenamefont {Eaves}, \citenamefont
  {Geim}, \citenamefont {Novoselov},\ and\ \citenamefont
  {Misra}}]{other2Dmagnets_4}%
  \BibitemOpen
  \bibfield  {author} {\bibinfo {author} {\bibfnamefont {D.}~\bibnamefont
  {Ghazaryan}}, \bibinfo {author} {\bibfnamefont {M.~T.}\ \bibnamefont
  {Greenaway}}, \bibinfo {author} {\bibfnamefont {Z.}~\bibnamefont {Wang}},
  \bibinfo {author} {\bibfnamefont {V.~H.}\ \bibnamefont {Guarochico-Moreira}},
  \bibinfo {author} {\bibfnamefont {I.~J.}\ \bibnamefont {Vera-Marun}},
  \bibinfo {author} {\bibfnamefont {J.}~\bibnamefont {Yin}}, \bibinfo {author}
  {\bibfnamefont {Y.}~\bibnamefont {Liao}}, \bibinfo {author} {\bibfnamefont
  {S.~V.}\ \bibnamefont {Morozov}}, \bibinfo {author} {\bibfnamefont
  {O.}~\bibnamefont {Kristanovski}}, \bibinfo {author} {\bibfnamefont {A.~I.}\
  \bibnamefont {Lichtenstein}}, \bibinfo {author} {\bibfnamefont {M.~I.}\
  \bibnamefont {Katsnelson}}, \bibinfo {author} {\bibfnamefont
  {F.}~\bibnamefont {Withers}}, \bibinfo {author} {\bibfnamefont
  {A.}~\bibnamefont {Mishchenko}}, \bibinfo {author} {\bibfnamefont
  {L.}~\bibnamefont {Eaves}}, \bibinfo {author} {\bibfnamefont {A.~K.}\
  \bibnamefont {Geim}}, \bibinfo {author} {\bibfnamefont {K.~S.}\ \bibnamefont
  {Novoselov}},\ and\ \bibinfo {author} {\bibfnamefont {A.}~\bibnamefont
  {Misra}},\ }\bibfield  {title} {\bibinfo {title} {Magnon-assisted tunnelling
  in van der {Waals} heterostructures based on {CrBr$_3$}},\ }\href
  {https://doi.org/10.1038/s41928-018-0087-z} {\bibfield  {journal} {\bibinfo
  {journal} {Nat. Electron.}\ }\textbf {\bibinfo {volume} {1}},\ \bibinfo
  {pages} {344} (\bibinfo {year} {2018})}\BibitemShut {NoStop}%
\bibitem [{\citenamefont {Vizner~Stern}\ \emph {et~al.}(2021)\citenamefont
  {Vizner~Stern}, \citenamefont {Waschitz}, \citenamefont {Cao}, \citenamefont
  {Nevo}, \citenamefont {Watanabe}, \citenamefont {Taniguchi}, \citenamefont
  {Sela}, \citenamefont {Urbakh}, \citenamefont {Hod},\ and\ \citenamefont
  {Ben~Shalom}}]{hBNFerro0}%
  \BibitemOpen
  \bibfield  {author} {\bibinfo {author} {\bibfnamefont {M.}~\bibnamefont
  {Vizner~Stern}}, \bibinfo {author} {\bibfnamefont {Y.}~\bibnamefont
  {Waschitz}}, \bibinfo {author} {\bibfnamefont {W.}~\bibnamefont {Cao}},
  \bibinfo {author} {\bibfnamefont {I.}~\bibnamefont {Nevo}}, \bibinfo {author}
  {\bibfnamefont {K.}~\bibnamefont {Watanabe}}, \bibinfo {author}
  {\bibfnamefont {T.}~\bibnamefont {Taniguchi}}, \bibinfo {author}
  {\bibfnamefont {E.}~\bibnamefont {Sela}}, \bibinfo {author} {\bibfnamefont
  {M.}~\bibnamefont {Urbakh}}, \bibinfo {author} {\bibfnamefont
  {O.}~\bibnamefont {Hod}},\ and\ \bibinfo {author} {\bibfnamefont
  {M.}~\bibnamefont {Ben~Shalom}},\ }\bibfield  {title} {\bibinfo {title}
  {Interfacial ferroelectricity by van der {Waals} sliding},\ }\href
  {https://doi.org/10.1126/science.abe8177} {\bibfield  {journal} {\bibinfo
  {journal} {Science}\ }\textbf {\bibinfo {volume} {372}},\ \bibinfo {pages}
  {1462} (\bibinfo {year} {2021})}\BibitemShut {NoStop}%
\bibitem [{\citenamefont {Woods}\ \emph {et~al.}(2021)\citenamefont {Woods},
  \citenamefont {Ares}, \citenamefont {Nevison-Andrews}, \citenamefont
  {Holwill}, \citenamefont {Fabregas}, \citenamefont {Guinea}, \citenamefont
  {Geim}, \citenamefont {Novoselov}, \citenamefont {Walet},\ and\ \citenamefont
  {Fumagalli}}]{hBNFerro1}%
  \BibitemOpen
  \bibfield  {author} {\bibinfo {author} {\bibfnamefont {C.~R.}\ \bibnamefont
  {Woods}}, \bibinfo {author} {\bibfnamefont {P.}~\bibnamefont {Ares}},
  \bibinfo {author} {\bibfnamefont {H.}~\bibnamefont {Nevison-Andrews}},
  \bibinfo {author} {\bibfnamefont {M.~J.}\ \bibnamefont {Holwill}}, \bibinfo
  {author} {\bibfnamefont {R.}~\bibnamefont {Fabregas}}, \bibinfo {author}
  {\bibfnamefont {F.}~\bibnamefont {Guinea}}, \bibinfo {author} {\bibfnamefont
  {A.~K.}\ \bibnamefont {Geim}}, \bibinfo {author} {\bibfnamefont {K.~S.}\
  \bibnamefont {Novoselov}}, \bibinfo {author} {\bibfnamefont {N.~R.}\
  \bibnamefont {Walet}},\ and\ \bibinfo {author} {\bibfnamefont
  {L.}~\bibnamefont {Fumagalli}},\ }\bibfield  {title} {\bibinfo {title}
  {Charge-polarized interfacial superlattices in marginally twisted hexagonal
  boron nitride},\ }\href {https://doi.org/10.1038/s41467-020-20667-2}
  {\bibfield  {journal} {\bibinfo  {journal} {Nat. Commun.}\ }\textbf {\bibinfo
  {volume} {12}},\ \bibinfo {pages} {347} (\bibinfo {year} {2021})}\BibitemShut
  {NoStop}%
\bibitem [{\citenamefont {Moore}\ \emph {et~al.}(2021)\citenamefont {Moore},
  \citenamefont {Ciccarino}, \citenamefont {Halbertal}, \citenamefont
  {McGilly}, \citenamefont {Finney}, \citenamefont {Yao}, \citenamefont {Shao},
  \citenamefont {Ni}, \citenamefont {Sternbach}, \citenamefont {Telford},
  \citenamefont {Kim}, \citenamefont {Rossi}, \citenamefont {Watanabe},
  \citenamefont {Taniguchi}, \citenamefont {Pasupathy}, \citenamefont {Dean},
  \citenamefont {Hone}, \citenamefont {Schuck}, \citenamefont {Narang},\ and\
  \citenamefont {Basov}}]{hBNFerro2}%
  \BibitemOpen
  \bibfield  {author} {\bibinfo {author} {\bibfnamefont {S.~L.}\ \bibnamefont
  {Moore}}, \bibinfo {author} {\bibfnamefont {C.~J.}\ \bibnamefont
  {Ciccarino}}, \bibinfo {author} {\bibfnamefont {D.}~\bibnamefont
  {Halbertal}}, \bibinfo {author} {\bibfnamefont {L.~J.}\ \bibnamefont
  {McGilly}}, \bibinfo {author} {\bibfnamefont {N.~R.}\ \bibnamefont {Finney}},
  \bibinfo {author} {\bibfnamefont {K.}~\bibnamefont {Yao}}, \bibinfo {author}
  {\bibfnamefont {Y.}~\bibnamefont {Shao}}, \bibinfo {author} {\bibfnamefont
  {G.}~\bibnamefont {Ni}}, \bibinfo {author} {\bibfnamefont {A.}~\bibnamefont
  {Sternbach}}, \bibinfo {author} {\bibfnamefont {E.~J.}\ \bibnamefont
  {Telford}}, \bibinfo {author} {\bibfnamefont {B.~S.}\ \bibnamefont {Kim}},
  \bibinfo {author} {\bibfnamefont {S.~E.}\ \bibnamefont {Rossi}}, \bibinfo
  {author} {\bibfnamefont {K.}~\bibnamefont {Watanabe}}, \bibinfo {author}
  {\bibfnamefont {T.}~\bibnamefont {Taniguchi}}, \bibinfo {author}
  {\bibfnamefont {A.~N.}\ \bibnamefont {Pasupathy}}, \bibinfo {author}
  {\bibfnamefont {C.~R.}\ \bibnamefont {Dean}}, \bibinfo {author}
  {\bibfnamefont {J.}~\bibnamefont {Hone}}, \bibinfo {author} {\bibfnamefont
  {P.~J.}\ \bibnamefont {Schuck}}, \bibinfo {author} {\bibfnamefont
  {P.}~\bibnamefont {Narang}},\ and\ \bibinfo {author} {\bibfnamefont {D.~N.}\
  \bibnamefont {Basov}},\ }\bibfield  {title} {\bibinfo {title} {Nanoscale
  lattice dynamics in hexagonal boron nitride moir{\'e} superlattices},\ }\href
  {https://doi.org/10.1038/s41467-021-26072-7} {\bibfield  {journal} {\bibinfo
  {journal} {Nat. Commun.}\ }\textbf {\bibinfo {volume} {12}},\ \bibinfo
  {pages} {5741} (\bibinfo {year} {2021})}\BibitemShut {NoStop}%
\bibitem [{\citenamefont {Weston}\ \emph {et~al.}(2022)\citenamefont {Weston},
  \citenamefont {Castanon}, \citenamefont {Enaldiev}, \citenamefont {Ferreira},
  \citenamefont {Bhattacharjee}, \citenamefont {Xu}, \citenamefont
  {Corte-Le{\'o}n}, \citenamefont {Wu}, \citenamefont {Clark}, \citenamefont
  {Summerfield}, \citenamefont {Hashimoto}, \citenamefont {Gao}, \citenamefont
  {Wang}, \citenamefont {Hamer}, \citenamefont {Read}, \citenamefont
  {Fumagalli}, \citenamefont {Kretinin}, \citenamefont {Haigh}, \citenamefont
  {Kazakova}, \citenamefont {Geim}, \citenamefont {Fal’ko},\ and\
  \citenamefont {Gorbachev}}]{TMDCFerro0}%
  \BibitemOpen
  \bibfield  {author} {\bibinfo {author} {\bibfnamefont {A.}~\bibnamefont
  {Weston}}, \bibinfo {author} {\bibfnamefont {E.~G.}\ \bibnamefont
  {Castanon}}, \bibinfo {author} {\bibfnamefont {V.}~\bibnamefont {Enaldiev}},
  \bibinfo {author} {\bibfnamefont {F.}~\bibnamefont {Ferreira}}, \bibinfo
  {author} {\bibfnamefont {S.}~\bibnamefont {Bhattacharjee}}, \bibinfo {author}
  {\bibfnamefont {S.}~\bibnamefont {Xu}}, \bibinfo {author} {\bibfnamefont
  {H.}~\bibnamefont {Corte-Le{\'o}n}}, \bibinfo {author} {\bibfnamefont
  {Z.}~\bibnamefont {Wu}}, \bibinfo {author} {\bibfnamefont {N.}~\bibnamefont
  {Clark}}, \bibinfo {author} {\bibfnamefont {A.}~\bibnamefont {Summerfield}},
  \bibinfo {author} {\bibfnamefont {T.}~\bibnamefont {Hashimoto}}, \bibinfo
  {author} {\bibfnamefont {Y.}~\bibnamefont {Gao}}, \bibinfo {author}
  {\bibfnamefont {W.}~\bibnamefont {Wang}}, \bibinfo {author} {\bibfnamefont
  {M.}~\bibnamefont {Hamer}}, \bibinfo {author} {\bibfnamefont
  {H.}~\bibnamefont {Read}}, \bibinfo {author} {\bibfnamefont {L.}~\bibnamefont
  {Fumagalli}}, \bibinfo {author} {\bibfnamefont {A.~V.}\ \bibnamefont
  {Kretinin}}, \bibinfo {author} {\bibfnamefont {S.~J.}\ \bibnamefont {Haigh}},
  \bibinfo {author} {\bibfnamefont {O.}~\bibnamefont {Kazakova}}, \bibinfo
  {author} {\bibfnamefont {A.~K.}\ \bibnamefont {Geim}}, \bibinfo {author}
  {\bibfnamefont {V.~I.}\ \bibnamefont {Fal’ko}},\ and\ \bibinfo {author}
  {\bibfnamefont {R.}~\bibnamefont {Gorbachev}},\ }\bibfield  {title} {\bibinfo
  {title} {Interfacial ferroelectricity in marginally twisted {2D}
  semiconductors},\ }\href {https://doi.org/10.1038/s41565-022-01072-w}
  {\bibfield  {journal} {\bibinfo  {journal} {Nat. Nanotech.}\ }\textbf
  {\bibinfo {volume} {17}},\ \bibinfo {pages} {390} (\bibinfo {year}
  {2022})}\BibitemShut {NoStop}%
\bibitem [{\citenamefont {Wang}\ \emph {et~al.}(2022)\citenamefont {Wang},
  \citenamefont {Yasuda}, \citenamefont {Zhang}, \citenamefont {Liu},
  \citenamefont {Watanabe}, \citenamefont {Taniguchi}, \citenamefont {Hone},
  \citenamefont {Fu},\ and\ \citenamefont {Jarillo-Herrero}}]{TMDCFerro1}%
  \BibitemOpen
  \bibfield  {author} {\bibinfo {author} {\bibfnamefont {X.}~\bibnamefont
  {Wang}}, \bibinfo {author} {\bibfnamefont {K.}~\bibnamefont {Yasuda}},
  \bibinfo {author} {\bibfnamefont {Y.}~\bibnamefont {Zhang}}, \bibinfo
  {author} {\bibfnamefont {S.}~\bibnamefont {Liu}}, \bibinfo {author}
  {\bibfnamefont {K.}~\bibnamefont {Watanabe}}, \bibinfo {author}
  {\bibfnamefont {T.}~\bibnamefont {Taniguchi}}, \bibinfo {author}
  {\bibfnamefont {J.}~\bibnamefont {Hone}}, \bibinfo {author} {\bibfnamefont
  {L.}~\bibnamefont {Fu}},\ and\ \bibinfo {author} {\bibfnamefont
  {P.}~\bibnamefont {Jarillo-Herrero}},\ }\bibfield  {title} {\bibinfo {title}
  {Interfacial ferroelectricity in rhombohedral-stacked bilayer transition
  metal dichalcogenides},\ }\href {https://doi.org/10.1038/s41565-021-01059-z}
  {\bibfield  {journal} {\bibinfo  {journal} {Nat. Nanotech.}\ }\textbf
  {\bibinfo {volume} {17}},\ \bibinfo {pages} {367} (\bibinfo {year}
  {2022})}\BibitemShut {NoStop}%
\bibitem [{\citenamefont {Liu}\ \emph {et~al.}(2022)\citenamefont {Liu},
  \citenamefont {Liu}, \citenamefont {Li}, \citenamefont {Yoo},\ and\
  \citenamefont {Hone}}]{FerroelectricWSe2}%
  \BibitemOpen
  \bibfield  {author} {\bibinfo {author} {\bibfnamefont {Y.}~\bibnamefont
  {Liu}}, \bibinfo {author} {\bibfnamefont {S.}~\bibnamefont {Liu}}, \bibinfo
  {author} {\bibfnamefont {B.}~\bibnamefont {Li}}, \bibinfo {author}
  {\bibfnamefont {W.~J.}\ \bibnamefont {Yoo}},\ and\ \bibinfo {author}
  {\bibfnamefont {J.}~\bibnamefont {Hone}},\ }\bibfield  {title} {\bibinfo
  {title} {Identifying the transition order in an artificial ferroelectric van
  der {Waals} heterostructure},\ }\href
  {https://doi.org/10.1021/acs.nanolett.1c04467} {\bibfield  {journal}
  {\bibinfo  {journal} {Nano Lett.}\ }\textbf {\bibinfo {volume} {22}},\
  \bibinfo {pages} {1265} (\bibinfo {year} {2022})}\BibitemShut {NoStop}%
\bibitem [{\citenamefont {Marmolejo-Tejada}\ \emph {et~al.}(2022)\citenamefont
  {Marmolejo-Tejada}, \citenamefont {Roll}, \citenamefont {Poudel},
  \citenamefont {Barraza-Lopez},\ and\ \citenamefont {Mosquera}}]{Juan}%
  \BibitemOpen
  \bibfield  {author} {\bibinfo {author} {\bibfnamefont {J.~M.}\ \bibnamefont
  {Marmolejo-Tejada}}, \bibinfo {author} {\bibfnamefont {J.~E.}\ \bibnamefont
  {Roll}}, \bibinfo {author} {\bibfnamefont {S.~P.}\ \bibnamefont {Poudel}},
  \bibinfo {author} {\bibfnamefont {S.}~\bibnamefont {Barraza-Lopez}},\ and\
  \bibinfo {author} {\bibfnamefont {M.~A.}\ \bibnamefont {Mosquera}},\
  }\bibfield  {title} {\bibinfo {title} {Slippery paraelectric transition-metal
  dichalcogenide bilayers},\ }\href
  {https://doi.org/10.1021/acs.nanolett.2c03373} {\bibfield  {journal}
  {\bibinfo  {journal} {Nano Lett.}\ }\textbf {\bibinfo {volume} {22}},\
  \bibinfo {pages} {7984} (\bibinfo {year} {2022})}\BibitemShut {NoStop}%
\bibitem [{\citenamefont {Sivadas}\ \emph {et~al.}(2018)\citenamefont
  {Sivadas}, \citenamefont {Okamoto}, \citenamefont {Xu}, \citenamefont
  {Fennie},\ and\ \citenamefont {Xiao}}]{sivadas}%
  \BibitemOpen
  \bibfield  {author} {\bibinfo {author} {\bibfnamefont {N.}~\bibnamefont
  {Sivadas}}, \bibinfo {author} {\bibfnamefont {S.}~\bibnamefont {Okamoto}},
  \bibinfo {author} {\bibfnamefont {X.}~\bibnamefont {Xu}}, \bibinfo {author}
  {\bibfnamefont {C.~J.}\ \bibnamefont {Fennie}},\ and\ \bibinfo {author}
  {\bibfnamefont {D.}~\bibnamefont {Xiao}},\ }\bibfield  {title} {\bibinfo
  {title} {Stacking-dependent magnetism in bilayer {CrI$_3$}},\ }\href
  {https://doi.org/10.1021/acs.nanolett.8b03321} {\bibfield  {journal}
  {\bibinfo  {journal} {Nano Lett.}\ }\textbf {\bibinfo {volume} {18}},\
  \bibinfo {pages} {7658} (\bibinfo {year} {2018})}\BibitemShut {NoStop}%
\bibitem [{\citenamefont {Wang}\ and\ \citenamefont
  {Sanyal}(2021)}]{CrI3-DFT-stacking}%
  \BibitemOpen
  \bibfield  {author} {\bibinfo {author} {\bibfnamefont {D.}~\bibnamefont
  {Wang}}\ and\ \bibinfo {author} {\bibfnamefont {B.}~\bibnamefont {Sanyal}},\
  }\bibfield  {title} {\bibinfo {title} {Systematic study of monolayer to
  trilayer {CrI$_3$}: Stacking sequence dependence of electronic structure and
  magnetism},\ }\href {https://doi.org/10.1021/acs.jpcc.1c04311} {\bibfield
  {journal} {\bibinfo  {journal} {J. Phys. Chem. C}\ }\textbf {\bibinfo
  {volume} {125}},\ \bibinfo {pages} {18467} (\bibinfo {year}
  {2021})}\BibitemShut {NoStop}%
\bibitem [{\citenamefont {Qiu}\ \emph {et~al.}(2021)\citenamefont {Qiu},
  \citenamefont {Holwill}, \citenamefont {Olsen}, \citenamefont {Lyu},
  \citenamefont {Li}, \citenamefont {Fang}, \citenamefont {Yang}, \citenamefont
  {Kashchenko},\ and\ \citenamefont {Lu}}]{R_CrI3_bilayer}%
  \BibitemOpen
  \bibfield  {author} {\bibinfo {author} {\bibfnamefont {Z.}~\bibnamefont
  {Qiu}}, \bibinfo {author} {\bibfnamefont {M.}~\bibnamefont {Holwill}},
  \bibinfo {author} {\bibfnamefont {T.}~\bibnamefont {Olsen}}, \bibinfo
  {author} {\bibfnamefont {P.}~\bibnamefont {Lyu}}, \bibinfo {author}
  {\bibfnamefont {J.}~\bibnamefont {Li}}, \bibinfo {author} {\bibfnamefont
  {H.}~\bibnamefont {Fang}}, \bibinfo {author} {\bibfnamefont {H.}~\bibnamefont
  {Yang}}, \bibinfo {author} {\bibfnamefont {K.~S.}\ \bibnamefont {Kashchenko},
  \bibfnamefont {Mikhailand~Novoselov}},\ and\ \bibinfo {author} {\bibfnamefont
  {J.}~\bibnamefont {Lu}},\ }\bibfield  {title} {\bibinfo {title} {Visualizing
  atomic structure and magnetism of {2D} magnetic insulators via tunneling
  through graphene},\ }\href {https://doi.org/10.1038/s41467-020-20376-w}
  {\bibfield  {journal} {\bibinfo  {journal} {Nat. Commun.}\ }\textbf {\bibinfo
  {volume} {12}},\ \bibinfo {pages} {70} (\bibinfo {year} {2021})}\BibitemShut
  {NoStop}%
\bibitem [{\citenamefont {McGuire}\ \emph {et~al.}(2015)\citenamefont
  {McGuire}, \citenamefont {Dixit}, \citenamefont {Cooper},\ and\ \citenamefont
  {Sales}}]{doi:10.1021/cm504242t}%
  \BibitemOpen
  \bibfield  {author} {\bibinfo {author} {\bibfnamefont {M.~A.}\ \bibnamefont
  {McGuire}}, \bibinfo {author} {\bibfnamefont {H.}~\bibnamefont {Dixit}},
  \bibinfo {author} {\bibfnamefont {V.~R.}\ \bibnamefont {Cooper}},\ and\
  \bibinfo {author} {\bibfnamefont {B.~C.}\ \bibnamefont {Sales}},\ }\bibfield
  {title} {\bibinfo {title} {Coupling of crystal structure and magnetism in the
  layered, ferromagnetic insulator {CrI$_3$}},\ }\href
  {https://doi.org/10.1021/cm504242t} {\bibfield  {journal} {\bibinfo
  {journal} {Chem. Mater.}\ }\textbf {\bibinfo {volume} {27}},\ \bibinfo
  {pages} {612} (\bibinfo {year} {2015})}\BibitemShut {NoStop}%
\bibitem [{\citenamefont {Li}\ and\ \citenamefont {Wu}(2017)}]{MoS2Ferro}%
  \BibitemOpen
  \bibfield  {author} {\bibinfo {author} {\bibfnamefont {L.}~\bibnamefont
  {Li}}\ and\ \bibinfo {author} {\bibfnamefont {M.}~\bibnamefont {Wu}},\
  }\bibfield  {title} {\bibinfo {title} {Binary compound bilayer and multilayer
  with vertical polarizations: Two-dimensional ferroelectrics, multiferroics,
  and nanogenerators},\ }\href {https://doi.org/10.1021/acsnano.7b02756}
  {\bibfield  {journal} {\bibinfo  {journal} {ACS Nano}\ }\textbf {\bibinfo
  {volume} {11}},\ \bibinfo {pages} {6382} (\bibinfo {year}
  {2017})}\BibitemShut {NoStop}%
\bibitem [{\citenamefont {Yang}\ \emph {et~al.}(2018)\citenamefont {Yang},
  \citenamefont {Wu},\ and\ \citenamefont {Li}}]{WTe2ferro}%
  \BibitemOpen
  \bibfield  {author} {\bibinfo {author} {\bibfnamefont {Q.}~\bibnamefont
  {Yang}}, \bibinfo {author} {\bibfnamefont {M.}~\bibnamefont {Wu}},\ and\
  \bibinfo {author} {\bibfnamefont {J.}~\bibnamefont {Li}},\ }\bibfield
  {title} {\bibinfo {title} {Origin of two-dimensional vertical
  ferroelectricity in {WTe$_2$} bilayer and multilayer},\ }\href
  {https://doi.org/10.1021/acs.jpclett.8b03654} {\bibfield  {journal} {\bibinfo
   {journal} {J. Phys. Chem. Lett.}\ }\textbf {\bibinfo {volume} {9}},\
  \bibinfo {pages} {7160} (\bibinfo {year} {2018})}\BibitemShut {NoStop}%
\bibitem [{\citenamefont {Zhou}\ \emph {et~al.}(2015)\citenamefont {Zhou},
  \citenamefont {Han}, \citenamefont {Dai}, \citenamefont {Sun},\ and\
  \citenamefont {Srolovitz}}]{landscapeGrapheneHBN}%
  \BibitemOpen
  \bibfield  {author} {\bibinfo {author} {\bibfnamefont {S.}~\bibnamefont
  {Zhou}}, \bibinfo {author} {\bibfnamefont {J.}~\bibnamefont {Han}}, \bibinfo
  {author} {\bibfnamefont {S.}~\bibnamefont {Dai}}, \bibinfo {author}
  {\bibfnamefont {J.}~\bibnamefont {Sun}},\ and\ \bibinfo {author}
  {\bibfnamefont {D.~J.}\ \bibnamefont {Srolovitz}},\ }\bibfield  {title}
  {\bibinfo {title} {van der {Waals} bilayer energetics: Generalized
  stacking-fault energy of graphene, boron nitride, and graphene/boron nitride
  bilayers},\ }\href {https://doi.org/10.1103/PhysRevB.92.155438} {\bibfield
  {journal} {\bibinfo  {journal} {Phys. Rev. B}\ }\textbf {\bibinfo {volume}
  {92}},\ \bibinfo {pages} {155438} (\bibinfo {year} {2015})}\BibitemShut
  {NoStop}%
\bibitem [{\citenamefont {Wang}\ \emph {et~al.}(2023)\citenamefont {Wang},
  \citenamefont {You}, \citenamefont {Cobden},\ and\ \citenamefont
  {Wang}}]{towards}%
  \BibitemOpen
  \bibfield  {author} {\bibinfo {author} {\bibfnamefont {C.}~\bibnamefont
  {Wang}}, \bibinfo {author} {\bibfnamefont {L.}~\bibnamefont {You}}, \bibinfo
  {author} {\bibfnamefont {D.}~\bibnamefont {Cobden}},\ and\ \bibinfo {author}
  {\bibfnamefont {J.}~\bibnamefont {Wang}},\ }\bibfield  {title} {\bibinfo
  {title} {Towards two-dimensional van der {Waals} ferroelectrics},\ }\bibfield
   {journal} {\bibinfo  {journal} {Nat. Mater.}\ }\href
  {https://doi.org/10.1038/s41563-022-01422-y} {10.1038/s41563-022-01422-y}
  (\bibinfo {year} {2023})\BibitemShut {NoStop}%
\bibitem [{\citenamefont {Niu}\ \emph {et~al.}(2022)\citenamefont {Niu},
  \citenamefont {Li}, \citenamefont {Han}, \citenamefont {Qu}, \citenamefont
  {Ding}, \citenamefont {Wang}, \citenamefont {Liu}, \citenamefont {Liu},
  \citenamefont {Han}, \citenamefont {Watanabe}, \citenamefont {Taniguchi},
  \citenamefont {Wu}, \citenamefont {Ren}, \citenamefont {Wang}, \citenamefont
  {Hong}, \citenamefont {Mao}, \citenamefont {Han}, \citenamefont {Liu},
  \citenamefont {Gan},\ and\ \citenamefont
  {Lu}}]{FerroelectricGrapheneBilayer2}%
  \BibitemOpen
  \bibfield  {author} {\bibinfo {author} {\bibfnamefont {R.}~\bibnamefont
  {Niu}}, \bibinfo {author} {\bibfnamefont {Z.}~\bibnamefont {Li}}, \bibinfo
  {author} {\bibfnamefont {X.}~\bibnamefont {Han}}, \bibinfo {author}
  {\bibfnamefont {Z.}~\bibnamefont {Qu}}, \bibinfo {author} {\bibfnamefont
  {D.}~\bibnamefont {Ding}}, \bibinfo {author} {\bibfnamefont {Z.}~\bibnamefont
  {Wang}}, \bibinfo {author} {\bibfnamefont {Q.}~\bibnamefont {Liu}}, \bibinfo
  {author} {\bibfnamefont {T.}~\bibnamefont {Liu}}, \bibinfo {author}
  {\bibfnamefont {C.}~\bibnamefont {Han}}, \bibinfo {author} {\bibfnamefont
  {K.}~\bibnamefont {Watanabe}}, \bibinfo {author} {\bibfnamefont
  {T.}~\bibnamefont {Taniguchi}}, \bibinfo {author} {\bibfnamefont
  {M.}~\bibnamefont {Wu}}, \bibinfo {author} {\bibfnamefont {Q.}~\bibnamefont
  {Ren}}, \bibinfo {author} {\bibfnamefont {X.}~\bibnamefont {Wang}}, \bibinfo
  {author} {\bibfnamefont {J.}~\bibnamefont {Hong}}, \bibinfo {author}
  {\bibfnamefont {J.}~\bibnamefont {Mao}}, \bibinfo {author} {\bibfnamefont
  {Z.}~\bibnamefont {Han}}, \bibinfo {author} {\bibfnamefont {K.}~\bibnamefont
  {Liu}}, \bibinfo {author} {\bibfnamefont {Z.}~\bibnamefont {Gan}},\ and\
  \bibinfo {author} {\bibfnamefont {J.}~\bibnamefont {Lu}},\ }\bibfield
  {title} {\bibinfo {title} {Giant ferroelectric polarization in a bilayer
  graphene heterostructure},\ }\href
  {https://doi.org/10.1038/s41467-022-34104-z} {\bibfield  {journal} {\bibinfo
  {journal} {Nat. Commun.}\ }\textbf {\bibinfo {volume} {13}},\ \bibinfo
  {pages} {6241} (\bibinfo {year} {2022})}\BibitemShut {NoStop}%
\bibitem [{\citenamefont {Zhong}\ \emph {et~al.}(2023)\citenamefont {Zhong},
  \citenamefont {Cheng}, \citenamefont {Ren},\ and\ \citenamefont {Wu}}]{CPL}%
  \BibitemOpen
  \bibfield  {author} {\bibinfo {author} {\bibfnamefont {T.}~\bibnamefont
  {Zhong}}, \bibinfo {author} {\bibfnamefont {L.}~\bibnamefont {Cheng}},
  \bibinfo {author} {\bibfnamefont {Y.}~\bibnamefont {Ren}},\ and\ \bibinfo
  {author} {\bibfnamefont {M.}~\bibnamefont {Wu}},\ }\bibfield  {title}
  {\bibinfo {title} {Theoretical studies of sliding ferroelectricity,
  magnetoelectric couplings, and piezo-multiferroicity in two-dimensional
  magnetic materials},\ }\href {https://doi.org/10.1016/j.cplett.2023.140430}
  {\bibfield  {journal} {\bibinfo  {journal} {Chem. Phys. Lett.}\ }\textbf
  {\bibinfo {volume} {818}},\ \bibinfo {pages} {140430} (\bibinfo {year}
  {2023})}\BibitemShut {NoStop}%
\bibitem [{\citenamefont {Wu}\ and\ \citenamefont {Li}(2021)}]{PNAS21}%
  \BibitemOpen
  \bibfield  {author} {\bibinfo {author} {\bibfnamefont {M.}~\bibnamefont
  {Wu}}\ and\ \bibinfo {author} {\bibfnamefont {J.}~\bibnamefont {Li}},\
  }\bibfield  {title} {\bibinfo {title} {Sliding ferroelectricity in {2D} van
  der {Waals} materials: Related physics and future opportunities},\ }\href
  {https://doi.org/10.1073/pnas.211570311} {\bibfield  {journal} {\bibinfo
  {journal} {Proc. Natl. Acad. Sci. (USA)}\ }\textbf {\bibinfo {volume}
  {118}},\ \bibinfo {pages} {e2115703118} (\bibinfo {year} {2021})}\BibitemShut
  {NoStop}%
\bibitem [{\citenamefont {Chang}\ \emph {et~al.}(2020)\citenamefont {Chang},
  \citenamefont {K{\"u}ster}, \citenamefont {Miller}, \citenamefont {Ji},
  \citenamefont {Zhang}, \citenamefont {Sessi}, \citenamefont {Barraza-Lopez},\
  and\ \citenamefont {Parkin}}]{doi:10.1021/acs.nanolett.0c02357}%
  \BibitemOpen
  \bibfield  {author} {\bibinfo {author} {\bibfnamefont {K.}~\bibnamefont
  {Chang}}, \bibinfo {author} {\bibfnamefont {F.}~\bibnamefont {K{\"u}ster}},
  \bibinfo {author} {\bibfnamefont {B.~J.}\ \bibnamefont {Miller}}, \bibinfo
  {author} {\bibfnamefont {J.-R.}\ \bibnamefont {Ji}}, \bibinfo {author}
  {\bibfnamefont {J.-L.}\ \bibnamefont {Zhang}}, \bibinfo {author}
  {\bibfnamefont {P.}~\bibnamefont {Sessi}}, \bibinfo {author} {\bibfnamefont
  {S.}~\bibnamefont {Barraza-Lopez}},\ and\ \bibinfo {author} {\bibfnamefont
  {S.~S.~P.}\ \bibnamefont {Parkin}},\ }\bibfield  {title} {\bibinfo {title}
  {Microscopic manipulation of ferroelectric domains in {SnSe} monolayers at
  room temperature},\ }\href {https://doi.org/10.1021/acs.nanolett.0c02357}
  {\bibfield  {journal} {\bibinfo  {journal} {Nano Lett.}\ }\textbf {\bibinfo
  {volume} {20}},\ \bibinfo {pages} {6590} (\bibinfo {year}
  {2020})}\BibitemShut {NoStop}%
\bibitem [{\citenamefont {Rivera}(1994)}]{Rivera}%
  \BibitemOpen
  \bibfield  {author} {\bibinfo {author} {\bibfnamefont {J.-P.}\ \bibnamefont
  {Rivera}},\ }\bibfield  {title} {\bibinfo {title} {On definitions, units,
  measurements, tensor forms of the linear magnetoelectric effect and on a new
  dynamic method applied to {Cr-Cl} boracite},\ }\href
  {https://doi.org/10.1080/00150199408213365} {\bibfield  {journal} {\bibinfo
  {journal} {Ferroelectrics}\ }\textbf {\bibinfo {volume} {161}},\ \bibinfo
  {pages} {165} (\bibinfo {year} {1994})}\BibitemShut {NoStop}%
\end{thebibliography}

%

\end{document}